\documentclass[12pt]{article}
\usepackage[margin=2.5cm]{geometry}
\usepackage[authoryear]{natbib}
\usepackage{graphicx}
\input epsf.sty

\usepackage{color}
\usepackage{amsmath,amsfonts,amssymb,amsthm}
\usepackage{booktabs}
\newcommand{\bs}{\boldsymbol}

\usepackage[ruled]{algorithm2e}
\usepackage{xcolor}
\usepackage{multirow}

\usepackage{hyperref}
\makeatletter
\newcommand{\lrarrow}{\mathrel{\mathpalette\lrarrow@\relax}}
\newcommand{\lrarrow@}[2]{%
  \vcenter{\hbox{\ooalign{%
    $\m@th#1\mkern6mu\rightarrow$\cr
    \noalign{\vskip2pt}
    $\m@th#1\leftarrow\mkern6mu$\cr
  }}}%
}
\makeatother

\usepackage{listings}

\lstdefinestyle{script} {stepnumber=1, 
                          numbers=left,%
                          numberstyle={\tiny \color{gray}},
                          numbersep=4pt, 
                          frame = shadowbox,
                          framexleftmargin=15pt,
                          }
                          
\lstdefinestyle{command}{frame=single}

\lstdefinestyle{ScriptStyle} {language=Matlab,style=script}
\lstdefinestyle{CommandStyle} {language=Matlab,style=command}

\definecolor{dkgreen}{rgb}{0,0.6,0}
\definecolor{mauve}{rgb}{0.58,0,0.82}

\title{Model based clustering of multinomial count data
}
\author{Panagiotis Papastamoulis\\
Department of Statistics\\
Athens University of Economics and Business\\
Athens, Greece}
\date{}

\begin{document}
\maketitle

\begin{abstract}
We consider the problem of inferring an unknown number of clusters in replicated multinomial data. Under a model based clustering point of view, this task can be treated by estimating finite mixtures of multinomial distributions with or without covariates. Both Maximum Likelihood (ML) as well as Bayesian estimation are taken into account. Under a Maximum Likelihood approach, we provide an Expectation--Maximization (EM) algorithm which exploits a careful initialization procedure combined with a ridge--stabilized implementation of the Newton--Raphson method in the M--step. Under a Bayesian setup, a stochastic gradient Markov chain Monte Carlo (MCMC) algorithm embedded within a prior parallel tempering scheme is devised. The number of clusters is selected according to the Integrated Completed Likelihood criterion in the ML approach and estimating the number of non-empty components in overfitting mixture models in the Bayesian case. Our method is illustrated in simulated data and applied to two real datasets. An {\tt R} package is available at \url{https://CRAN.R-project.org/package=multinomialLogitMix}.
\end{abstract}

\abstract{We consider the problem of inferring an unknown number of clusters in multinomial count data, by estimating finite mixtures of multinomial distributions with or without covariates. Both Maximum Likelihood (ML) as well as Bayesian estimation are taken into account. Under a Maximum Likelihood approach, we provide an Expectation--Maximization (EM) algorithm which exploits a careful initialization procedure combined with a ridge--stabilized implementation of the Newton--Raphson method in the M--step. Under a Bayesian setup, a stochastic gradient Markov chain Monte Carlo (MCMC) algorithm embedded within a prior parallel tempering scheme is devised. The number of clusters is selected according to the Integrated Completed Likelihood criterion in the ML approach and estimating the number of non-empty components in overfitting mixture models in the Bayesian case. Our method is illustrated in simulated data and applied to two real datasets. The proposed methods are implemented in a contributed {\tt R} package, available online.}




\maketitle

\section{Introduction}\label{sec1}

Multinomial count data arise in various applications (see e.g.~\cite{yu2014efficient, nowicka2016drimseq}) and clustering them is a task of particular interest \citep{https://doi.org/10.1111/j.1467-842X.2004.00325.x,govaert2007clustering, portela2008clustering, 4407701, zamzami2020sparse, chen2020single}.  Finite mixture models \citep{mclachlan2004finite, Marin2005459, fruhwirth2006finite, fruhwirth2019handbook} are widely used for clustering heterogeneous datasets. Their applicability is extended beyond the model-based clustering framework, by also providing a means for semiparametric inference, see e.g.~\cite{morel1993finite}, where mixtures of multinomial distributions model extra multinomial variation in count data. 

In many instances, the resulting inference can be improved by taking into account the  presence of covariates, when available. Naturally, the framework of mixtures of multinomial logistic regressions \citep{grun2008identifiability} can be used for dealing with such data, under a model-based clustering point of view. These models belong to the broader family of mixtures of generalized linear models \citep{flexmix1, flexmix2, flexmix3}, which are estimated either under a maximum likelihood approach via the EM algorithm \citep{dempster1977maximum}, or in a Bayesian fashion using MCMC sampling \citep{doi:10.1080/01621459.1993.10476321, hurn2003estimating}. 

Various latent class models are based on mixtures of multinomial distributions. \cite{durante2019nested} cluster multivariate categorical data by estimating mixtures of products of multinomial distributions,  under the presence of covariates in the mixing proportions.  \cite{galindo2006avoiding} estimate latent class models using Bayesian Maximum A Posteriori estimation and illustrate via simulations that the Bayesian approach is more accurate than maximum likelihood estimation. More general latent class models based on multinomial distributions include hidden Markov models \citep{zuanetti2017generalized} and Markov random fields \citep{li2011spectral}.

In this paper, our goal is to cluster multinomial count data using finite mixtures of multinomial logistic regression models. Before proceeding we introduce some notation.  Let $\bs Y = (Y_1,\ldots,Y_J;Y_{J+1})^\top$ denote a random vector distributed according to a multinomial distribution  
\[\bs Y\sim\mathcal M_{J+1}(S, \bs\theta).\]
$S\in\mathbb Z_+$ corresponds to the number of independent replicates of the multinomial experiment, while the vector $\bs\theta = (\theta_1,\ldots,\theta_{J};\theta_{J+1})$, with $0<\theta_j< 1$ and $\sum_{j=1}^{J+1}\theta_j = 1$ contains the probabilities of observing each category. 

Under the presence of $K$ heterogeneous sub-populations in the multinomial experiment, we typically model the outcome using a finite mixture model as follows. Let $\bs Z=(Z_{1},\ldots,Z_{K})^{\top}\sim\mathcal M_{K}(1,\bs\pi)$
denote a latent multinomial random variable with $K$ categories, $\bs\pi=(\pi_1,\ldots,\pi_{K-1};\pi_K)$ is such that $0<\pi_{k}< 1$ and $\sum_{k=1}^{K}\pi_{k}=1$. Conditional on $Z_k = 1$ we assume that 
\[
\bs Y\vert Z_k = 1\sim\mathcal M_{J+1}(S, \bs\theta_k) 
\]
where $\bs\theta_k = (\theta_{k1},\ldots,\theta_{kJ};\theta_{k(J+1)})$, with $0<\theta_{kj}< 1$ and $\sum_{j=1}^{J+1}\theta_{kj} = 1$ contains the probabilities of observing each category for the corresponding multinomial experiment. It follows  that $\bs Y$ is drawn from a finite mixture of $K$ multinomial distributions, so the probability mass function of $\bs Y$ can be written as 
\begin{equation}
\sum_{k = 1}^{K}\pi_kf(\bs y\vert\bs\theta_k).
\end{equation}
The weights $\pi_1,\ldots,\pi_K$ correspond to the mixing proportions. Finally, $f(\cdot\vert\bs\theta_k)$ denotes the probability mass function of the $(J+1)$-dimensional multinomial distribution $\mathcal M_{J+1}(S,\bs\theta)$, that is,
\begin{equation}
\label{eq:mult}
f(\bs y\vert\bs\theta_k)=\dfrac{S!}{\prod_{j=1}^{J+1}y_{j}!}
\prod_{j=1}^{J+1}y_j^{\theta_{kj}}\mathrm{I}_{\mathcal Y_{S,J}}(\bs y),
\end{equation}
where 
$$\mathcal Y_{S,J}=\left\{y_1,\ldots,y_{J}\in\mathbb Z_+:0\leqslant \sum_{j\leqslant J} y_j\leqslant S; y_{J+1}:=S-\sum_{j\leqslant J}y_{j}\right\}$$ and $S\in\mathbb Z_+$. A necessary and sufficient condition for the generic identifiability of finite mixtures of multinomial distributions is the restriction $S\geqslant 2K - 1$ \citep{teicher1963, doi:10.1080/01621459.1964.10482176, titterington1985statistical, grun2008identifiability}. 

Given a vector of $P$ covariates $\bs x = (x_{1},\ldots,x_{P})$ and assuming that category $J+1$ is the baseline (in general, this can be any of the $J+1$ categories), we express the log-odds  as
\begin{equation}
\label{eq:logit}
\mathrm{logit}\theta_{j}=\log\frac{\theta_j}{\theta_{J+1}}= \bs\beta_j^\top\bs x,
\end{equation}
for $j = 1,\ldots,J$. The vector $\bs\beta_j = (\beta_{j1},\ldots,\beta_{jP})^\top\in\mathbb R^P$ contains the regression coefficients for category $j$. It follows from \eqref{eq:logit} that
\begin{equation}
\label{eq:eta}
\theta_{j} = \dfrac{\exp\{\bs\beta_{j}^\top\bs x\}}{1+\sum_{\ell\leqslant J}\exp\{\bs\beta_{\ell}^\top\bs x\}},
\end{equation}
for $j = 1,\ldots,J$.

Extending the previous model to the case of $K$ latent groups, Equation \eqref{eq:logit} becomes
\begin{equation}
\label{eq:logit_k0}
\mathrm{logit}\theta_{kj} = \bs\beta_{kj}^\top\bs x,
\end{equation}
for category $j = 1,\ldots,J$ and group-specific parameters $\bs\beta_{kj}=(\beta_{kj1},\ldots,\beta_{kjP})^\top$, $k = 1,\ldots,K$. In analogy to \eqref{eq:eta}, define
\begin{equation}
\label{eq:eta_k0}
\theta_{kj}= \begin{cases}
\dfrac{\exp\{\bs\beta_{kj}^\top\bs x\}}{1+\sum\limits_{\ell\leqslant J}\exp\{\bs\beta_{k\ell}^\top\bs x\}},&\quad j\leqslant J\\
\dfrac{1}{1+\sum\limits_{\ell\leqslant J}\exp\{\bs\beta_{k\ell}^\top\bs x\}},&\quad j = J+1
\end{cases}
\end{equation}
for  $k = 1,\ldots,K$.

We assume that we observe $n$ independent pairs $(\bs y_i, \bs x_i)$; $i = 1,\ldots,n$, where the joint-probability mass function of $\bs Y = (\bs Y_1,\ldots,\bs Y_n)^\top$ given $\bs x = (\bs x_1,\ldots,\bs x_n)^{\top}$  is written as 
\begin{align}
\nonumber
f(\bs y\vert\bs\pi, \bs \beta, \bs x) &= \prod_{i=1}^{n}f(\bs y_i\vert\bs\pi, \bs \beta, \bs x_i)\\
\label{eq:logit_mixture}
&=\prod_{i=1}^{n}\sum_{k=1}^{K}\pi_k \dfrac{S_i!}{\prod_{j=1}^{J+1}y_{ij}!}
\prod_{j=1}^{J+1}y_{ij}^{g_{ikj}}\mathrm{I}_{\mathcal Y_{S_i, J}}(\bs y_i).
\end{align}
where
\begin{equation}
\label{eq:eta_k}
g_{ikj}= \begin{cases}
\dfrac{\exp\{\bs\beta_{kj}^\top\bs x_i\}}{1+\sum\limits_{\ell\leqslant J}\exp\{\bs\beta_{k\ell}^\top\bs x_i\}},&\quad j\leqslant J\\
\dfrac{1}{1+\sum\limits_{\ell\leqslant J}\exp\{\bs\beta_{k\ell}^\top\bs x_i\}},&\quad j = J+1
\end{cases}
\end{equation}
for  $i = 1,\ldots,n$; $k = 1,\ldots,K$. In practice, $S_i$ is derived given $\bs y_i$, $i=1,\ldots,n$. 

The {\tt R} package {\tt mixtools} \citep{mixtools} can be used to estimate mixtures of multinomial distributions (among numerous other functionalities), under a maximum likelihood approach using the EM algorithm. However, the usage of covariates is not considered. On the other hand, the {\tt flexmix} package \citep{flexmix1, flexmix2, flexmix3} can estimate mixtures of multinomial logistic regression models using the {\tt FLXMRmultinom()} function, which also implements the EM algorithm. The package allows the user to run the EM algorithm repeatedly for different numbers of components and returns the maximum likelihood solution for each. However, alternative --and perhaps more efficient-- initialization schemes are not considered. Finally, a fully Bayesian implementation is not available in both packages.

The contribution of the present study it to offer an integrated approach to the problem of clustering multinomial count data using mixtures of multinomial logit models.  For this purpose we use frequentist as well as Bayesian methods. Both the EM algorithm (for the frequentist approach) as well as the MCMC sampler (for the Bayesian approach) are carefully implemented in order to deal with  various computational and inferential challenges imposed by the complex nature of mixture likelihoods/posterior distributions \citep{celeux2000computational}. At first, it is well known that the EM algorithm may converge to local modes of the likelihood surface. We tackle this problem by extending the initialization of the EM algorithm for mixture of Poisson regression models as suggested in \cite{papastamoulis2016estimation}. Second, we implement a ridge-stabilized version of the Newton-Raphson algorithm in the M-step. This adjustment is based on a quadratic approximation of the function of interest on a suitably chosen spherical region  and effectively avoids many of the pitfalls of standard Newton-Raphson iterations \citep{goldfeld1966maximization}. In the presented applications and simulation studies, our interest lies in cases where the multinomial data consists of a large number of replications for each multinomial observation. When the number of replicates is small, identifiability of the model is not guaranteed (see \cite{grun2008identifiability}).

Under a Bayesian   approach, traditional Bayesian methods estimate the number of clusters using the reversible jump MCMC \citep{green1995reversible,richardson1997bayesian} or the birth-death MCMC technique \citep{stephens2000bayesian}.  In multivariate settings, however, the practical application of these methods is limited. More recently, alternative Bayesian methods for estimating the number of clusters focus on the use of overfitting mixture models \citep{rousseau2011asymptotic}, information theoretic techniques which allow to post-process  MCMC samples of partitions to summarize the posterior in Bayesian clustering models \citep{10.1214/17-BA1073}, and  generalized mixtures of finite mixtures \citep{fruhwirth2021generalized}. Our Bayesian model combines recent advances on overfitting mixture models \citep{rousseau2011asymptotic, overfitting,papastamoulis2020clustering} with stochastic gradient MCMC sampling \citep{10.2307/3318418, doi:10.1080/01621459.2020.1847120} and running parallel MCMC chains which can exchange states. Moreover, we efficiently deal with the label switching problem, using the Equivalence Classes Representatives (ECR) algorithm \citep{Papastamoulis:10}. In such a way,  the returned MCMC output is directly interpretable and provides various summaries of marginal posterior distributions such as posterior means and Bayesian credible intervals of the parameters of interest.  

The combination of Maximum Likelihood and Bayesian estimation provides additional insights: it is demonstrated that the best-performing approach is to initialize the MCMC algorithm using information from the solution obtained in the EM implementation. Therefore, our proposed method provides a  powerful and practical approach that allows to easily estimate the unknown number of clusters and related parameters in  multinomial count datasets. 

The rest of the paper is organized as follows. Maximum likelihood estimation of finite mixtures  of multinomial distributions with or without covariates via the EM algorithm is described in Section \ref{sec:em}.  The careful treatment of the M-step  is extensively described in Section \ref{sec:mstep}. Section \ref{sec:em_init} discusses initialization issues in the EM implementation. Section \ref{sec:em_select} describes the selection of the number of clusters under the EM algorithm. The Bayesian formulation is described in Section \ref{sec:bayes}. The proposed MCMC sampler is introduced in Section \ref{sec:mcmc}. Section \ref{sec:overfitting} describes the estimation of the number of clusters using overfitting mixture models as well as how we deal with the label switching problem. Applications are illustrated in Section \ref{sec:app}. The paper concludes with a Discussion in Section \ref{sec:disc}. An Appendix contains further implementation details, additional simulation results and  comparisons with alternative approaches ({\tt flexmix}). 

\section{Maximum Likelihood estimation via the EM algorithm}\label{sec:em}

In this section we describe the Expectation--Maximization (EM) algorithm \citep{dempster1977maximum} for estimating mixtures of multinomial logistic regressions. 
For the case of covariates, the complete log-likelihood is written as
\begin{equation}
\label{eq:zFullCondPost}
\log f(\bs y\vert\bs\pi, \bs\beta,\bs x,\bs z) = \sum_{i=1}^{n}\sum_{k=1}^{K}z_{ik}\left\{\log \pi_k + \log c_i + \sum_{j = 1}^{J+1}y_{ij}\log g_{ikj}\right\},
\end{equation}
where $c_i = S_i!/\prod_{j=1}^{J+1}y_{ij}!$.

The EM algorithm proceeds by computing the expectation of the complete log-likelihood (see Section \ref{sec:cll} with respect to the latent allocation variables $\bs Z$ (given $\bs y$ and $\bs x$). Then, the expected complete log-likelihood is maximized with respect to the parameters $\bs \pi, \bs \beta$ (see Section \ref{sec:mstep}), given the current expected values of missing data. In the case of mixtures of multinomial logistic regressions this task can become quite challenging, since typical numerical implementations (such as the standard Newton-Raphson algorithm) may fail. For this reason, it is crucial to apply more robust numerical implementations \citep{goldfeld1966maximization} as discussed in Section  \ref{sec:mstep}. In Section \ref{sec:em_init} special attention is given to the important issue of initialization of the EM algorithm. Section \ref{sec:em_select} describes the selection of the number of clusters under the EM algorithm.

\subsection{Expectation step}\label{sec:cll}

The expectation step (E-step) consists of evaluating the expected complete log-likelihood, with respect to the conditional distribution of $\bs Z$ given the observed data $\bs y$ (and $\bs x$ in the covariates case), as well as  a current estimate of the parameters $(\bs\pi^{(t)},\bs\beta^{(t)})$. Define the posterior membership probabilities $w_{ik}$ as
\[
w_{ik} = \mathrm{P}(Z_{ik} = 1\vert\bs y_i, \bs x_i, \bs\pi,\bs \beta)= \frac{\pi_k f(\bs y_i\vert\bs g_{ik})}{\sum_{\ell=1}^{K}\pi_\ell f(\bs y_i\vert\bs g_{i\ell})},\quad i = 1,\ldots,n; k = 1,\ldots,K.
\]
Note that, according to the Maximum A Posteriori rule, the estimated clusters are obtained as
\[
c_i = \mathrm{argmax}_{k\in\{1,\ldots,K\}}\{w_{ik}; k = 1,\ldots,K\},\quad i =1,\ldots,n.
\]

The expected complete log-likelihood is equal to
\begin{align}
\nonumber
Q(\bs\pi,\bs\beta\vert\bs\pi^{(t)},\bs\beta^{(t)})&:=\mathrm{E}_{\bs Z\vert\bs y, \bs x, \bs\pi^{(t)}, \bs \beta^{(t)}}\left\{\log f(\bs y\vert\bs\pi, \bs\beta,\bs x,\bs Z)\right\} \\
&=\sum_{i=1}^{n}\sum_{k=1}^{K}w_{ik}\left\{\log \pi_k + \log c_i + \sum_{j = 1}^{J+1}y_{ij}\log g_{ikj}\right\}
\label{eq:complete_logistic}
\end{align}
where the current parameter values $(\bs\pi^{(t)},\bs\beta^{(t)})$ are used to compute $w_{ik}$.


\subsection{Maximization step}\label{sec:mstep}

In the maximization step (M-step), \eqref{eq:complete_logistic} is maximized with respect to the parameters $\bs\pi,\bs\beta$, that is,
\[
\left(\bs\pi^{(t+1)},\bs\beta^{(t+1)}\right)=
\underset{\bs\pi,\bs\beta}{\mbox{argmax}}Q(\bs\pi,\bs\beta\vert\bs\pi^{(t)},\bs\beta^{(t)})
\]

The maximization of the expected complete log-likelihood with respect to the mixing proportions leads to 
\[
\pi_{k}^{(t+1)} = \frac{1}{n}\sum_{i=1}^{n}w_{ik},\quad k = 1,\ldots,K.
\] 

The maximization with respect to $\bs\beta$ is analytically tractable only when $P = 1$ (that is, a model with just a constant term). Recall that when no covariates are present then the model is reparameterized with respect to the multinomial probabilities, that is,
\[
\theta_{kj} = \frac{e^{\beta_{kj}}}{1+e^{\beta_{kj}}}.
\]
The expected complete log-likelihood is maximized with respect to $\theta_{kj}$. The analytical solution of the M-step in this case is 
\begin{equation}
\theta_{kj}^{(t+1)} = \frac{\sum_{i=1}^{n}w_{ik}y_{ij}}{\sum_{i=1}^{n}w_{ik}S_i},
\end{equation}
for $k = 1,\ldots,K$ and $j = 1,\ldots,J+1$.

In case where $P\geqslant 2$ numerical methods are implemented. We have used two optimization techniques: the typical Newton-Raphson algorithm, as well as a ridge-stabilized version introduced by \cite{goldfeld1966maximization}.  It is easy to show that the partial derivative of \eqref{eq:complete_logistic} with respect to $\beta_{kjp}$ is
\begin{equation}
\label{eq:derivative}
\frac{\partial Q}{\partial\beta_{kjp}}=\sum_{i=1}^{n}w_{ik}\left\{y_{ij}-S_ig_{ikj}\right\}x_{ip},
\end{equation}
$k = 1,\ldots,K$, $j = 1,\ldots,J$ and $p = 1,\ldots,P$. Thus, the gradient vector can be expressed as
\begin{equation}
\label{eq:gradient}
\nabla Q(\bs\beta):= \left(
\sum_{i=1}^{n}w_{i1}\left\{\bs y_i - S_i\bs g_{i1}\right\}\otimes\bs x_i,\ldots,
\sum_{i=1}^{n}w_{iK}\left\{\bs y_i - S_i\bs g_{iK}\right\}\otimes\bs x_i,
\right)^\top,
\end{equation}
where $\otimes$ denotes the Kronecker product and we have also defined $\bs g_{ik}:=(g_{ik1},\ldots,g_{ikJ})^\top$, $k = 1,\ldots,K$.

The second partial derivative of the log-likelihood function \eqref{eq:complete_logistic} with respect to $\beta_{kjp}$ and $\beta_{k'j'p'}$ is
\[
\frac{\partial^2 Q}{\partial\beta_{kjp}\partial\beta_{k'j'p'}} = -\delta_{kk'}\sum_{i=1}^{n}S_iw_{ik}x_{ip}x_{ip'}g_{ikj}(\delta_{jj'}-g_{ikj'}),
\]
where $\delta_{ij}$ denotes the Kronecker delta, for $k, k' = 1,\ldots,K$, $j, j' = 1,\ldots,J$ and $p,p' = 1,\ldots,P$. Note that the corresponding Hessian $$H(\bs\beta)=\begin{pmatrix}
H_1(\bs\beta_1) & 0 & \ldots & 0\\
0 & H_2(\bs\beta_2) &\ldots & 0\\
\vdots & \vdots &\ddots & \vdots\\
0 & 0 &\ldots & H_K(\bs\beta_K)\\
\end{pmatrix}$$ is a block diagonal matrix consisting of $K$ blocks $H_k$, where each one of them being a $JP\times JP$-dimensional matrix, with $$H_{k} =\left\{ \frac{\partial^2 Q}{\partial\beta_{kjp}\partial\beta_{kj'p'}}\right\}_{j=1,\ldots,J; p=1,\ldots,P}.$$ This is particularly useful because the inverse of this $KJP\times KJP$-dimensional matrix is the corresponding block diagonal matrix of the inverse matrices, that is,
 $$H^{-1}(\bs\beta)=\begin{pmatrix}
H^{-1}_1(\bs\beta_1) & 0 & \ldots & 0\\
0 & H^{-1}_2(\bs\beta_2) &\ldots & 0\\
\vdots & \vdots &\ddots & \vdots\\
0 & 0 &\ldots & H^{-1}_K(\bs\beta_K)\\
\end{pmatrix}.$$
Consequently, the Newton-Raphson update can be performed independently for each $k = 1,\ldots,K$, as described in the sequel. 

In order to maximize the expected complete log-likelihood with respect to $\bs\beta$ we used a ridge-stabilized version \citep{goldfeld1966maximization} of the Newton-Raphson algorithm. Denote by $\bs\beta^{(t,1)}$ the initial value of $\bs\beta$ at the M-step of iteration $t$ of the EM algorithm. Then,  the typical Newton-Raphson update at the $m+1$-th iteration takes the form
\begin{equation}\label{eq:nr}
\bs\beta^{(t,m+1)} = \bs\beta^{(t,m)} -  H^{-1}(\bs\beta^{(t,m)})\nabla Q(\bs\beta^{(t,m)}),\quad m = 1,2,\ldots.
\end{equation}
Let $M$ denote the last iteration of the sequence of Newton-Raphson updates. The updated value of $\bs\beta$ for iteration $t$ of the EM algorithm is then equal to 
\[
\bs\beta^{(t)} := \bs\beta^{(t,M)}.
\]

In case that a second-order Taylor expansion  is a good approximation of the underlying function around a maximum, the Newton-Raphson method will converge rapidly \citep{crockett1955gradient}. However, in general settings, it may happen that the step of the basic update in Equation \eqref{eq:nr} will be too large, or $-H$ will be negative definite, in which case the quadratic approximation has no validity. 

The following technique addresses these issues by maximizing a quadratic approximation to the function on a suitably chosen spherical region.  The algorithm of \cite{goldfeld1966maximization} is based on the updates
\begin{equation}\label{eq:nr_ridge}
\bs\beta^{(t,m+1)} = \bs\beta^{(t,m)} -  H_\alpha^{-1}(\bs\beta^{(t,m)})\nabla Q(\bs\beta^{(t,m)}),\quad m = 1,2,\ldots,
\end{equation}
where, 
\begin{align}
\label{eq:ridge_step}
\alpha &= \lambda_1+R\lvert\lvert\nabla Q(\bs\beta^{t-1})\rvert\rvert\\
H_\alpha(\bs\beta) &=\begin{cases}
H(\bs\beta) - \alpha I,&\quad\mbox{if }\alpha > 0\\
H(\bs\beta),&\quad\mbox{if }\alpha \leqslant 0,
\end{cases}
\end{align}
while $\lambda_1$ and $\lvert\lvert x\rvert\rvert$ denote the largest eigenvalue of $H$ and the length of vector $x$, respectively.
The parameter $R$ controls the step size of the update: smaller values result to larger step sizes. This parameter is adjusted according to the procedure described \cite{goldfeld1966maximization}: the step size tends to increase when the quadratic approximation appears to be satisfactory. 

\begin{figure}[t]
\centering
\includegraphics[scale=0.65]{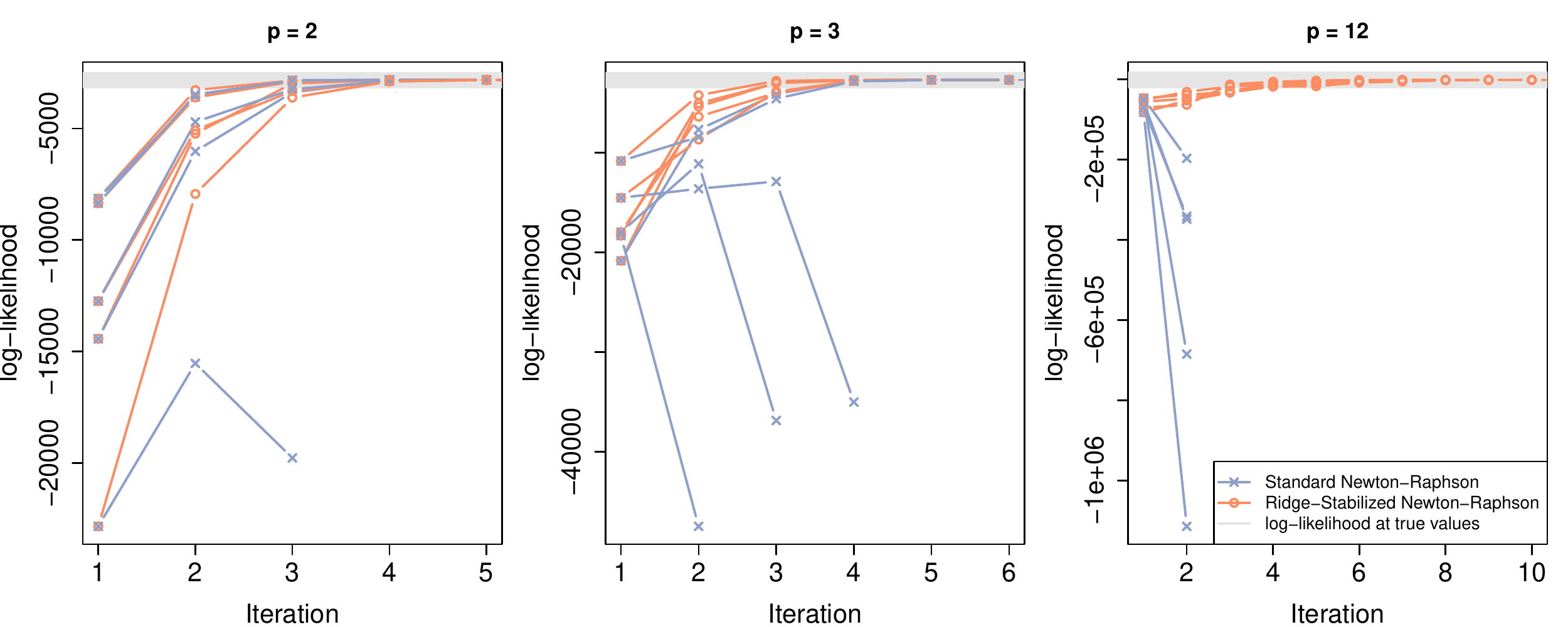}
\caption{Estimation of a typical ($K=1$) multinomial logit model with $D = 6$ categories and $p$ covariates (including constant term): Log-likelihood values per iteration  of the standard Newton-Raphson algorithm and the ridge-stabilized version, based on 5 random starting values. }
\label{fig:nr}
\end{figure}

Figure \ref{fig:nr} illustrates the two algorithms using simulated data of $n=250$  observations from a typical ($K = 1$) multinomial logit model with $D = 6$ categories and varying number of explanatory variables $p$. In each case, the same random starting value was used for both the standard Newton-Raphson as well as the modified version. Observe that, especially as the number of parameters increases, the standard Newton-Raphson updates may decrease the log-likelihood function. On the other hand, the ridge stabilized version produces a sequence of updates which converge to the mode of the log-likelihood function (as indicated by the gray line). 

\subsection{EM Initialization}\label{sec:em_init}
Careful selection of initial values for the EM algorithm is crucial  \citep{biernacki2003choosing, karlis2003choosing, baudry2015mixtures, papastamoulis2016estimation} in order to avoid convergence to minor modes of the likelihood surface. Following \cite{papastamoulis2016estimation}, a small-EM  \citep{biernacki2003choosing} procedure is used. A small-EM initialization refers to the strategy of starting the main EM algorithm from values arising by  a series of short runs of the EM. Each run consists of a small number of  iterations (say 5 or 10), under different starting values. The selected values that will be used to initialize the main EM algorithm are the ones that correspond to the largest log-likelihood across all small-EM runs. The starting values of each small-EM are selected according to three alternative strategies, namely: \textit{random}, \textit{split} and \textit{shake} small-EM schemes, described in detail in the sequel.

In what follows, we will use the notation 
\begin{equation}
\label{eq:postProb}
\widehat{w}^{(K)}_{ik} = \mathrm{P}(Z_{ik} = 1\vert\bs y, \bs x, \widehat{\bs \pi}, \widehat{\bs \beta}), \quad k = 1,\ldots,K; i = 1,\ldots,n.
\end{equation}
in order to explicitly refer to the estimated membership probabilities arising from a mixture model with $K$ components, where $\widehat{\bs\pi}$ and $\widehat{\bs\beta}$ are the parameter estimates obtained at the last iteration of the EM algorithm.

\paragraph{Random small-EM} This strategy corresponds to the random selection of $M_{\mathrm{random}}$ starting values and running the EM for a small number (say $T = 5$ or 10) iterations. The parameters of the run which  results to the largest log-likelihood value in the last ($T$-th) iteration are used to initialize the main EM algorithm. The random selection can refer to  either choosing random values for the coefficients of the multinomial logit model or for the posterior membership probabilities. The latter scheme is followed in our approach, in particular each row of the $n\times K$ matrix of posterior probabilities is generated according to the $\mathcal U(0,1)$ distribution. Each row is then normalized according to the unity sum constraint.

\paragraph{Split small-EM}

\begin{figure}[t]
\centering
\includegraphics[scale=0.65]{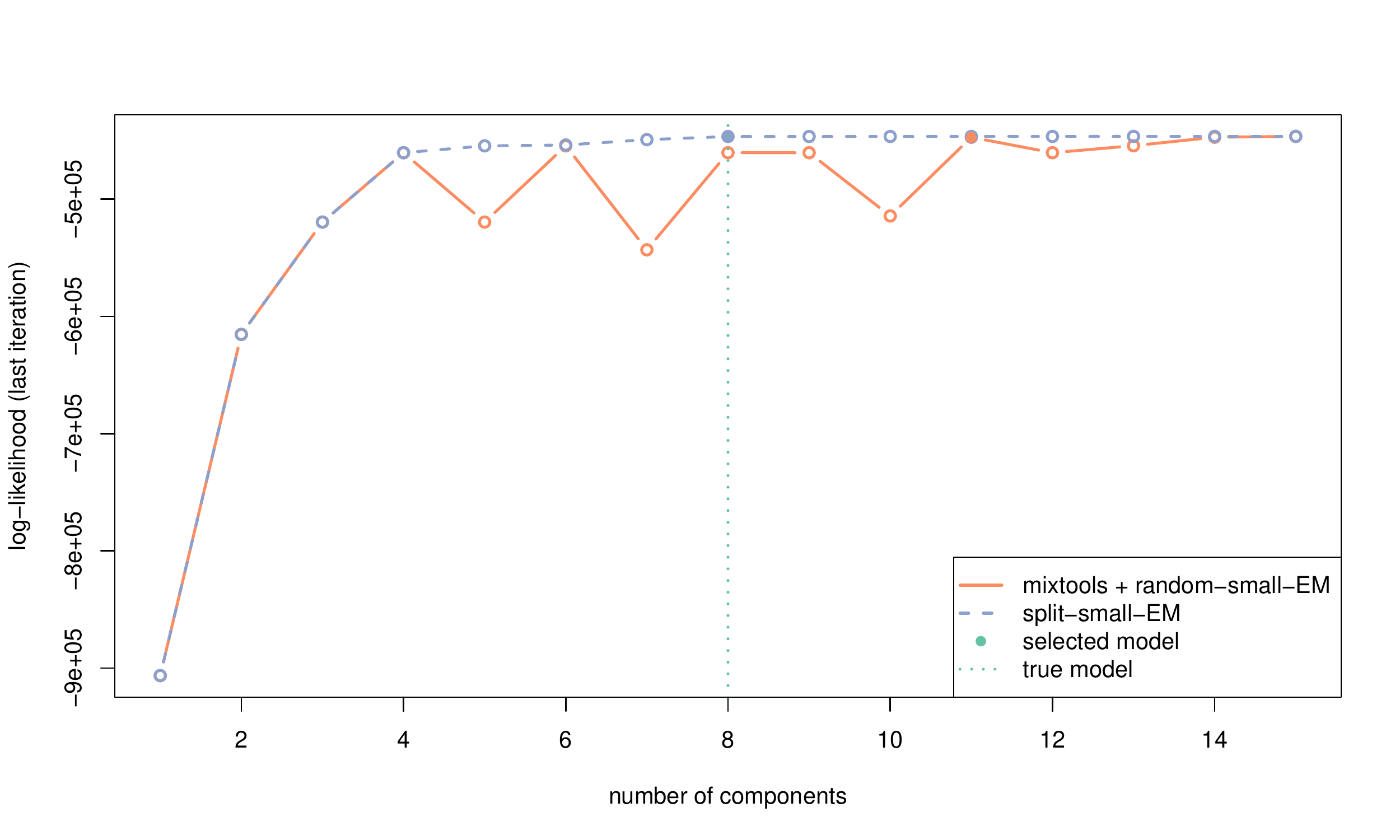}
\caption{Estimation of a multinomial  mixture (without covariates) with $K=8$ components: Log-likelihood values obtained at the last iteration of the EM algorithm for each value of $K$. The information criterion used to select $K$ is the ICL. The small-EM scheme details are: $M_{\mathrm{split}}=M_{\mathrm{random}}=10$ repetitions, each one consisting of $T = 5$ iterations.} 
\label{fig:logL}
\end{figure}

\cite{fraley2005incremental, papastamoulis2016estimation} proposed to  begin the EM algorithm from a model that underestimates the number of clusters and consecutively adding one component using a splitting procedure among the previously estimated clusters. In our setup, this procedure begins with estimating the one-component ($K = 1$) mixture model. Then, for $g = 2,\ldots,K$, we estimate a $g$-component mixture by proposing to randomly split clusters obtained by the estimated model corresponding to $g-1$ components. The way that clusters are split is determined by a random transformation of the estimated posterior classification probabilities $\widehat{w}^{(g)}_{ik}$, defined in Equation \eqref{eq:postProb}. Given $\widehat{w}^{(g-1)}_{ik}$, denote by $I_1,\ldots,I_{g-1}$ the clusters obtained applying the Maximum A Posteriori rule on the estimated model with $g-1$ components. First, a non-empty component $I_{g^\star}$ is chosen at random among $\{I_1,\ldots,I_{g-1}\}$. Second, a new  component labelled as $g$ is formed by splitting the selected cluster $I_{g^\star}$ into two new ones, via a random transformation of the estimated posterior probabilities:
\begin{align*}
\widehat w_{ik}^{(g)} &= \widehat w_{ik}^{(g-1)},\quad k\notin\{g^\star,g\}\\
\widehat w_{ig^\star}^{(g)} &= u_i\widehat{w}_{ig^{\star}}^{(g-1)}\\
\widehat w_{ig}^{(g)} &= (1 - u_i)\widehat{w}_{ig^{\star}}^{(g-1)},
\end{align*}
where $u_i\sim \mathcal \mathrm{Beta}(a,b)$, for $i = 1,\ldots,n$, with $\mathrm{Beta}(a,b)$ denoting the Beta distribution with parameters $a>0$ and $b>0$. In our examples we have used $a=b=1$, that is,  a Uniform distribution in $(0,1)$. Another valid option would be to set $a=b < 1$ in order to enforce greater cluster separation. Finally, the EM algorithm for a mixture with $g$ components starts by plugging in $\{w_{ik}^{(g)}, i =1,\ldots,n;k = 1,\ldots,g\}$ as starting values for the posterior membership probabilities. This procedure is repeated $M_\mathrm{split}$ times by running small EM algorithms and the one resulting to largest log-likelihood value is chosen to start the main EM for model $g$. We will refer to this strategy as a split-small-EM initialization scheme. A comparison between the random-small-EM strategy using {\tt mixtools} \citep{mixtools} and the split-small-EM scheme for a mixture of 8 multinomial distributions is shown in Figure \ref{fig:logL}. More detailed comparisons between the random small-EM initializations are reported in the simulation study of Section \ref{sec:sim} and in Appendix \ref{sec:flexmix}.

\paragraph{Shake small-EM}

Assume that there are at least $K\geqslant 2$  clusters in the fitted model and that the estimated posterior membership probabilities are equal to $\widehat w^{(K)}_{ik}$, $i=1,\ldots,n$, $k = 1,\ldots,K$. We randomly select 2 of them (say $k_1$ and $k_2$) and propose to randomly re-allocate the assigned observations within those 2 clusters. More specifically, let $I_{k_1}$ and $I_{k_2}$ denote the observations assigned (according to the MAP rule) to clusters $k_1$ and $k_2$, respectively. A small-EM algorithm is initialized by a state which uses a matrix $(\widehat w^{'(K)}_{ik})$ obtained by a random perturbation of the posterior probabilities as follows
\begin{eqnarray*}
\widehat w^{'(K)}_{i{k}} &=& \widehat w^{(K)}_{ik},\quad k\notin \{k_1,k_2\} \\
\widehat w^{'(K)}_{i{k_1}} &=& u_i (w_{ik_{1}}+w_{ik_{2}})\\
\widehat w^{'(K)}_{i{k_2}} &=& (1-u_i) (w_{ik_{1}}+w_{ik_{2}}).
\end{eqnarray*}
Note that in this way only the posterior probabilities of those observations assigned to components $k_1$ and $k_2$ are affected. 
This procedure is repeated $M_{\mathrm{shake}}$ times
and the one leading to the highest log-likelihood value after $T$ EM iterations is selected in order to initialize the algorithm. We will refer to this strategy as a shake small-EM initialization. 

The aforementioned initialization schemes will be compared in our simulation study in Section \ref{sec:sim} (see also Appendix \ref{sec:flexmix}). We will use the notation $$\mbox{EM}\left(M_{\mathrm{split}},  M_{\mathrm{shake}},  M_{\mathrm{random}}\right)$$  to refer to a small-EM algorithm initialization consisting of $M_{\mathrm{split}}$ split-small-EM rounds, which are then followed by a sequence of $M_{\mathrm{shake}}$ shake-small-EM rounds and $M_{\mathrm{random}}$ random-small-EM rounds.

\subsection{Estimation of the number of clusters under the EM algorithm}\label{sec:em_select}

There is a plethora of techniques in order to select the number of components in a mixture model, see e.g.~Chapter 6 in \cite{mclachlan2004finite}. One of the most popular choices is the Bayesian Information Criterion \citep{schwarz1978estimating}, defined as 
\[
\mathrm{BIC}(K) = -2\log f(\bs y\vert\bs x, \widehat{\bs\theta}_K) + d_K\log n,
\]
where $\widehat{\bs\theta}_K$ and $d_K$ denote the Maximum Likelihood estimate and the number of parameters of the mixture model with $K$ components, respectively. Another criterion which is particularly suited to the task of model-based clustering is the Integrated Complete Likelihood (ICL) criterion \citep{biernacki2000assessing}. 
\[
\mathrm{ICL}(K) = -2\log f(\bs y\vert\bs x, \widehat{\bs\theta}_K) + d_K\log n -2 \sum_{i = 1}^{n}\sum_{k=1}^{K}\widehat{w}_{ik}\log\widehat{w}_{ik}.
\]

It has been demonstrated that BIC may overestimate the number of clusters (see e.g.~\cite{10.1093/bioinformatics/btu845, papastamoulis2016estimation}). In what follows, the number of clusters in the EM approach is selected according to the ICL criterion.

\section{Bayesian Formulation}\label{sec:bayes}

We assume that the mixing proportions of the mixture model \eqref{eq:logit_mixture} follow a Dirichlet prior distribution, that is,
\begin{equation}
\label{eq:dirichlet}
\bs\pi\sim\mathcal D(\alpha_1,\ldots,\alpha_K)
\end{equation}
for some fixed hyper-parameters $\alpha_k >0$, $k = 1,\ldots,K$. Usually, there is no prior information which separates the components between each other so typically \citep{Marin2005459} we set $\alpha_1=\ldots=\alpha_K = \alpha > 0$ (see  also Section \ref{sec:overfitting}).

The prior distribution of the coefficients $\beta_{kjp}$ is normal centered on zero, that is,
\begin{equation}\label{eq:betaPrior}
\beta_{kjp}\sim \mathcal N(0,\nu^2), \quad\mbox{independent for}\quad k=1,\ldots,K; j= 1,\ldots,J, p = 1,\ldots,P.
\end{equation}
The prior variance $\nu^2$ is assumed constant. A default value of $\nu^2 = 100$ was used in all of all our examples presented in subsequent sections, which corresponds essentialy to a vague\footnote{This depends on the scale of the covariates, but in our simulations we are using standardized values in all cases.} prior distribution, however  we will also consider more informative choices ($\nu = 1$), in order to penalize large values of the coefficients. 
We furthermore assume that $\bs \beta$, $\bs\pi$ are a-priori independent random variables, thus the joint prior distribution of the parameters and latent allocation variables is written as
\[
f(\bs z, \bs\pi,\bs\beta\vert K, \bs\alpha, \nu) =
f(\bs z\vert\bs\pi,K)f(\bs\pi\vert K,\bs\alpha)f(\bs\beta\vert K,\nu).
\]

The joint posterior distribution of $\bs z, \bs\pi,\bs\beta\vert \bs y,\bs x, K$ is written as
\[
f(\bs z, \bs\pi,\bs\beta\vert \bs y,\bs x, K, \bs\alpha, \tau) \propto 
f(\bs y\vert \bs x, \bs z,\bs\beta,K)f(\bs z\vert \bs\pi, K)f(\bs\pi\vert K, \bs\alpha)f(\bs\beta\vert K,\nu)
\]

\subsection{A Hybrid Metropolis-Adjusted-Langevin within Gibbs MCMC Algorithm}\label{sec:mcmc}

From Equation \eqref{eq:zFullCondPost} follows that the full conditional posterior distribution of the latent allocation vector for observation $i$ is
\begin{equation}\label{eq:z_fc}
\bs Z_{i}\vert \cdots\sim\mathcal M(1; w_{i1},\ldots,w_{iK}),  
\end{equation}
independent for $i = 1,\ldots,n$.

The full conditional posterior distribution of mixing proportions is a Dirichlet distribution with parameters
\begin{equation}\label{eq:w_fc}
\bs \pi\vert \cdots \sim \mathcal D(\alpha_1 + n_1,\ldots,\alpha_K+n_K), 
\end{equation}
where $n_k = \sum_{i=1}^{n}z_{ik}$.

For the regression coefficients we use a Metropolis--Hastings step,  although other approaches  which are based on the Gibbs sampler have been proposed \citep{dellaportas1993bayesian, holmes2006bayesian, gramacy2012}. Note however that these approaches impose additional augmentation steps in the hierarchical model and have been applied only for simple (that is, $K = 1$) logistic regression models.  

One could use a random walk for proposing updates to $\bs\beta$, but it is well known that the large number of parameters would lead to slow-mixing and poor convergence of the MCMC sampler. In order to overcome this issue, we used a proposal distribution which is based on the gradient information of the full conditional distribution. The Metropolis Adjusted Langevin Algorithm (MALA) \citep{10.2307/3318418, https://doi.org/10.1111/1467-9868.00123, girolami2011riemann} is based on the following proposal mechanism 
\begin{equation}
\label{eq:proposal}
\widetilde{\bs\beta} = \bs\beta^{(t)} + \tau\nabla\log f(\bs\beta^{(t)}\vert \bs y, \bs x, \bs z, \bs \pi) + \sqrt{2\tau}\bs\varepsilon, 
\end{equation}
where $\bs\varepsilon\sim\mathcal N(\bs 0,\mathrm{\bs I})$ and $\nabla\log f(\bs\beta^{(t)}\vert \bs y, \bs x, \bs z, \bs \pi)$ denotes the gradient vector of the logarithm of the full conditional of $\bs\beta$, evaluated at $\bs\beta^{(t)}$. In order to select a value of $\tau$ with a reasonable acceptance rate betweeen proposed moves the MCMC sampler runs for an initial warm-up period. During this period $\tau$ is adaptively tuned as the MCMC sampler progresses in order to achieve acceptance rates of the proposed updates between user-specified limits (see Appendix \ref{sec:mcmc_details} for details). The final value of $\tau$ is then selected as the one that will be used in the subsequent main MCMC sampler. 

The derivative of the logarithm of the joint posterior distribution of $\bs \beta$, conditional on $\bs z$ and $\bs \pi$ is equal to 
\begin{equation}\label{eq:posterior_derivative}
\frac{\partial \log f(\bs\beta\vert \bs y, \bs x, \bs z, \bs \pi)}{\partial\beta_{kjp}} = \sum_{i=1}^{n}z_{ik}(y_{ij} - S_ig_{ikj})x_{ip} - \frac{\beta_{kjp}}{\nu^2}
\end{equation}
Note that the first term on the right-hand side of the previous expression corresponds to the log-derivative of the complete log-likelihood (that is, given $\bs z$), while the second term corresponds to the derivative of the prior distribution in \eqref{eq:betaPrior}. 

The proposal in \eqref{eq:proposal} is accepted according to the usual Metropolis-Hastings probability, that is,
\begin{align}
\label{eq:map}
\alpha(\bs\beta^{(t)}, \widetilde{\bs\beta}\vert \bs z^{(t)},\bs\pi^{(t)}) = \min\left\{1, 
\frac{f(\bs y\vert \bs x, \bs z^{(t)}, \widetilde{\bs \beta}, \bs \pi^{(t)})\pi(\widetilde{\bs\beta})}{f(\bs y\vert\bs x, \bs z^{(t)}, \bs \beta^{(t)}, \bs \pi^{(t)})\pi(\bs\beta^{(t)})}
\frac{
\mathrm{P}\left(\widetilde{\bs\beta}\rightarrow\bs\beta^{(t)}\right)
}{
\mathrm{P}\left(\bs\beta^{(t)}\rightarrow\widetilde{\bs\beta}\right)
}
\right\},
\end{align}
where $\mathrm{P}\left(a\rightarrow b\right)$ denotes the probability density of proposing state $b$ while in $a$.  From \eqref{eq:proposal} we have that $\mathrm{P}\left(\bs\beta^{(t)}\rightarrow\widetilde{\bs\beta}\right)$ is the density of the $$\mathcal N_{KJP}\left(\bs\beta^{(t)} + \tau\nabla\log f(\bs\beta^{(t)}\vert\bs y, \bs x, \bs z^{(t)}, \bs \pi^{(t)}), 2\tau I_{KJP}\right)$$
distribution, evaluated at $\widetilde{\bs\beta}$. The density of the reverse transition $\left(\widetilde{\bs\beta}\rightarrow\bs\beta^{(t)}\right)$ is equal to the density of the distribution 
$$\mathcal N_{KJP}\left(\widetilde{\bs\beta} + \tau\nabla\log f(\widetilde{\bs\beta}\vert\bs y, \bs x, \bs z^{(t)}, \bs \pi^{(t)}), 2\tau I_{KJP}\right)$$
evaluated at $\bs\beta^{(t)}$.

The overall procedure is summarized at Algorithm \ref{alg:mala}.

\begin{algorithm*}[p]
\colorbox{gray!25}{\parbox{0.86\textwidth}{

\caption{Metropolis-Adjusted Langevin Within Gibbs MCMC}
\SetKwInOut{Input}{Input}
\SetKwInOut{Output}{Output}
\SetKwBlock{blocknotext}{~}{end}
\SetKwBlock{step}{Step}{~}
\SetKwBlock{stepo}{Step 0:  Initialization}{~}
\SetKwBlock{stepi}{Step 1:  Gibbs sampling for $\bs Z$}{~}
\SetKwBlock{stepii}{Step 2:  Gibbs sampling for $\bs \pi$}{~}
\SetKwBlock{stepiii}{Step 3:  MALA proposal for $\bs \beta$}{~}
\SetKwBlock{stepiv}{Step 4:  While $t < M$}{~}
\SetKwFor{For}{for}{~}{endfor}
\SetKwFor{ForAll}{for~all}{~}{endfor}
\SetKwFor{ForEach}{for~each}{~}{endfor}

\Input{data $\bs y, \bs x$\\ 
number of components $K$\\ 
prior hyper-parameters $\nu^2,\bs \alpha$\\
scale of the MALA proposal $\tau>0$\\
number of MCMC iterations $M$\\
optional vector of starting values $(\bs \pi^{(0)}, \bs z^{(0)}, \bs \beta^{(0)})$}
\Output{
MCMC sample $\{\bs z^{(t)}, \bs \pi^{(t)}, \bs \beta^{(t)}; t = 1,\ldots,M\}$}
and acceptance rate $r$ of the MALA proposal. 

\stepo{
Initialize acceptance rate counter: $r = 0$.\\
\SetAlgoVlined\If{Starting values not supplied}{
Obtain random starting values $(\bs \pi^{(0)}, \bs z^{(0)}, \bs \beta^{(0)})$}
}
\For{$t=1$ \KwTo $M$}{
\stepi{
\For{$i=1$ \KwTo $n$}{
Simulate $\bs z^{(t)}_i$ from $\bs Z_i\vert\bs y, \bs x, \bs 
\beta^{(t-1)}, \bs \pi^{(t-1)}$ in Equation \eqref{eq:z_fc}
}
}
\stepii{
Simulate $\bs\pi^{(t)}$ from  $\bs\pi\vert\bs z^{(t)}$ in Equation \eqref{eq:w_fc}
}
\stepiii{
\begin{enumerate}
\item[3.1] compute $\nabla\log f(\bs\beta^{(t-1)}\vert\bs y, \bs x, \bs z^{(t)}, \bs \pi^{(t)})$ according to \eqref{eq:posterior_derivative}

\item[3.2] propose $\widetilde{\bs\beta}$ according to \eqref{eq:proposal}

\item[3.3] Compute $\alpha(\bs\beta^{(t-1)},\widetilde{\bs\beta}\vert\bs z^{(t)}, \bs\pi^{(t)})$ in \eqref{eq:map}

\item[3.4] generate $u\sim\mathcal U(0,1)$ 
\end{enumerate}

\SetAlgoVlined\eIf{$u < \alpha(\bs\beta^{(t-1)},\widetilde{\bs\beta}\vert\bs z^{(t)}, \bs\pi^{(t)})$}{set $\bs\beta^{(t)} = \widetilde{\bs\beta}$  and $r \leftarrow r+1$}{set $\bs\beta^{(t)} =\bs\beta^{(t-1)}$}

}
}
Set $r\leftarrow r/M$.\\

END     of algorithm
\label{alg:mala}
}}
\end{algorithm*}

\subsection{Estimation of the number of clusters using overfitting Bayesian mixtures}\label{sec:overfitting}

The Bayesian setup allows to estimate the number of clusters by using overfitting mixture models, that is, models where the number of mixture components is much larger than the number of clusters. Let $K_{\max} > K$ denote an upper bound on the number of clusters and define the overfitting mixture model
\[
f(\bs y\vert\bs\theta,K_{\max})=\sum_{k=1}^{K_{\max}}\pi_k f_k(\bs y\vert\bs\theta_k)
\]
where $f_k\in\mathcal F_\Theta = \{f(\cdot\vert\bs\theta); \theta\in\Theta\}$ denotes a member of a parametric family of distributions. Let also $d$ denote the dimension of free parameters in the distribution $f_k(\cdot)$. For instance, in the case of a mixture of multinomial logistic regression models with $J+1$ categories and $P$ covariates (including constant term) $d = JP$.


\cite{rousseau2011asymptotic} show that the asymptotic behavior of the $K_{\max} - K$ extra components depends on the prior distributions of the mixing proportions $\bs\pi = (\pi_1,\ldots,\pi_{K_{\max}})$.  For the case of a Dirichlet prior as in Equation \eqref{eq:dirichlet}, if $\max\{\alpha_k; k = 1,\ldots,K_{\max}\} < d/2$, the posterior weight of the extra components  will tend to zero as $n\rightarrow\infty$ and force the posterior distribution to put all its mass in the sparsest way to approximate the true density. 

Following \cite{PAPASTAMOULIS2018220}, we set $\alpha_1=\ldots=\alpha_K = \alpha$, thus the distribution of mixing proportions in Equation \eqref{eq:dirichlet} becomes
\begin{equation}\label{eq:dirichlet_prior_same}
 \bs \pi\sim\mathcal D\left(\alpha,\ldots,\alpha\right)
\end{equation}
where $0 < \alpha < d/2$ denotes a pre-specified positive number.

Therefore, the inference on the number of mixture components can be based on the posterior distribution of the  ``alive'' components of the overfitted model, that is, the components which contain at least one allocated observation. In order to estimate the number of clusters we only have to keep track of the number of components with at least one allocated observation, across the MCMC run. This reduces to record the variable $K_0^{(t)} = \vert\vert\{k = 1,\ldots,K_{\max}:\sum_{i=1}^{n}z_{ik}^{(t)} > 0\}\vert\vert$, where $\bs z^{(t)}_{i}$ denotes the simulated allocation vector for observation $i$ at  MCMC iteration $t = 1, 2, \ldots$.

In order to produce a MCMC sample from the joint posterior distribution of the parameters of the overfitting mixture model (including the number of clusters), we embed the scheme described in Section \ref{sec:mcmc} within a prior parallel tempering scheme \citep{geyer1991, geyer1995, Altekar12022004}. Each heated chain ($c = 1,\ldots,C$) corresponds to a model with identical likelihood as the original, but with a different prior distribution. Although the prior tempering can be imposed on any subset of parameters, it is only applied to the Dirichlet prior distribution of mixing proportions \citep{overfitting, PAPASTAMOULIS2018220, papastamoulis2020clustering}. 
The inference is based on the output of the first chain  ($c = 1$) of the prior parallel tempering scheme \citep{overfitting}. 

Let us denote by $f_c(\bs\varphi\vert\bs x)$ and $f_c(\bs\varphi)$; $c=1,\ldots,C$, the posterior and prior distribution of the $c$-th chain, respectively. Obviously, $f_c(\bs\varphi\vert\bs x) \propto f(\bs x\vert\bs\varphi)f_c(\bs\varphi)$. Let $\bs\varphi^{(t)}_c$ denote the state of chain $c$ at iteration $t$ and assume that a swap between chains $c_1$ and $c_2$ is proposed. The proposed move is accepted with probability $\min\{1,A(\bs\pi_{c_1},\bs\pi_{c_2})\}$ where
\begin{equation}\label{eq:mh_ar}A(\bs\pi_{c_1},\bs\pi_{c_2}) = \frac{f_{c_1}(\bs\varphi_{c_2}^{(t)}\vert\bs x)f_{c_2}(\bs\varphi_{c_1}^{(t)}\vert\bs x)}{f_{c_1}(\bs\varphi_{c_1}^{(t)}\vert\bs x)f_{c_2}(\bs\varphi_{c_2}^{(t)}\vert\bs x)}=
\frac{f_{c_1}(\bs\varphi_{c_2}^{(t)})f_{c_2}(\bs\varphi_{c_1}^{(t)})}{f_{c_1}(\bs\varphi_{c_1}^{(t)})f_{c_2}(\bs\varphi_{c_2}^{(t)})}=
\frac{\widetilde{f}_{c_1}( \bs \pi_{c_2}^{(t)})\widetilde{f}_j( \bs \pi_{c_1}^{(t)})}{\widetilde{f}_{c_1}( \bs \pi_{c_1}^{(t)})\widetilde{f}_{c_2}( \bs \pi_{c_2}^{(t)})},\end{equation}
and $\widetilde{f}_{c}(\cdot)$ corresponds to the probability density function of the Dirichlet prior distribution related to chain $c = 1,\ldots,C$. According to Equation \eqref{eq:dirichlet_prior_same}, this is
\begin{equation}\label{eq:dir_prior_tempering}
 \bs \pi\sim D\left(\alpha_{(c)},\ldots,\alpha_{(c)}\right),
\end{equation} 
for a pre-specified set of parameters $\alpha_{(c)}>0$ for chain $c = 1,\ldots,C$.

When estimating a Bayesian mixture model, a well known problem stems from the label switching phenomenon \citep{jasra2005}, which arises from the fact that both the likelihood and prior distribution are invariant to permutation of the labels of mixture componets. The posterior distribution of the parameters will also be invariant, thus the parameters are not marginally identifiable. We deal with this problem by post-processing the MCMC output of the overfitting mixture via  the ECR algorithm  \citep{Papastamoulis:10, papastamoulis2016label}. 
Note that after post-processing the MCMC output for correcting label switching, the estimated classification for observation $i$ is obtained as the mode of the (reordered) simulated values of $\bs Z_i$ in Equation \eqref{eq:z_fc} across the MCMC run (after discarding the draws corresponding to the burn-in period of the sampler), $i = 1,\ldots,n$. For more details the reader is referred to the {\tt label.switching} package \citep{papastamoulis2016label}.
The overall procedure is summarized in Algorithm \ref{alg:overfittingmala}.

\begin{algorithm*}[p]
\colorbox{gray!25}{\parbox{0.9\textwidth}{
\caption{Prior Parallel Tempering MALA-within-Gibbs MCMC for Overfitting Mixtures of Multinomial Logistic Regressions}
\SetKwInOut{Input}{Input}
\SetKwInOut{Output}{Output}
\SetKwBlock{blocknotext}{~}{end}
\SetKwBlock{step}{Step}{~}
\SetKwBlock{stepo}{Step 0:  Warm up the parallel chains}{~}
\SetKwBlock{stepi}{Step 1: perform a cycle of MALA within Gibbs per chain}{~}
\SetKwBlock{stepii}{Step 2: Chain swaping}{~}
\SetKwBlock{stepiii}{Step 3: Undo label switching}{~}
\SetKwFor{For}{for}{~}{endfor}
\SetKwFor{ForAll}{for~all}{~}{endfor}
\SetKwFor{ForEach}{for~each}{~}{endfor}

\Input{data $\bs y, \bs x$\\ 
upper bound on the number of clusters $K_{\max}$\\ 
number of parallel chains $C$\\
prior hyper-parameters $\nu^2, \alpha_{(1)}, \ldots, \alpha_{(C)}$\\
number of MCMC cycles $T$\\
MCMC iterations per cycle $m_{1}$\\
number of iterations that will be used for warm-up $m_{0}$\\
initial value of the scale of the MALA proposal $\tau$
}
\Output{identified MCMC sample $\{\bs z^{(t)}, \bs \pi^{(t)}_{\mbox{alive}}, \bs \beta^{(t)}_{\mbox{alive}}; t = 1,\ldots,T\}$}
\stepo{
\For{chain $c=1$ \KwTo $C$}{
\begin{enumerate}
\item[0.0] Set $\tau_{(c)} = \tau$ for all  $c=1,\ldots,C$ and run Algorithm \ref{alg:mala} with $M = m_0$  and $\alpha_k =\alpha_{(c)}$; $k = 1,\ldots,K_{\mbox{max}}$.
\item[0.1] Adjust the scale of the MALA proposal ($\tau_{(c)}$) every $m_{check}$ iterations
\item[0.2] Set $\bs\theta^{(0)}_{(c)} \leftarrow \left(\bs z^{(m_0)}, \bs \pi^{(m_0)}, \bs\beta^{(m_0)}\right)$
\end{enumerate}
}
}

\For{MCMC cycle $t=1$ \KwTo $T$}{
\stepi{
\For{chain $c=1$ \KwTo $C$}{
\begin{enumerate}
\item[1.1] Run Algorithm \ref{alg:mala} with $M = m_{\mbox{1}}$, $\alpha_k = \alpha_{(c)}$; $k = 1,\ldots,K_{\mbox{max}}$\\ starting value $\bs\theta^{(t - 1)}_{(c)}$ and scale of the MALA proposal equal to $\tau = \tau_{(c)}$.
\item[1.2] Set $\bs\theta^{(t)}_{(c)} \leftarrow \left(\bs z^{(m_1)}, \bs \pi^{(m_1)}, \bs\beta^{(m_1)}\right)$
\end{enumerate}
}
}
\stepii{
\begin{enumerate}
\item[2.1] Randomly choose $1\leqslant c\leqslant C - 1$ and set $c_1 = c$, $c_2 = c + 1$ 
\item[2.2] Generate $u\sim\mathcal U(0,1)$
\end{enumerate}

\SetAlgoVlined\If{$u < A(\bs\pi_{c_1},\bs\pi_{c_2})$ in Equation \eqref{eq:mh_ar}}{set $\bs\theta_{c_1}^{(t)}\lrarrow \bs\theta_{c_2}^{(t)}$}

}
}
\stepiii{
Apply the ECR algorithm on the output of chain $c = 1$ and return an identifiable MCMC sample $\{\bs z^{(t)}, \bs \pi^{(t)}_{\mbox{alive}}, \bs \beta^{(t)}_{\mbox{alive}}; t = 1,\ldots,T\}$ 
}
END of algorithm
\label{alg:overfittingmala}
}}
\end{algorithm*}

Regarding the initialization of the overfitting mixture model (see Step 0) in Algorithm \ref{alg:overfittingmala}, we use two alternative approaches. The first initialization is based on random starting values (``MCMC-RANDOM'' scheme) and the second initialization scheme uses a more elaborate scheme, by exploiting the output of the EM algorithm under the split-small-EM scheme (``MCMC-EM'' scheme). As expected, the latter scheme performs better as illustrated in the simulation studies. 
The overall procedure of the MCMC sampler is summarized in Algorithm \ref{alg:mala} and Algorithm \ref{alg:overfittingmala}. The typical  choices of the parameters as well as further details on the prior parallel tempering scheme and initialization schemes are given in Section \ref{sec:mcmc_details} of the Appendix.

\begin{table}[h]
\caption{Values for sample size ($n$), number of clusters ($K$), covariates ($P$) and number of categories ($J+1$) in the simulation study.}
\centering
\begin{tabular}{cccc}
$n$ & $K$  & $P$ & $J+1$\\
\hline
$\{125, 250,500,1000\}$& 
$\{1,\ldots,8\}$ & 
$\{2,4,6\}$ &
$\{6,9,12\}$
\end{tabular}
\label{tab:sim_setup}
\end{table}

\section{Applications}\label{sec:app}

In Section \ref{sec:sim} we use a simulation study in order to evaluate and rank the proposed methods in terms of their ability in clustering multinomial data.  Next, we present two applications on real data: in Section \ref{sec:phthiotis} our method is used to identify clusters of age profiles within a regional unit in Greece and in Section \ref{sec:fb} we study clusters of Facebook engagement metrics in Thailand. Further simulation results and comparisons with {\tt flexmix} are reported in Appendix \ref{app:more}.

\subsection{Simulation study}\label{sec:sim}

\begin{figure}[ht]
\centering
\includegraphics[scale=0.5]{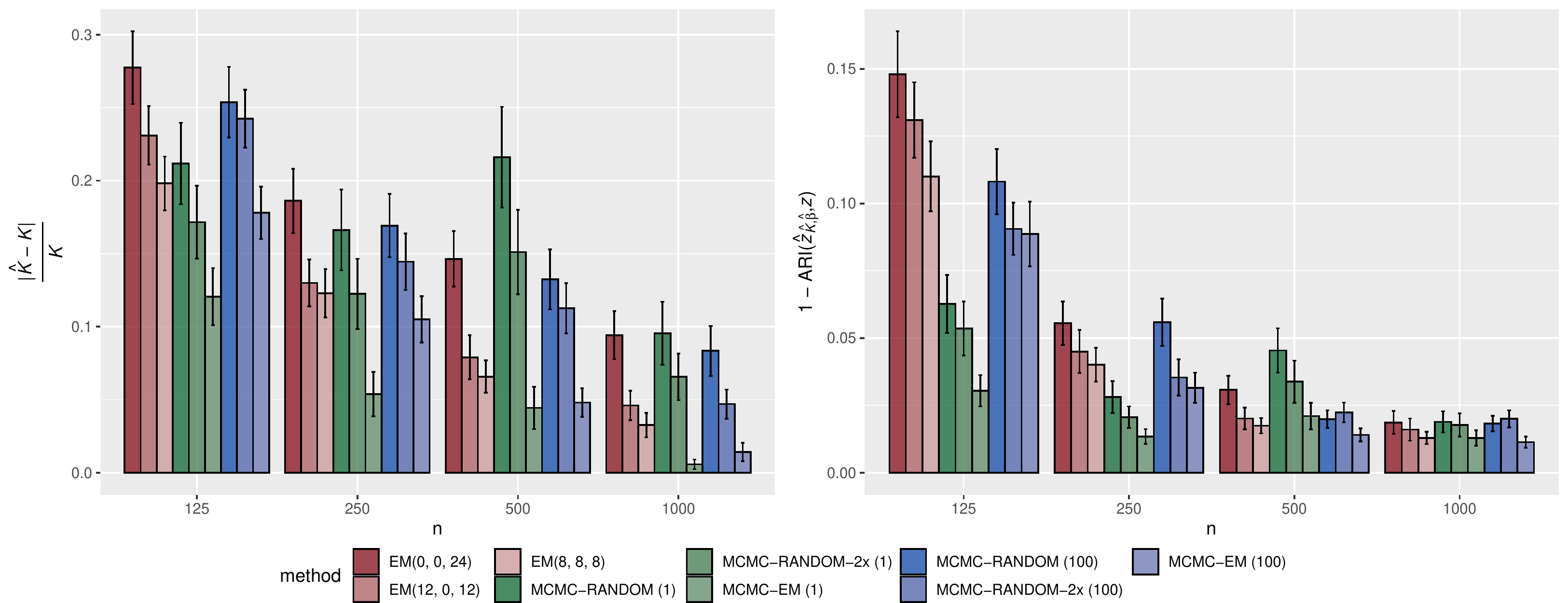}

\caption{Simulation study summary.  See Table \ref{tab:sim_setup} for the simulation study set-up. For the EM implementations: the three numbers in parenthesis refer to the number of split, shake and random small-EM runs. For the MCMC implementations, the number in parenthesis denotes the prior variance ($\nu^2$) of the coefficients $\beta_{kjp}$ in Equation \eqref{eq:betaPrior}.}
\label{fig:sim}
\end{figure}

In order to evaluate the ability of the proposed methods in clustering multinomial count data, we considered synthetic datasets generated from a mixture of multinomial logistic regression models \eqref{eq:logit_mixture}. The number of multinomial replicates ($S_i$) per observation is drawn from a negative binomial distribution: $S_i\sim\mathcal{NB}(r, p)$ with number of successful trials $r = 20$ and probability of success $p = 0.025$. We simulated 500 datasets in total where the values of $n, K, P, J$ are uniformly drawn in the range of values shown in Table \ref{tab:sim_setup}. Given $K$, the weight of each cluster was equal to $\pi_k \propto k$, $k = 1,\ldots,K$. Notice that this setup gives rise to mixture models with total number of free parameters ranging from 10 up to 535. 

The true values of the regression coefficients were simulated according to
\begin{align*}
\sigma&\sim\mathcal U(1, 5)\\
\beta_{kjp}\vert\sigma&\sim 0.5\mathrm{I}_{\{0\}}(\beta) + 0.5\phi(\beta;0, \sigma^2)
\end{align*}
(conditionally) independent for $k = 1,\ldots,K$; $j = 1,\ldots,J$; $p = 1,\ldots,P$, where $\mathrm{I}_{\{0\}}(\beta)$ denotes a discrete distribution degenerate at $0$ and $\phi(\beta;\mu, \sigma^2)$ denotes the density function of the normal distribution with mean $\mu$ and variance $\sigma^2$.

We applied the proposed methodology in order to estimate mixtures of multinomial logit models. In particular we compared the EM algorithm under three initialization schemes, as well the MCMC sampling scheme under random initialization and an initialization based on the output of the EM algorithm (under the split-shake-random small-EM scheme). In total we considered 24 different starts in the small-EM schemes: a random small-EM with 24 starts: EM(0, 0, 24), a combination of split and random small-EM with 12 starts each: EM(12, 0, 12) and finally, a combination of split, random and shake small-EM with 8 starts each: EM(8, 8, 8). The total number of MCMC iterations is held fixed at 100000. We also present  results when considering the double amount of iterations (both in the warm-up period as well as the main MCMC sampler) under the MCMC-RANDOM scheme, which we will denote by MCMC-RANDOM-2x. Finally, we considered two different values of prior variance of the coefficients in Equation \eqref{eq:betaPrior}: $\nu = 1$ and $\nu^2 = 100$. The first choice corresponds to an informative prior distribution, heavily penalizing large values of $\lvert\beta_{kjp}\rvert$.  The second choice corresponds to a vague prior distribution. The chosen value of $\nu$ will be denoted in a parenthesis, that is, MCMC-RANDOM ($\nu^2$) and MCMC-EM ($\nu^2$) will indicate the output of MCMC algorithm with random and EM  initialization schemes (respectively) and prior variance equal to $\nu^2$.  
See Appendix \ref{sec:mcmc_details} for further details of various other parameters for the EM and MCMC algorithms.

Figure \eqref{fig:sim} illustrates a graphic summary of the simulation study findings, based on our 500 synthetic datasets. The metrics we are focusing are the following: The left graph shows the mean of relative absolute error $\frac{\vert \hat K - K\vert}{K}$ between the estimated number of clusters ($\hat K$) and the correspoding true value ($K$). The right graph  displays  the mean of the adjusted Rand index  (with respect to the ground-truth classification) subtracted from 1. In all cases we conclude that the EM algorithm with a random small-EM initialization (denoted as EM(0,0,24)) is worse compared to the split-shake-random small EM initialization (denoted as EM(8,8,8)). Regarding the MCMC sampler we see that the random initialization scheme (MCMC-RANDOM) is worse than the EM-based initialization (MCMC-EM), when both MCMC-RANDOM and MCMC-EM run for the same number of iterations. However, as the number of iterations increases in the randomly initialized MCMC sampler (MCMC-RANDOM-2x), the results are improved, particularly for the mean relative absolute error of the estimation of the number of clusters.  Overall, the MCMC algorithm initialized by the (split-small) EM  solution is the best performing method, closely followed  by the EM algorithm under the split-small EM scheme. Naturally, the informative prior distribution ($
\nu = 1$, corresponding to the green-coloured bars in Figure \eqref{fig:sim}) outperforms the vague prior distribution ($\nu^2 = 100$, corresponding to the red-coloured bars). More detailed summaries of the resulting estimates are given in Appendix \ref{sec:more1}.

\subsection{Phthiotis Population Dataset}\label{sec:phthiotis}

\begin{figure}[t]
\centering
\begin{tabular}{cc}
\includegraphics[scale=0.5]{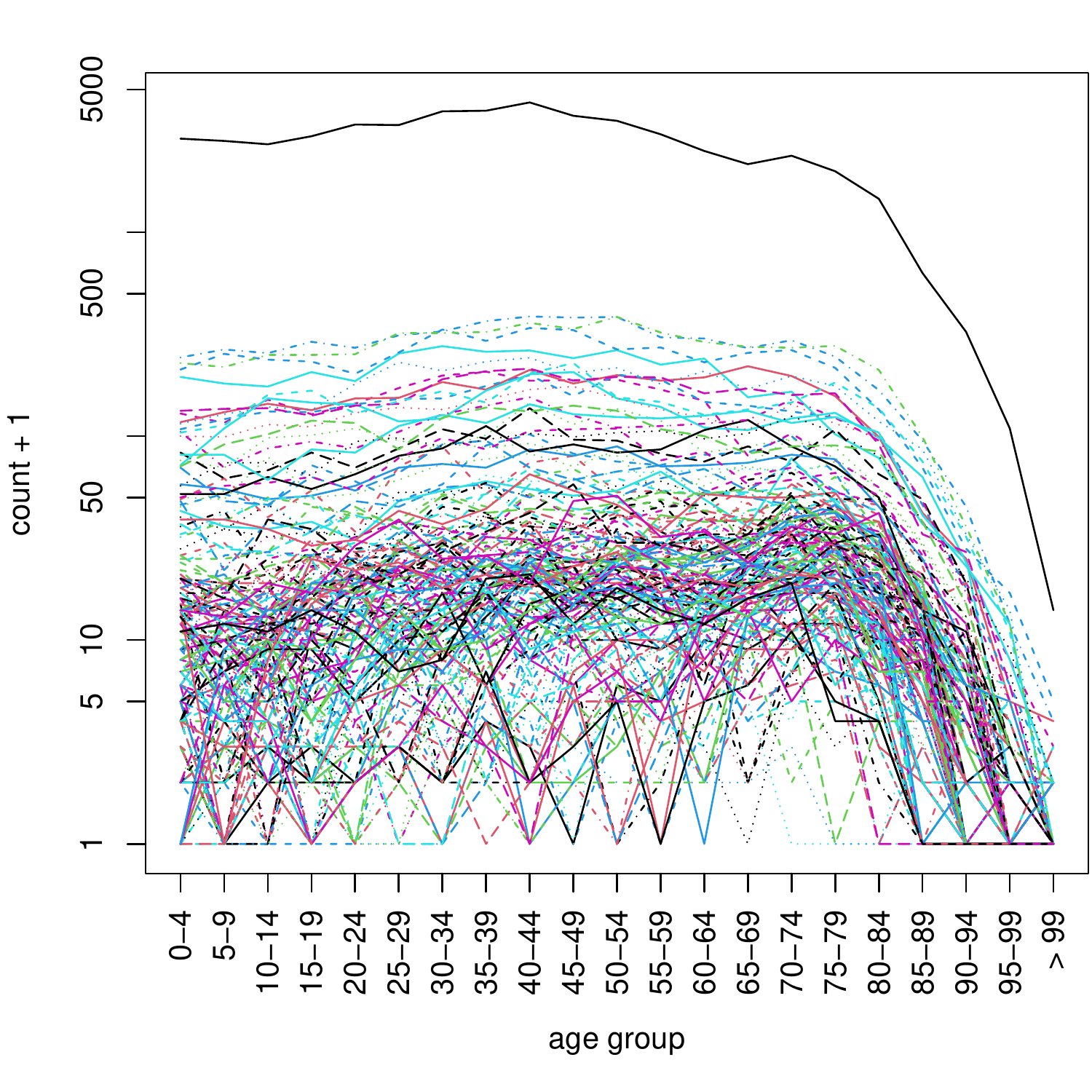} &
\includegraphics[scale=0.5]{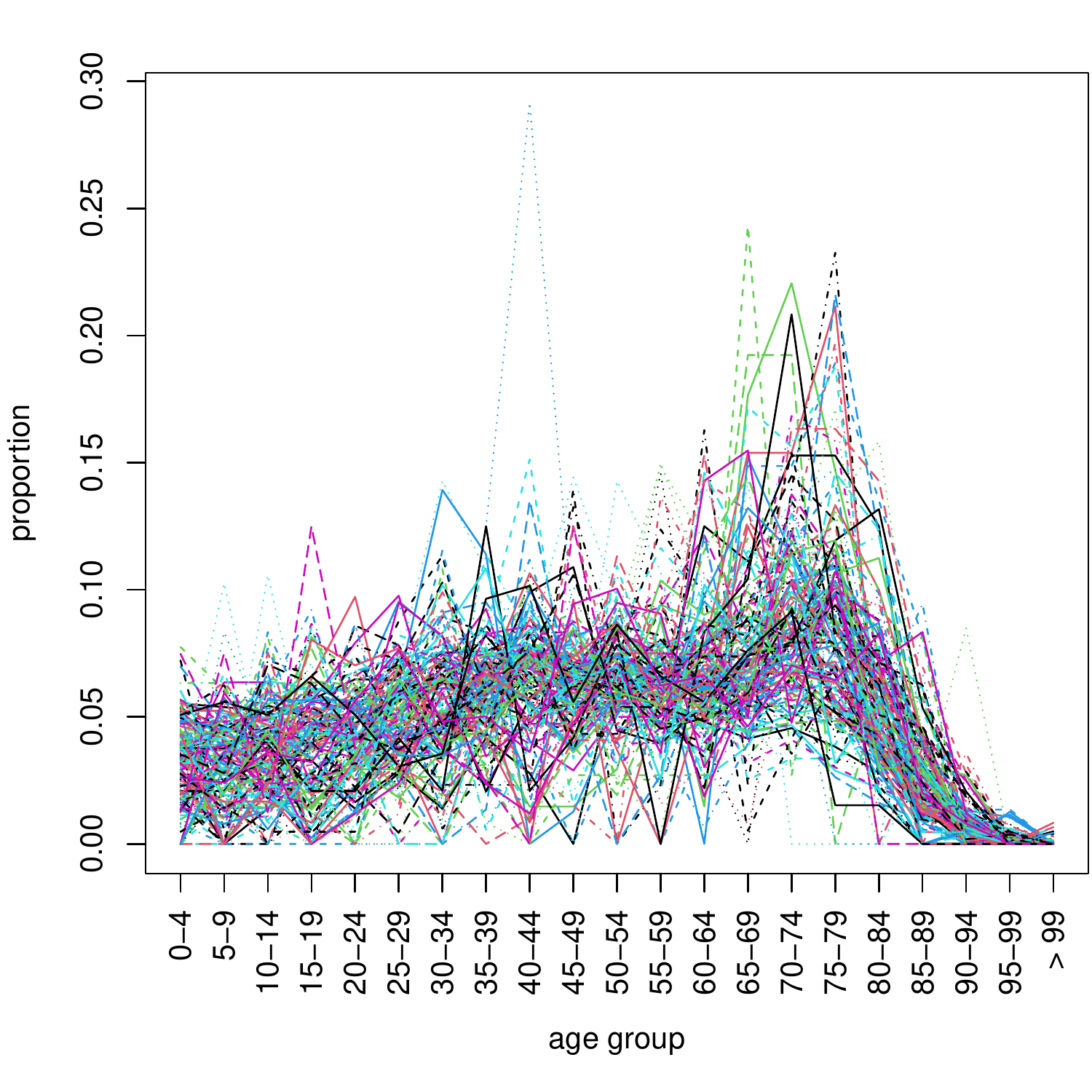}\\
(a) & (b)
\end{tabular}
\caption{Age profiles for $n = 187$ settlements in the Phthiotis regional unit according to the 2011 census of Eurostat. (a): Population counts (increased by 1) displayed in log-scale in the $y$ axis and (b): relative frequency of population counts. }
\label{fig:fthiotis_data}
\end{figure}

In this example we present an application of our methodology in clustering areas within a certain region with respect to the age profiles of their population, taking also into account geographical covariate information. For this purpose we considered population  data based on the 2011 census of Eurostat\footnote{https://ec.europa.eu/eurostat/web/main/data/database}. We considered the Phthiotis area, a regional unit located in central Greece. Our extracted dataset consists of number of people per age group ($21$ groups: $0-4, 5-9, \ldots,95-99, >99$ years old) for a total of $n = 187$ settlements (such as villages, towns, and  the central city of the regional unit, \textit{Lamia}), as displayed in Figure \ref{fig:fthiotis_data}. The separated line in the upper part of Figure \ref{fig:fthiotis_data}.(a) corresponds to \textit{Lamia}. Observe that there are various regions where there is a peak in the older population groups (between 65-85), as vividly displayed when looking at the plot of relative frequencies per age group at Figure \ref{fig:fthiotis_data}.(b). A different behaviour is obvious for \textit{Lamia} where we see that the dominating age groups are between $30-50$. The research question is to cluster these settlements based on the age profiles of their population. 

If we cluster the raw dataset of age counts within each group using a mixture of multinomial distributions without any covariate information, then a large number of clusters is found. In particular, when using {\tt mixtools} \citep{mixtools}, a relatively large number of clusters is found ($\hat K = 8$ using ICL). Therefore, we opt to apply our method using the following covariate information for settlement $i=1,\ldots,n$: 
\begin{itemize}
\item[] $x_{i1}$: distance (in Km) from \textit{Lamia} (capital city of the regional unit) 
\item[] $x_{i2}$: logarithm of the altitude (elevation, in m).
\end{itemize}
Both covariates were scaled to zero mean and unit variance. 
Let us denote by $y_{ij}$ the number of people in age group $j = 1,\ldots,21$ for settlement $i =1, \ldots,n$. The probability $\theta_{kj}^{(i)}$ denotes the proportion of population  being in age group $j$ conditional on the event that settlement $i$ belongs to  cluster $k$, where $\sum_{j=1}^{21}\theta_{kj}=1$ for all $k$. Conditional on cluster $k=1,\ldots,K$, the random vector $\bs Y_i = (Y_{i1},\ldots,Y_{i,21})^\top$ is distributed according to a multinomial distribution 
\begin{align*}
\bs Y_i\vert Z_{ik} = 1&\sim\mathcal M_{21}(S_i,\bs\theta_k^{(i)})\\\log\frac{\theta_{kj}^{(i)}}{\theta_{k,1}^{(i)}}&= \beta_{kj0} + \beta_{kj1} x_{i1} + \beta_{kj} x_{i2},\quad j=2,\ldots,21
\end{align*}
where $\bs\theta_k^{(i)} = (\theta_{k1}^{(i)},\ldots,\theta_{k,21}^{(i)})$ for $k=1,\ldots,K$ and $S_i$ denotes the total population  for settlement $i$, for $i = 1,\ldots,n$. Note that we have used the 1st category (ages between $0$ and $4$) as baseline in order to express the log-odds of the remaining groups. The  distribution of counts per age group is written as a mixture of multinomial distributions
\[
\bs Y_i \sim \sum_{k=1}^{K}\pi_k\mathcal M_{21}(S_i,\bs\theta_k^{(i)}),\quad \mbox{independent for} \quad  i = 1,,\ldots,n
\]
where $\pi_k$ denotes the weight of cluster $k$. 
Hence, each cluster represents areas with different age profile as reflected by the corresponding vector of multinomial probabilities. The total number of clusters ($K$) is unknown. 

At first we used the EM algorithm under the proposed initialization scheme to estimate mixtures of multinomial logistic regression models for a series of $K=1,2,\ldots,K_{\max} = 10$ components. According to the ICL criterion, the selected number of clusters is equal to $K = 3$. Next we estimated an overfitting Bayesian mixture of $K_{\max} = 10$ components, using a prior parallel tempering scheme based on 12 chains. The MCMC algorithm was initialized from the EM solution, while all remaining parameters were initialized from a zero value. The MCMC sampler ran for a warm-up period of 100000 iterations, followed by 400000 iterations. A thinned MCMC sample of 20000 iterations was retained for inference. In almost all MCMC draws the number of non-empty mixture components was equal to  $K_0 = 3$ (estimated posterior probability equal to $99.5\%$). The retained MCMC sample was then post-processed according to the ECR algorithm \citep{Papastamoulis:10, papastamoulis2016label} in order to undo label switching. The confusion matrix of the single best clusterings between the two methods (EM and MCMC) is displayed in Table \ref{tab:Fthiotis}. The corresponding adjusted Rand index is equal to $0.67$ indicating that the two resulting partitions have strong agreement.

\begin{table}[ht]
\caption{Confusion matrix between the single best clustering of the Phthiotis Population Dataset arising from the EM and MCMC algorithms (after post-processing the MCMC output for correcting label switching).}
\centering
\begin{tabular}{rrrrr}
  \toprule
  &&\multicolumn{3}{c}{MCMC}\\  
  \midrule
& & 1 & 2 & 3 \\ 
  \midrule
\multirow{3}{*}{EM}&1 &  29 &   2 &   0\\ 
 & 2 &  1 &  73 &   5\\ 
  &3 &  0 &  12 &  65\\ 
   \bottomrule
\end{tabular}
\label{tab:Fthiotis}
\end{table}

\begin{figure}[p]
\centering
\begin{tabular}{ccc}
\includegraphics[scale=0.35]{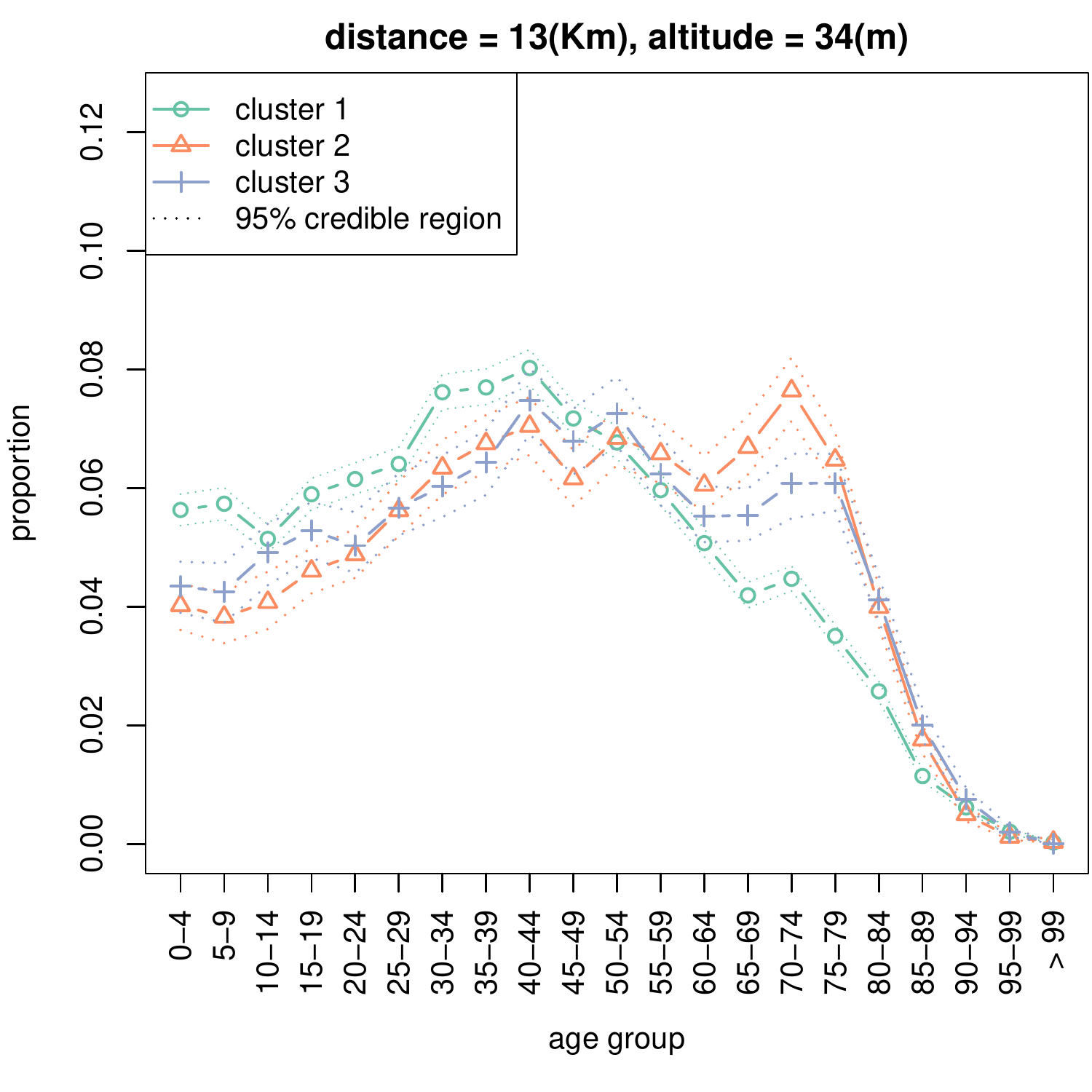}&
\includegraphics[scale=0.35]{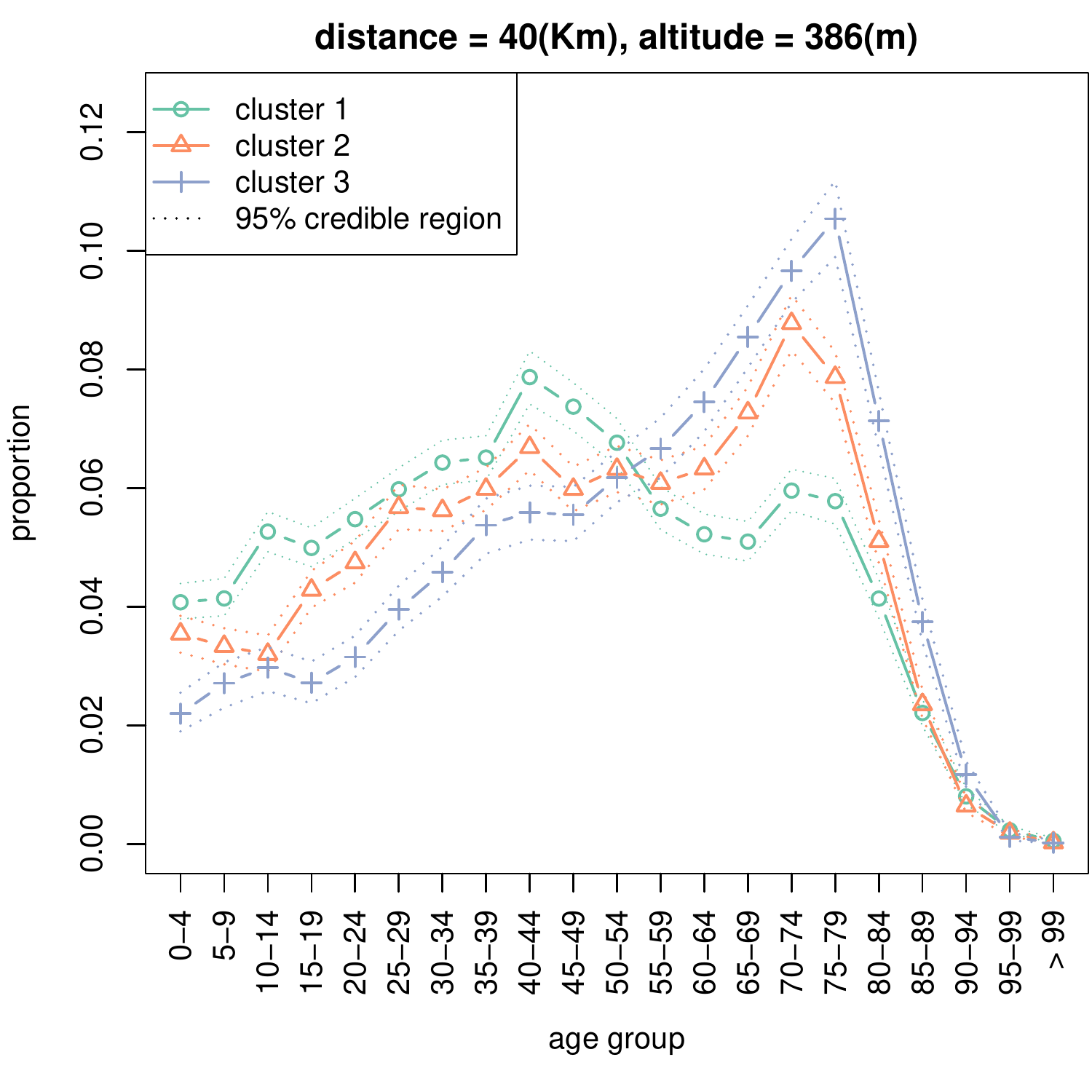}&
\includegraphics[scale=0.35]{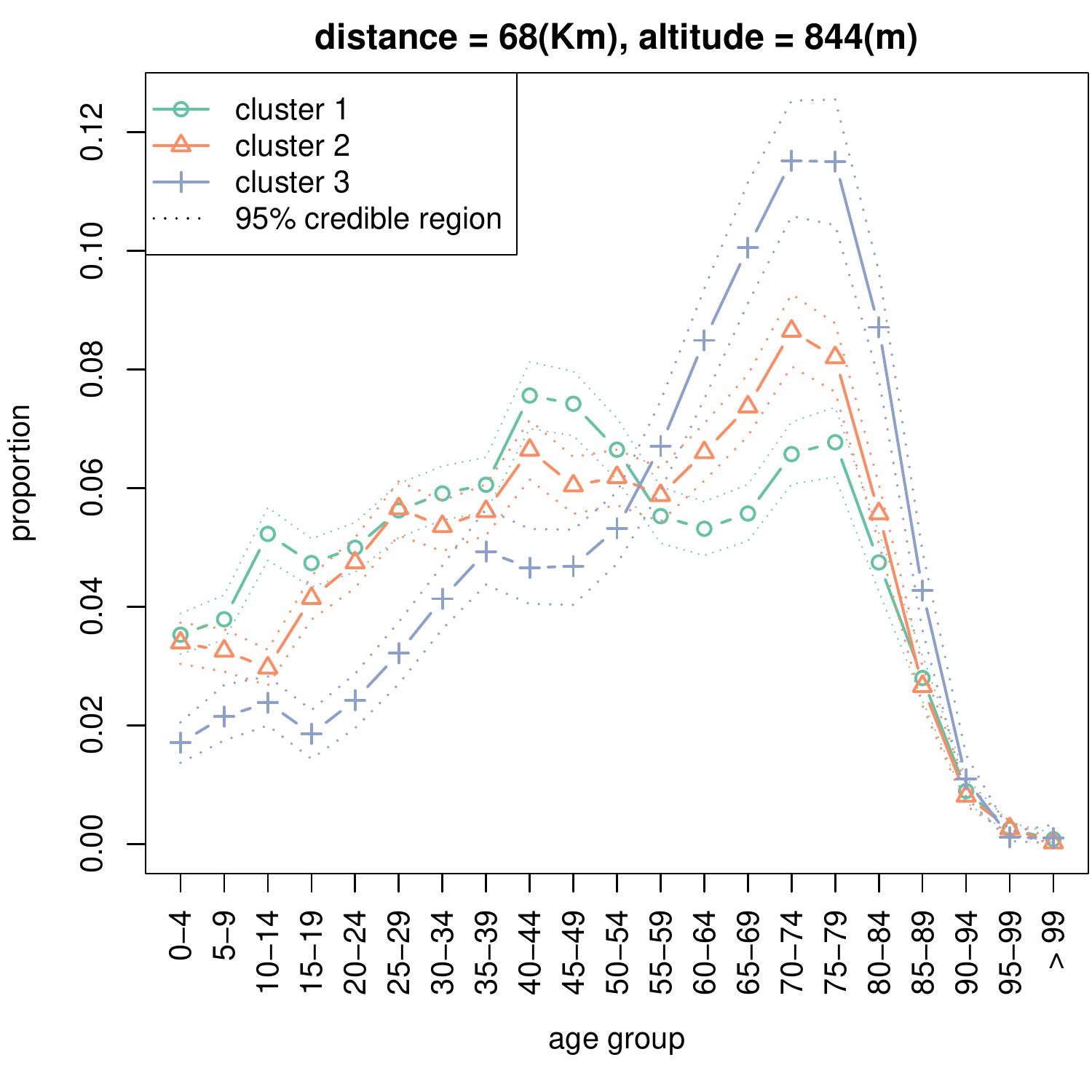}\\
(a)  & (b)&(c)\\
\end{tabular}
\caption{Posterior mean and $95\%$ credible region of age profiles per cluster for the Phthiotis population data. The two  covariates (distance from Lamia and altitude) are set equal to the corresponding $0.1$ (a), $0.5$ (b) and $0.9$ (c) percentiles.}
\label{fig:fthiotis_clusters}
\end{figure}

\begin{figure}[p]
\centering
\hspace{-3ex}\includegraphics[scale=0.5]{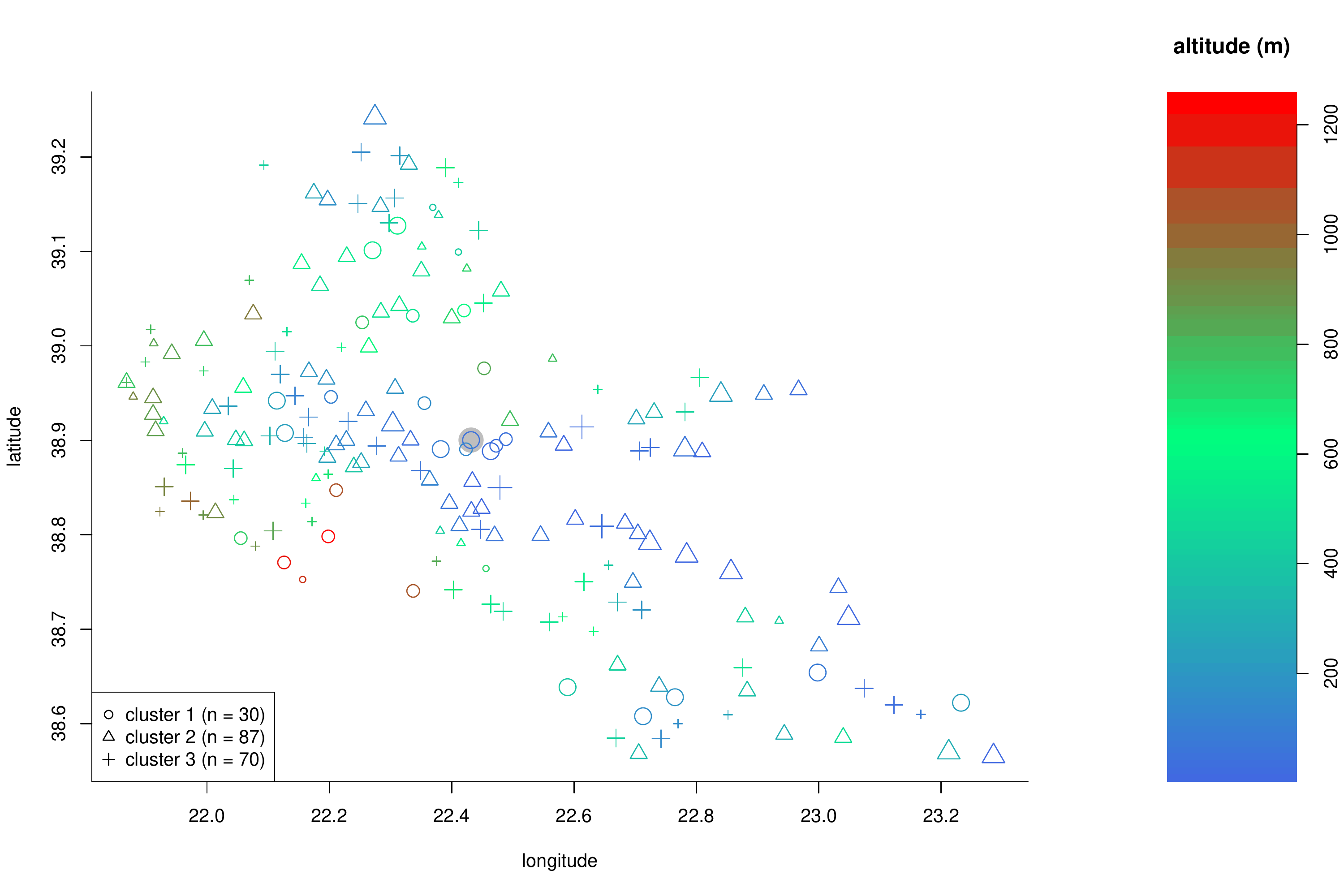}
\caption{Geographical coordinates of the settlements and inferred cluster membership according to the Maximum A Posteriori rule on the output of the MCMC sampler. The gray circle indicates Lamia, that is, the central city of the Phthiotis region. Different point sizes are used according to the total population of each settlement: small ($S_i<150$), medium $150 \leqslant S_i\leqslant 999$ and larger ($S_i > 999$). }
\label{fig:fthiotis_map}
\end{figure}

\begin{figure}[ht]
\centering
\includegraphics[scale=0.4]{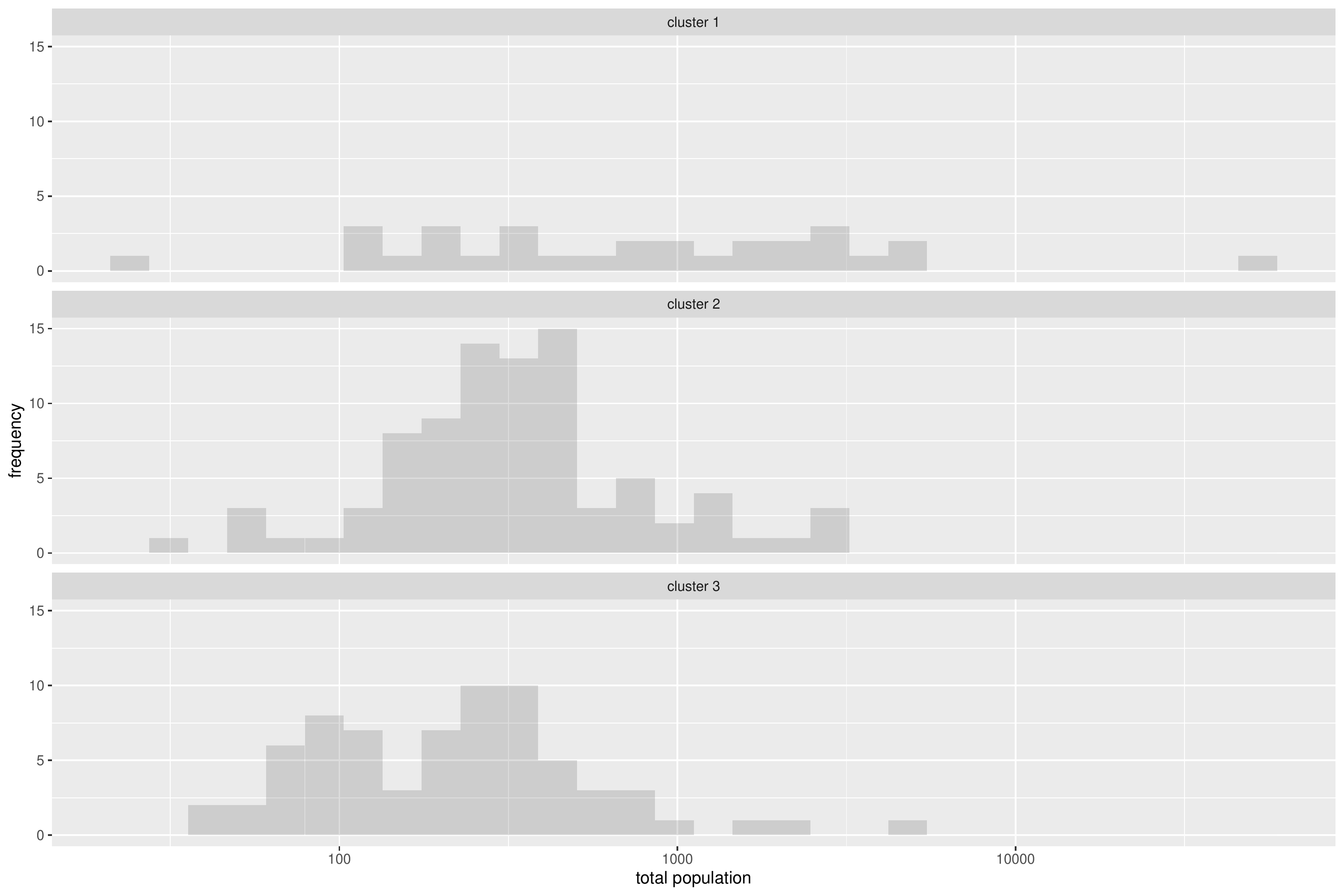}
\caption{Total population counts per cluster of the Phthiotis  dataset.}
\label{fig:popCounts}
\end{figure}

Next we focus on the results according to the MCMC algorithm (after post-processing). Figure \ref{fig:fthiotis_clusters} illustrates the posterior mean (and $95\%$ credible region) of the probability $\theta_{kj}^{(i)}$ for age group $j$ in  cluster $k=1,\ldots,3$. Three characteristic configurations of covariate levels were used, that is, the 0.1, 0.5 and 0.9 percentiles of the two covariates. In all cases we see distinct age group characteristics and what is evident is the presence of a group which contains places with younger age profiles (cluster 1). In cluster 3, notice a strong peak at the group of ages between $76$ to $84$, which emerges in cases of moderate to large values of the two covariates. In cluster 2, the peak is also  located at the older age groups however it is less pronounced compared to cluster 3. Figure \ref{fig:fthiotis_map} visualizes the three clusters on the map of the regional unit. We may conclude that cluster 1 (the ``younger'' cluster) mainly consists of settlements that are either located close to \textit{Lamia} (gray spot on the map), including \textit{Lamia} itself, or their total population is larger than 1000 (towns such as \textit{Makrakomi}, \textit{Malessina}, \textit{Sperchiada}, \textit{Atalanti}, \textit{Domokos}, \textit{Stavros} and the central city of \textit{Lamia}). However this younger group of age profiles is also present in some of the most distant and mountainous southwestern  areas (\textit{Dafni}, \textit{Neochori Ypatis}, \textit{Kastanea}, \textit{Pavliani} and \textit{Anatoli}: the altitude of these small villages is larger than 1000 m). In general, however, as we move further away from \textit{Lamia} the ``older'' and ``eldest'' clusters  dominate, particularly for areas with a small number of population. See also the histogram of settlement populations per cluster in Figure \ref{fig:popCounts}.  Note that the majority of smaller villages (population of 100 citizens, approximately) are mainly assigned to the third cluster (the eldest group). 

Finally, we have to mention that this specific application involves an ordinal and not a nominal response. Therefore, one could use alternative techniques to model the data,  such as proportional odds models, or smoothing the changes between adjacent categories. 

\subsection{Facebook Live Sellers in Thailand Data Set}\label{sec:fb}

The dataset of \cite{DEHOUCHE2020105661} (see also \cite{wongkitrungrueng2020live}) contains engagement metrics  of  Facebook pages for Thai fashion and cosmetics retail sellers. We consider the number of emoji reactions for each Facebook post, which are known as ``like'', ``love'', ``wow'', ``haha'', ``sad'' and ``angry''. The aim of our analysis is to cluster posts based on the reaction profiles, using additional covariate information. Each post can  be of a different nature (``video'', ``photo'', ``status''), a categorical variable which we are taking into account as a categorical predictor. In addition, we  also use as covariate the number of shares per post (in log-scale). The dataset is available at the UCI machine learning repository\footnote{\url{https://archive.ics.uci.edu/ml/datasets/Facebook+Live+Sellers+in+Thailand}}.

\begin{figure}[ht]
\centering
\begin{tabular}{cc}
\includegraphics[scale=0.5]{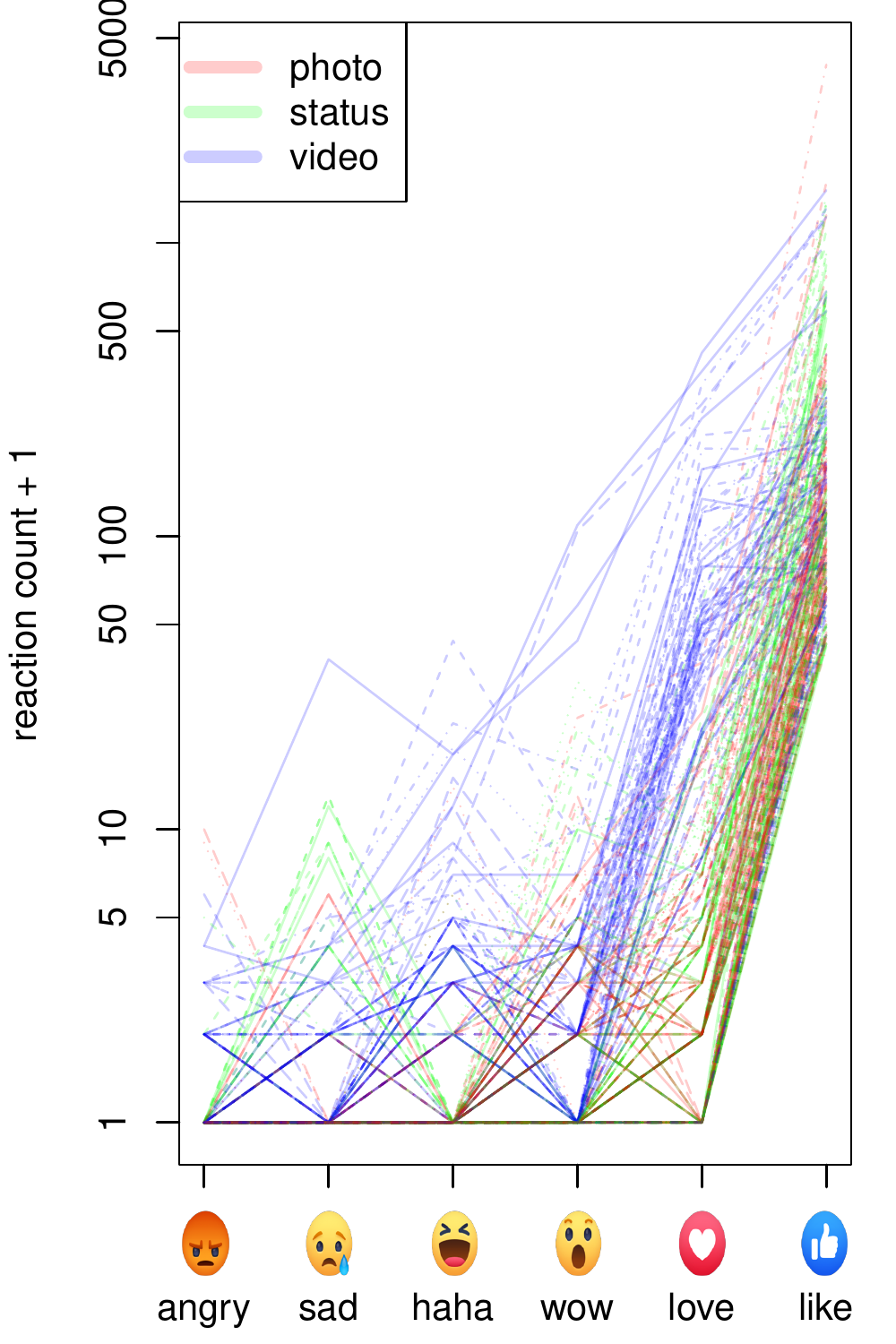}
&
\includegraphics[scale=0.5]{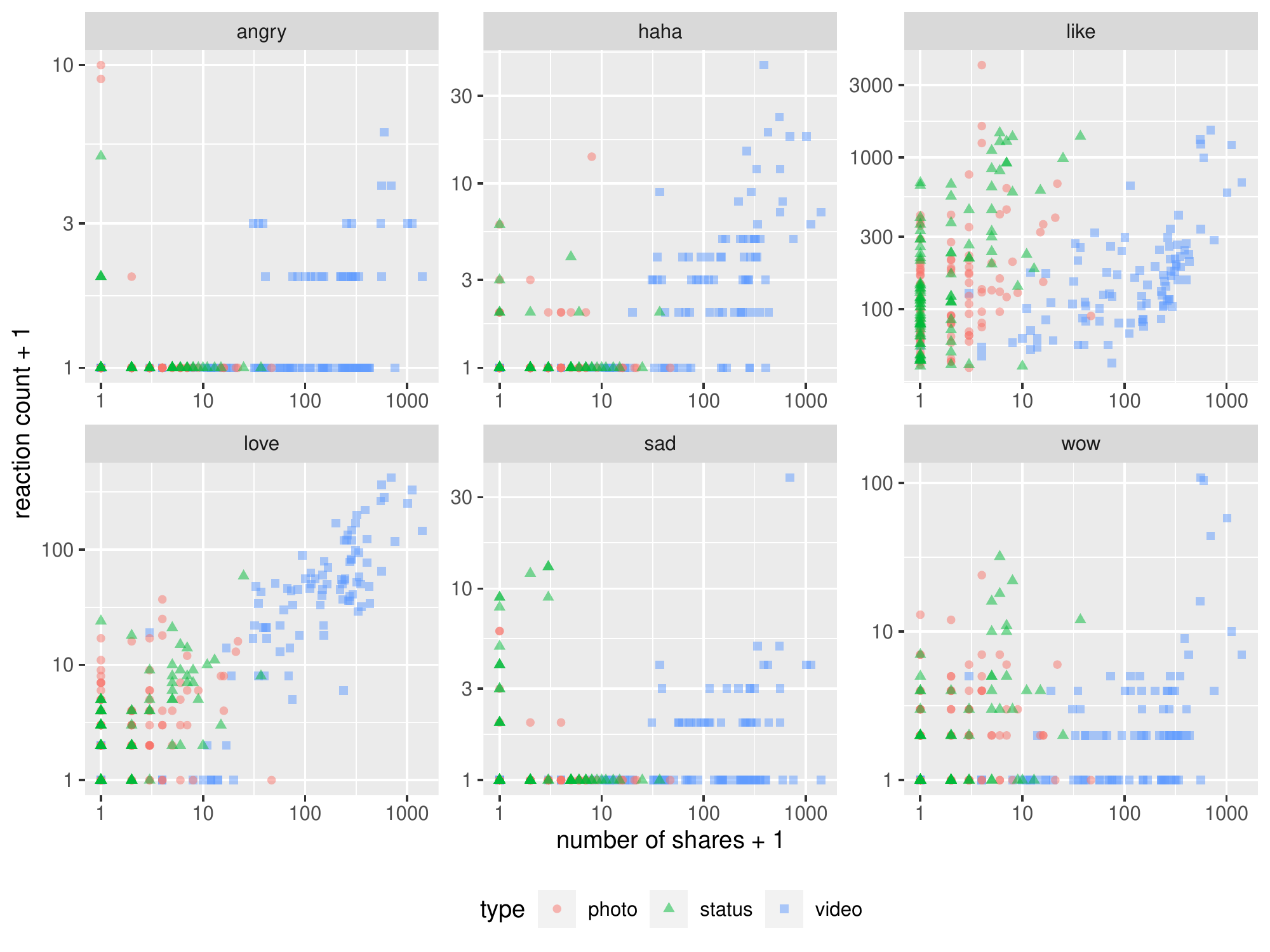}
\end{tabular}
\caption{Reaction counts for 300 posts of the Facebook Live Sellers Dataset. A different colour displays the type of each post (100 video, 100 photos and 100 statuses). Note that the $y$ axis on both graphs as well as the $x$ axis of the right graph  are displayed in log-scale after increasing each observed count by one.}
\label{fig:fb_dataset}
\end{figure}

We considered the period between 2017-01-01 and 2018-12-31 taking into account posts with a minimum overall number of reactions equal to 40. We then randomly selected 100 posts per type (100 videos, 100 photos and 100 statuses), that is, 300 posts in total. 
The observed data is displayed in Figure \ref{fig:fb_dataset} (note that for the sole purpose of visualization in the log-scale, each count is increased by 1). It is evident that most reactions correspond to ``loves'' and ``likes''. There is also some visual evidence that videos may result to a larger number of ``loves'' compared to photos or statuses. On the other hand, many posts result to zero counts for any kind of reaction other than ``like''. So we might expect that such a dataset exhibits heterogeneity, due to zero inflation in the first five categories. Thus, it makes sense to cluster posts according to the reaction profiles, i.e.~reaction probability. 

Let us denote by  $\bs y_i = (y_1,y_2,y_3,y_4,y_5,y_6)^\top$ the observed vector of reaction counts for post $i=1,\ldots,n$ ($n=300$). We assume that $\bs y_i$, conditional on post type and number of shares (as well as the total number of reactions for that particular post),  is distributed according to a mixture of multinomial distributions with $J + 1 = 6$ categories, where $y_j$ denotes the number of reactions of type $j$ for post $i$, $i = 1,\ldots,n$. The type of each post serves as a categorical predictor with three levels (``video'', ``photo'' and ``status''). Selecting the probability of ``like'' as the reference category and conditional on cluster $k = 1,\ldots,K$, the multinomial logit model is written as 
\[
\log\frac{\theta_{kj}^{(i)}}{\theta_{k6}^{(i)}} = \beta_{kj0} + \beta_{kj1} x^{\mathrm{status}}_{i} + \beta_{kj2} x^{\mathrm{photo}}_{i} + \beta_{kj3}\log(1+x_i^{\mathrm{shares}}),\quad j=1,2,3,4,5
\]
where $\theta_{kj}^{(i)}$ denotes the probability of reaction $j$ corresponding to ``angry'' ($j=1$), ``sad'' ($j=2$), ``haha'' ($j=3$), ``wow'' ($j=4$), ``love'' ($j=5$) and ``like'' ($j=6$). Note that the categorical predictor consists of three levels, thus, we created the two dummy variables 
\[
x^{\mathrm{status}}_{i} = \begin{cases}
1,& \mbox{if post $i$ is ``status''}\\
0,& \mbox{otherwise}\\
\end{cases},
x^{\mathrm{photo}}_{i} = \begin{cases}
1,& \mbox{if post $i$ is ``photo''}\\
0,& \mbox{otherwise}\\
\end{cases} 
\]
after selecting the ``video'' type as the baseline. In addition, $x_i^{\mathrm{shares}}$ denotes the number of shares for post $i$.

\begin{figure}[ht]
\centering
\vspace{-8ex}
\begin{tabular}{ cc}
\includegraphics[width = 0.4\textwidth]{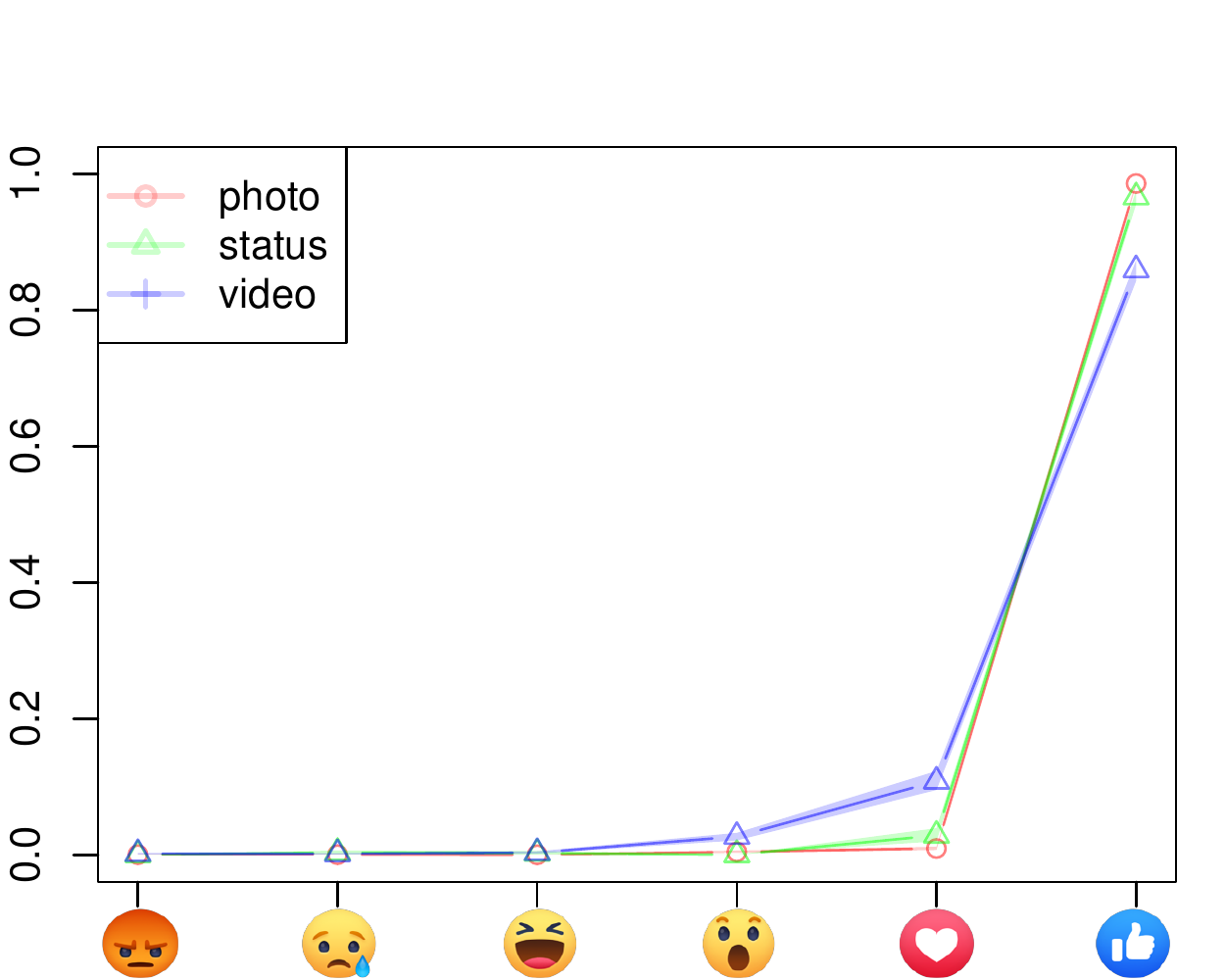} & 
\includegraphics[width = 0.4\textwidth]{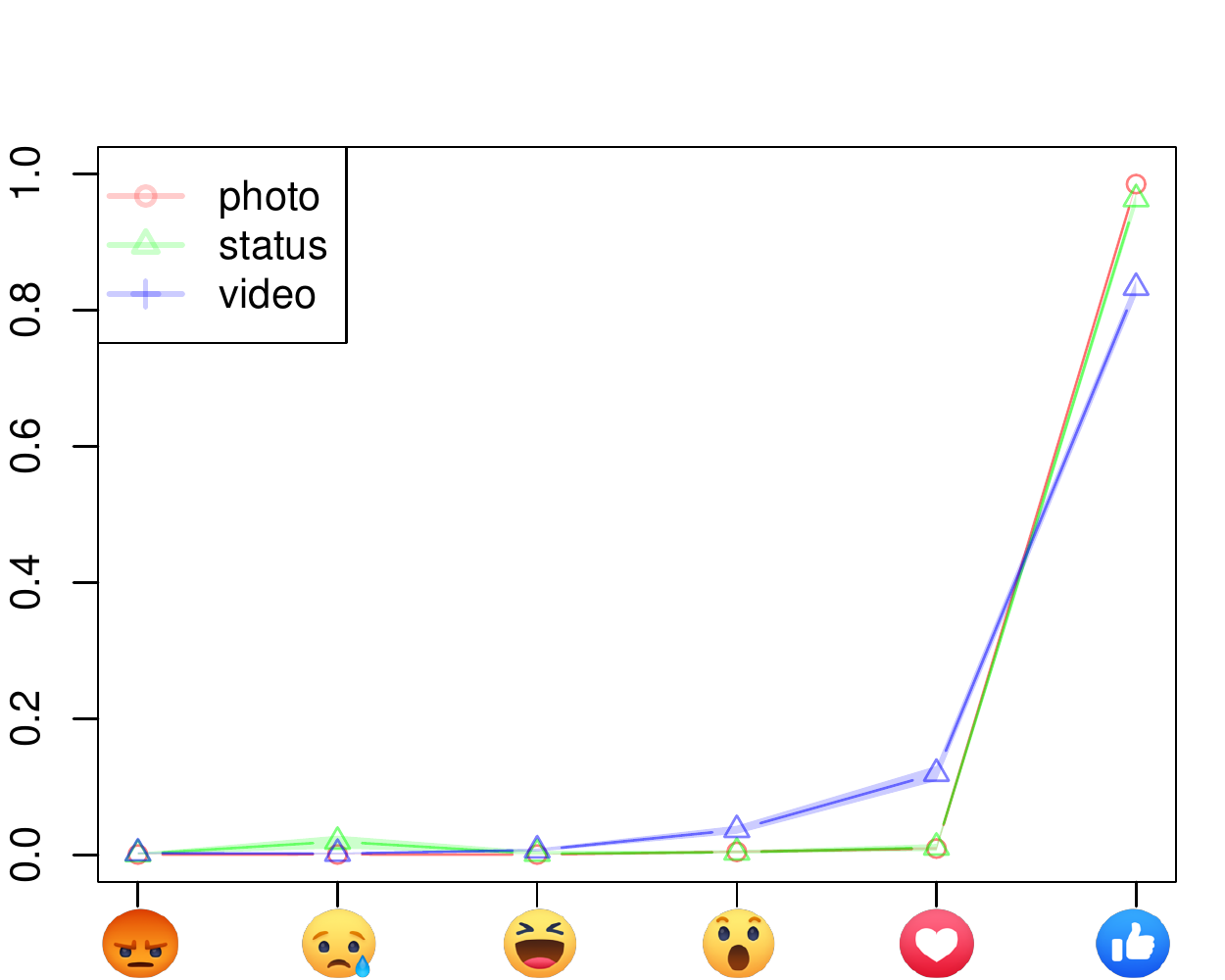} \\
$\nu^2=100$: cluster 1 ($n_1 = 47$) & $\nu^2=1$: cluster 1 ($n_1 = 50$) \\
\includegraphics[width = 0.4\textwidth]{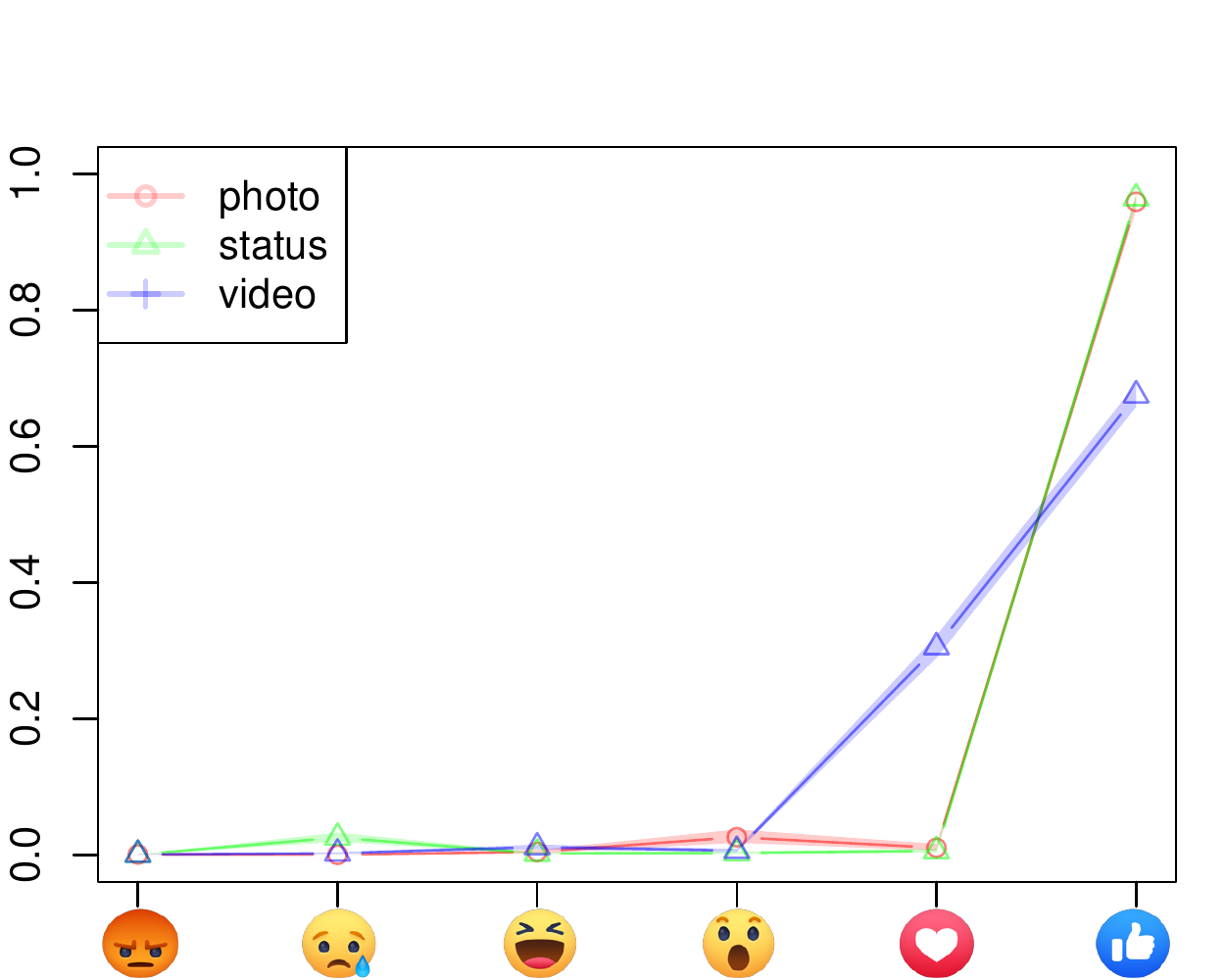} & 
\includegraphics[width = 0.4\textwidth]{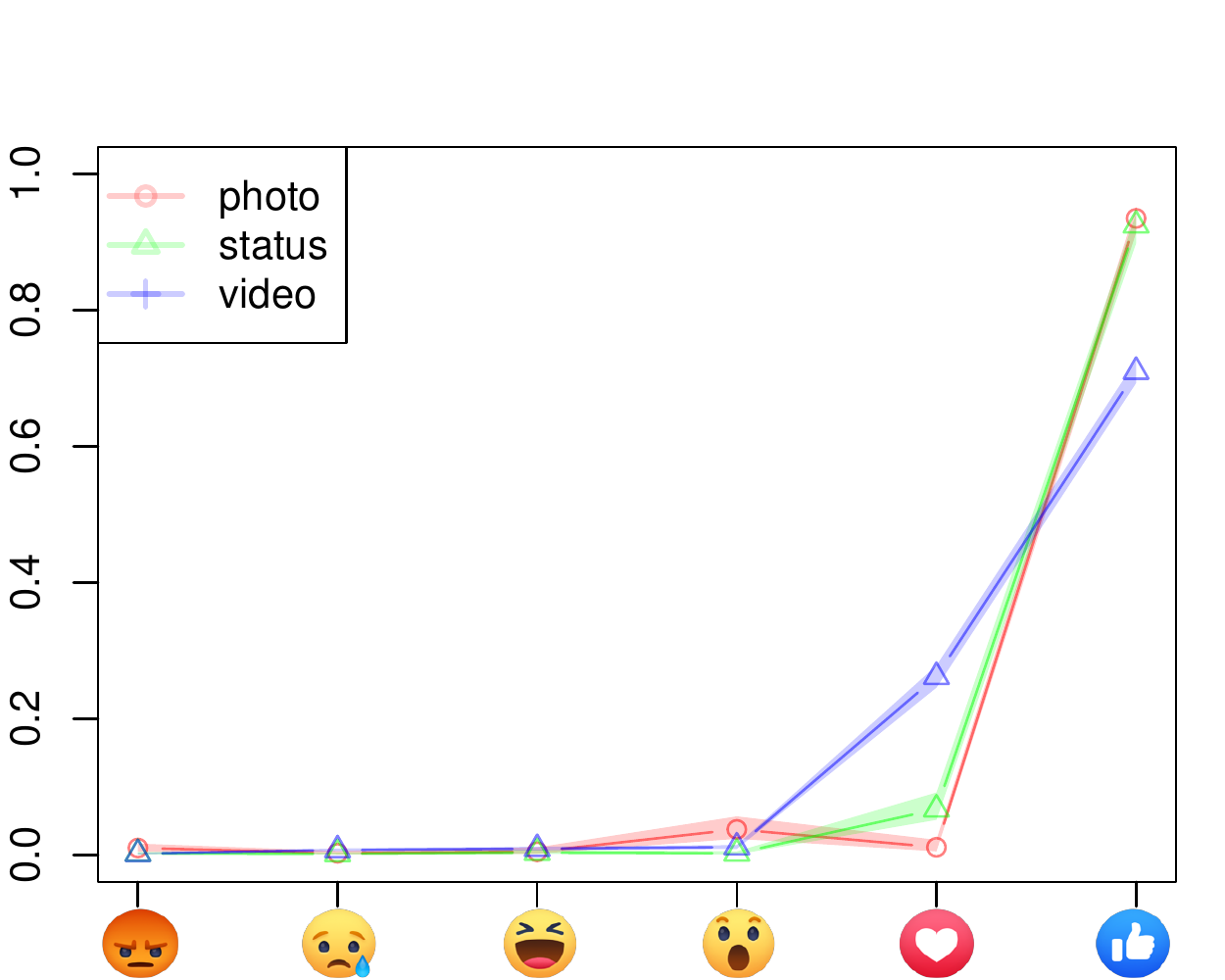}\\
$\nu^2=100$: cluster 2 ($n_2 = 44$)& $\nu^2=1$: cluster 2 ($n_2 = 30$)\\
\includegraphics[width = 0.4\textwidth]{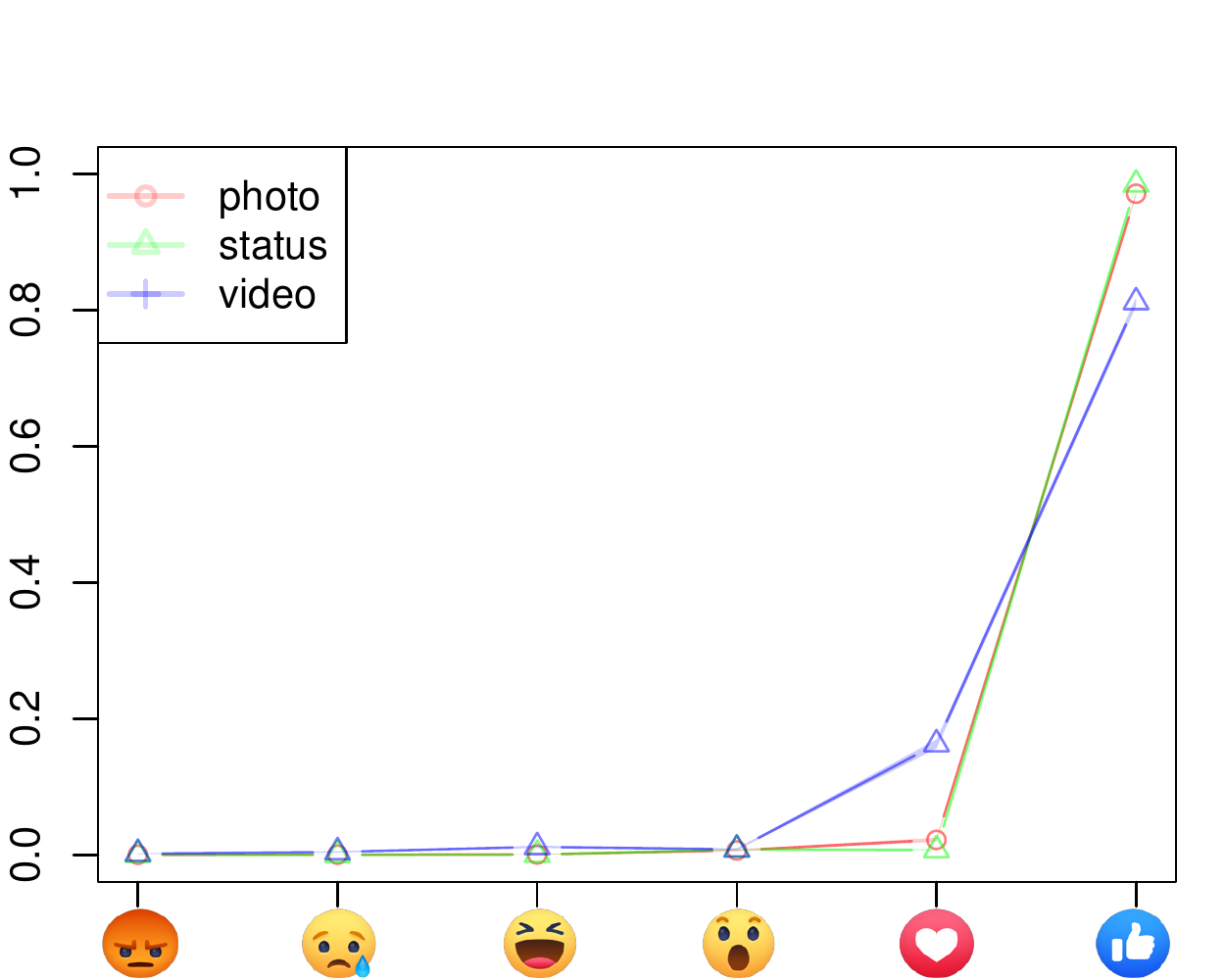}&
\includegraphics[width = 0.4\textwidth]{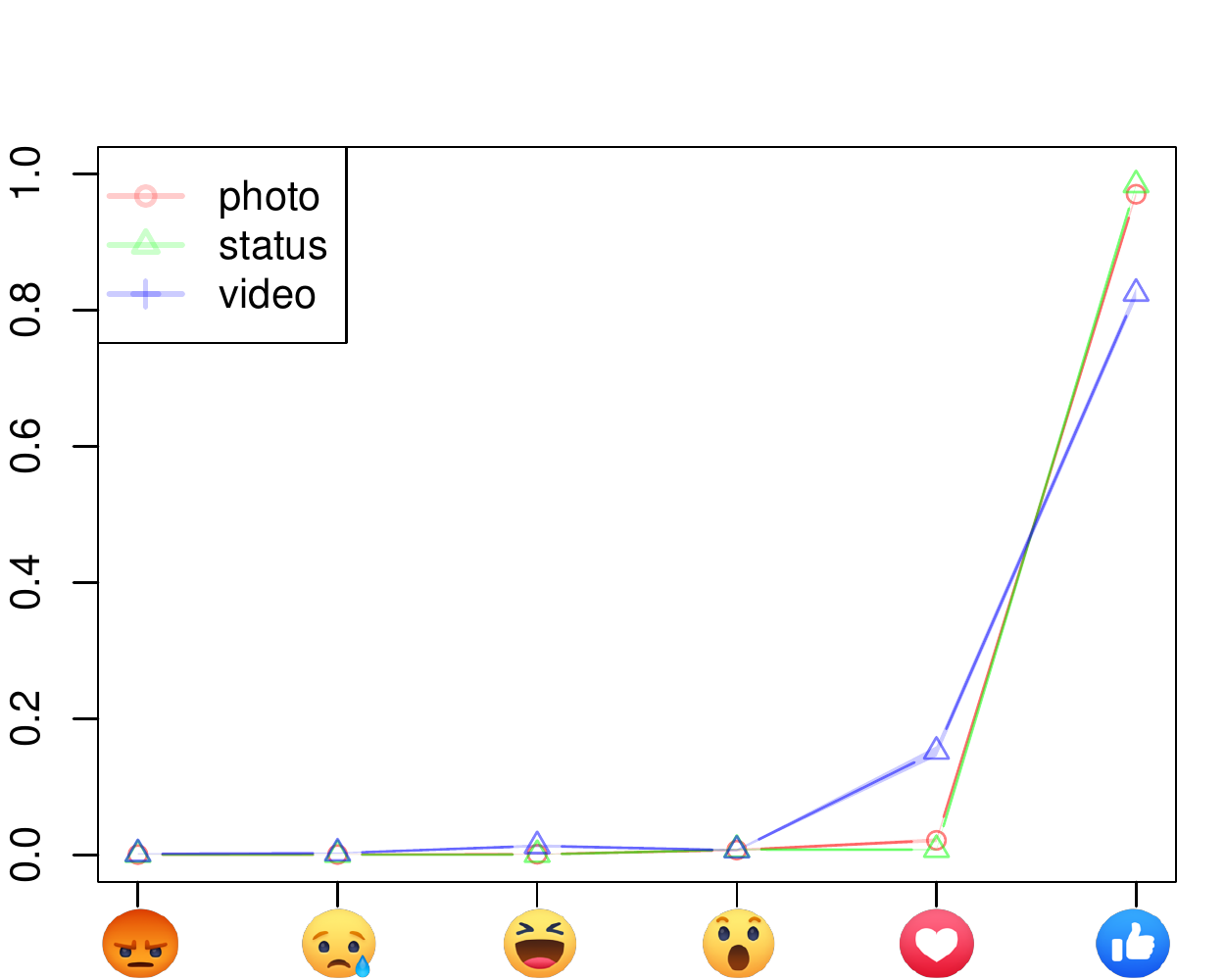}\\
$\nu^2=100$: cluster 3 ($n_3 = 187$) & $\nu^2=1$: cluster 3 ($n_3 = 193$) \\
\includegraphics[width = 0.4\textwidth]{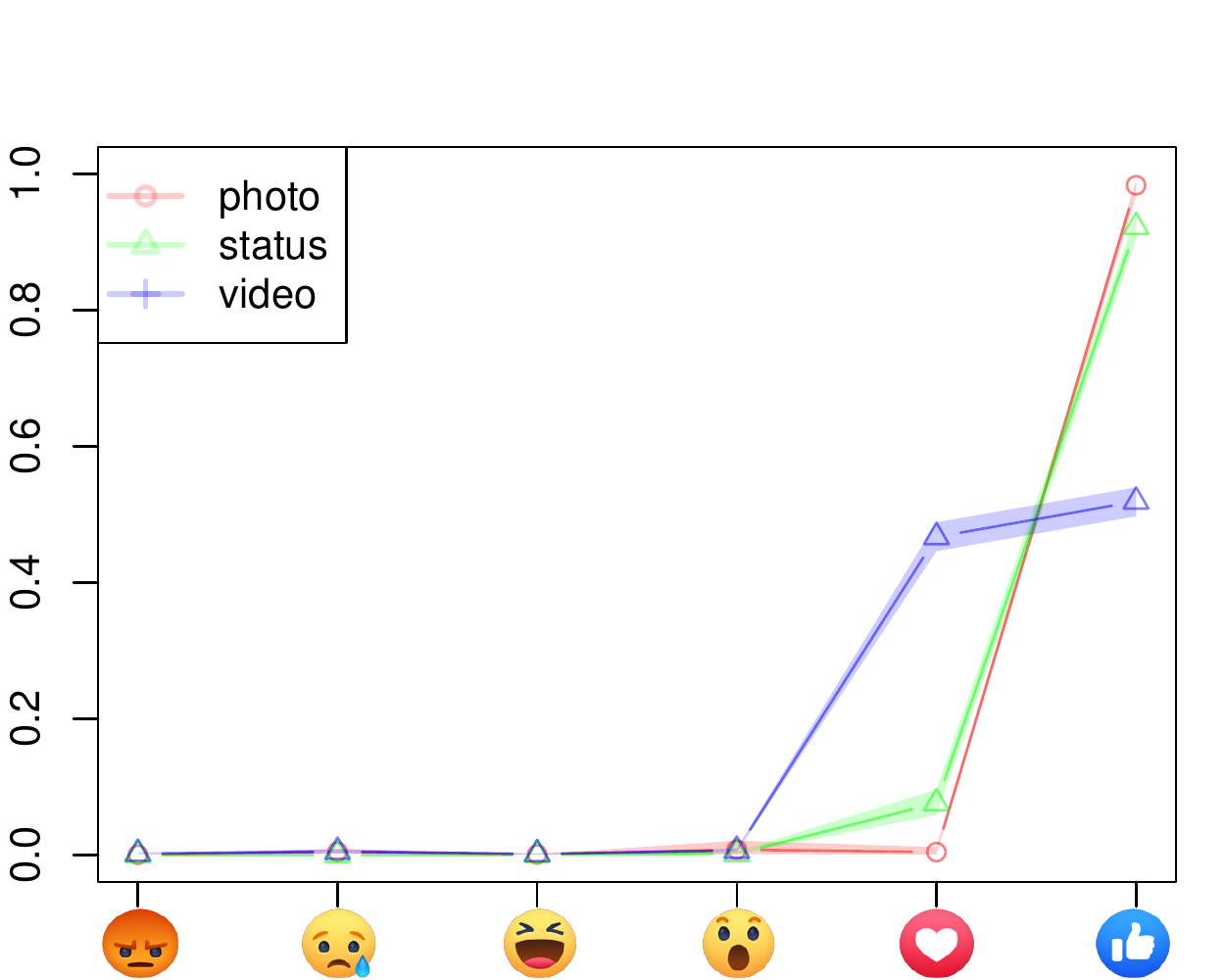}&
\includegraphics[width = 0.4\textwidth]{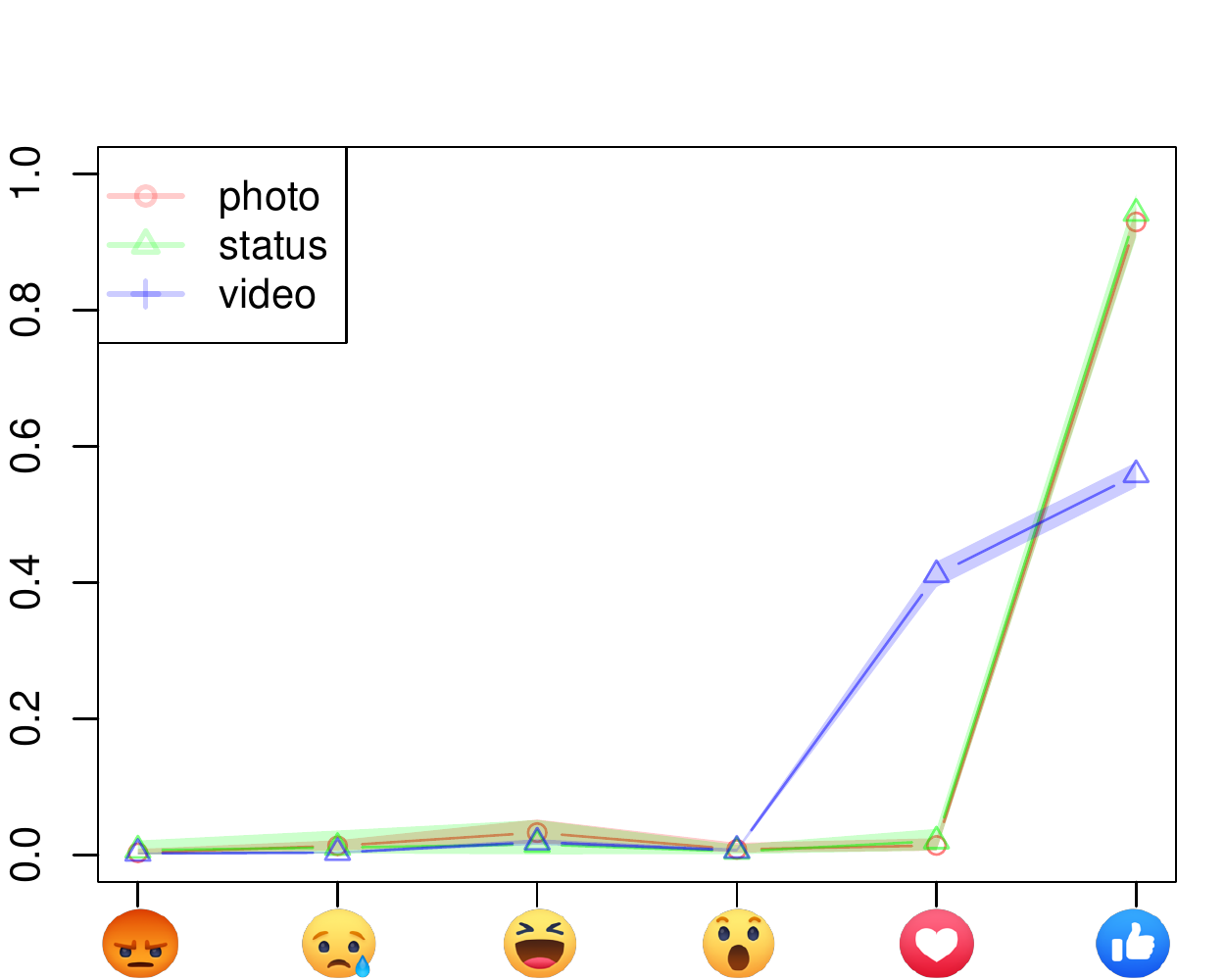}\\
$\nu^2=100$: cluster 4 ($n_4 = 22$)& $\nu^2=1$: cluster 4 ($n_4 = 27$)\\
\end{tabular}
\caption{Posterior mean and $95\%$ Credible Region of the reaction probabilities $(\theta_1,\theta_2,\theta_3,\theta_4,\theta_5,\theta_6)$ per cluster for the Facebook Sellers data, when setting the continuous covariate (number of shares) equal to its mean per post type. The left and right columns correspond to the MCMC samplers with the large ($\nu^2 = 100$) and small ($\nu^2 = 1$) prior variance of the regression coefficients in Equation \eqref{eq:betaPrior}, respectively. }
\label{fig:fb300}
\end{figure}

We applied our method using the EM algorithm with the proposed initialization scheme as well as the MCMC sampler using an overfitting mixture model with $K_{\max} = 10$ components. A total of 12 chains under the prior parallel tempering scheme were considered. The MCMC sampler ran for an initial warm-up period of 100000 iterations, followed by 400000 iterations. A thinned MCMC sample of 20000 iterations was retained for inference. Both methods select $K  = 4$ clusters.  In the MCMC sampler we have considered two different levels of the prior variance of the regression coefficients, that is, $\nu^2 = 100$ (vague prior distribution) and $\nu^2=1$ (informative prior). 

More specifically, for the EM algorithm the minimum value of ICL   is equal to $4610.89$ (corresponding to a model with $K=4$ clusters) while for the MCMC sampler the mode of the posterior distribution of the number of non-empty mixture components corresponds to $K_0 = 4$, with $\hat P(K_0 = 4\vert\rm{data}) = 0.67$ for the sampler with $\nu^2 = 100$. The same number of components is also selected when considering the prior distribution with $\nu^2 =1$, where $\hat P(K_0 = 4\vert\rm{data}) = 0.80$. The confusion matrix of the single best clusterings arising after applying the Maximum A Posteriori rule is displayed in Table \ref{tab:fb}. The corresponding adjusted Rand Indices  between the two partitions are equal to $\mathrm{ARI}(\mathrm{EM}, \mathrm{MCMC}(100)) \approx 0.96$, $\mathrm{ARI}(\mathrm{EM}, \mathrm{MCMC}(1)) \approx 0.74$ and $\mathrm{ARI}(\mathrm{EM}, \mathrm{MCMC}(100)) \approx 0.76$, indicating a high level of agreement between the three approaches. 

Next we are concerned with the identifiability of the selected model with $K = 4$ components. Recall that in our extracted dataset the minimum number of reactions is equal to 40, thus,  condition 1.(a) in Theorem 2 of \cite{grun2008identifiability} (see also \cite{hennig2000identifiablity}) is satisfied. If we were considering only the categorical predictor (video type)  in our model, the number of distinct hyperplanes (lines in this case) needed to cover the covariates of each cluster would be equal to 2: one line covering the points $(0,0)$ (origin) and $(x_i^{\mathrm{status}}, x_i^{\mathrm{photo}}) = (1,0)$ and another line covering the points $(0,0)$ and $(x_i^{\mathrm{status}}, x_i^{\mathrm{photo}}) = (0,1)$. This number is less than the selected number of clusters and the coverage condition (see condition 1.(b) in Theorem 2 of \cite{grun2008identifiability}) is violated. However, this condition is satisfied after including a continuous covariate with cluster-specific effect (number of shares). Finally, the generated MCMC sample has been post-processed according to the ECR algorithm \citep{Papastamoulis:10, papastamoulis2016label}  in order to deal with the label switching issue.

\begin{table}[ht]
\caption{Confusion matrix between the single best clustering of the Facebook Live Sellers Dataset arising from the EM and MCMC algorithms with a prior variance of regression coefficients equal to $\nu^2 = 100$ and $\nu^2=1$ (after post-processing the MCMC output for correcting label switching). }
\centering
\begin{tabular}{rrrrrrrrrr}
  \toprule
  &&\multicolumn{4}{c}{MCMC ($\nu^2 = 100$)} &\multicolumn{4}{c}{MCMC ($\nu^2 = 1$)}\\
  \midrule
  & & 1 & 2 & 3 & 4 & 1 & 2 & 3 & 4\\ 
  \midrule
\multirow{4}{*}{EM} & 1& 47 &   0 &   0 &   0 & 37 &   2 &   8 &   0\\ 
&  2 & 0 &  44 &   3 &   0 & 10 &  20 &   6 &  11\\ 
 & 3 &   0 &   0 & 183 &   0 & 2 &   3 & 178 &   0\\ 
 & 4 &   0 &   0 &   1 &  22&   1 &   5 &   1 &  16\\ 
   \bottomrule
\end{tabular}
\label{tab:fb}
\end{table}

Figure \ref{fig:fb300} displays the posterior mean of reaction probability per cluster and the corresponding (equally tailed) $95\%$ credible interval, for both prior setups. The continuous covariate (number of shares) is set equal to the observed mean per post type. In all cases, there is an increased probability of a ``love'' reaction when the post is a video, compared to a photo or a status. However, the average probability of such a reaction is different between the clusters, with the most notable difference obtained in cluster ``4''. Finally, notice the similarity of cluster profiles for both prior distributions.

\section{Discussion}\label{sec:disc}

The problem of clustering multinomial count data under the presence of covariates has been treated using a frequentist as well as a Bayesian approach. Our simulations showed that our proposed models perform well, provided that the suggested estimation and initialization schemes are selected. The application of our method in clustering real count datasets reveal the interpretability of our approach in real-world data. Our contributed package in {\tt R} makes our method directly available to the research community.

Under a frequentist approach we have demonstrated that an efficient initialization (i.e.~the split-shake-random small-EM scheme in Section \ref{sec:em_init}) yields improved results, when compared to a more standard random small-EM initialization scheme. Furthermore, a crucial point for the implementation of the maximization step of the EM algorithm is the control of the step size of the Newton-Raphson iterations, something that was achieved using the ridge-stabilized version in Section \ref{sec:mstep}. 

We did not address the issue of estimating standard errors in our EM implementation. However, these can be obtained by approximating the covariance matrix of the estimates by the inverse of the observed information matrix \citep{louis1982finding, meng1991using, jamshidian2000standard} or using bootstrap approaches \citep{basford1997standard, mclachlan1999emmix, grun2004bootstrapping, galindo2006avoiding}. Maximum likelihood estimation with the EM algorithm can be modified in order to provide Maximum A Posteriori estimates under a regularized likelihood approach, as implemented in \cite{galindo2006avoiding}. 

The Bayesian framework of Section \ref{sec:bayes} has clear benefits over the frequentist approach, but of course, under the cost of increased computing time. As demonstrated in our simulations, the proposed MCMC scheme outperforms the EM algorithm in terms of estimation of the number of clusters as well the clustering of the observed data in terms of the Adjusted Rand Index. Moreover, the Bayesian setup allows for even greater flexibility in the resulting inference, such as the calculation of Bayesian credible intervals from the MCMC output which provide a direct assessment of the uncertainty in our point-estimates.  For this purpose we used state-of-the-art algorithms that deal with the label switching problem in mixture, suitably adjusted to the special framework of overfitting mixture models.

A natural and interesting extension of our research is to consider the problem of variable selection in model based clustering \citep{maugis2009variable, dean2010latent, yau2011hierarchical, 10.1214/18-SS119}. In the Bayesian setting, one could take into account alternative prior distributions of the multinomial logit coefficients per cluster, e.g.~spike and slab or shrinkage prior distributions \citep{malsiner2016model, vavra2022clusterwise} that encourage sparsity in the model. Another direction for future research is to combine our mixture model with alternative Bayesian logistic regression models that exploit data augmentation schemes \citep{held2006bayesian, fruhwirth2010data, polson2013bayesian, choi2013polya} and assess whether MCMC inference is improved.

\section*{Declarations}

\paragraph{Acknowledgements} This study was funded by the  program ``DRASI 1/Research project: 11338201'', coordinated by the Research Center of the Athens University of Economics and Business (RC/AUEB). The author wishes to thank Prof.~Dimitris Karlis (Athens University of Economics and Business) for fruitful discussions and three anonymous reviewers for their valuable comments and suggestions. 

\paragraph{Availability} All datasets and software used in this paper are available online at \url{https://github.com/mqbssppe/multinomialLogitMix}.

\bibliographystyle{apalike}
\bibliography{main}

\appendix

\section*{Appendix}

\renewcommand{\thesubsection}{\Alph{subsection}}
\addcontentsline{toc}{section}{Appendices}
\renewcommand{\theequation}{\thesubsection.\arabic{equation}}
\renewcommand\thefigure{\thesubsection.\arabic{figure}}
\setcounter{equation}{0}
\setcounter{figure}{0}

\section{Details of the MCMC sampler}\label{sec:mcmc_details}

This section describes the values of the parameters for the EM and MCMC sampler, which were used in order to produce the results reported in our simulation study in Section \ref{sec:sim}.

\paragraph{Prior parameters and number of parallel chains:} We implemented the prior parallel tempering MALA-within-Gibbs algorithm described in the previous section  with the following set-up. We used 8 parallel chains and overfitted mixtures with a  total of $K_{\max}=20$ components. For chain $c = 1,\ldots,C$, the parameters of the Dirichlet prior $\mathcal D(\alpha_c,\ldots,\alpha_c)$  were set equal to 
\begin{align}
\alpha_1 &=\frac{1}{200}\quad\mbox{(target chain)} \label{eq:dirPriorSim1}\\
\alpha_c &=\frac{1}{200} + \frac{1}{4000}\exp\left\{2+12\frac{c-2}{C-2}\right\},\quad \mbox{for chain}\quad c=2,\ldots,C. \label{eq:dirPriorSim2}
\end{align}
The specific set of values worked reasonably well in our simulations and applications on real datasets, however, we do not claim that they are in any way ``optimal'' choices. Our guide for choosing this set of values is to achieve  average acceptance rate between chain swaps around $10\%-30\%$.
The variance of the normal prior distribution of the coefficients in \eqref{eq:betaPrior} is set to $\tau^2 = 100$. Notice that in our applications we have also presented results with a much smaller prior variance ($\nu^2 = 1$), a choice which can be seen as a regularized estimation approach. 

\paragraph{Warm-up period:} The warm-up period of the MCMC sampler consists of $m_0 = 48000$ iterations. As the sampler progresses, it keeps track of the proposal acceptance ratio within the last $500$ iterations. The parameter $\nu$ (which controls the scale of the MALA proposal) is adaptively tuned in order the proposal acceptance rate stays within the range $15\% - 25\%$. In case that a sequence of $500$ iterations the acceptance rate is less than $15\%$ then $\nu \leftarrow 0.9\nu$. On the other hand, if the acceptance rate is too large (larger than $25\%$), then $\nu \leftarrow \nu/0.9$. The value of $\nu$ obtained at the last iteration of this initial phase is used at the main MCMC sampler. The initial value is set equal to $\nu =  0.00035$. 

\paragraph{Main MCMC sampler:} After the warm-up period, the MCMC sampler runs for a total of $T = 2600$ cycles. Each cycle consists of $m_1 = 20$ iterations. A chain swap is attempted at the end of each MCMC cycle. The results are obtained retaining the last $2500$ cycles (and discarding the first $100$ cycles as burn-in period) of the MCMC sampler. 

Note that the initial warm-up period and the main MCMC sampler consist  of $48000 + 2600\times 20 = 100000$ MCMC iterations in total. 

\paragraph{MCMC sampler initialization:} Each chain was initialized under two different schemes: a scheme based on randomly selected started values (which we will be referring to as ``MCMC-RANDOM'') and a more elaborate initialization scheme based on the output of the EM algorithm (``MCMC-EM'' starting scheme). More specifically, in ``MCMC-RANDOM'' all parameters are initialized by simulating from the prior distributions. In ``MCMC-EM'' we first estimate the number of clusters as well as the model parameters according to the EM algorithm under our split-small-EM scheme. Let us denote by $\hat K^{(EM)}$ the selected number of clusters according to the EM algorithm (using ICL), under the split-small-EM initialization. Next, consider a Bayesian overfitting mixture with $K_{\max} > K^{(EM)}$ components. The parameters of the first $K^{(EM)}$ components are all set equal to the values of the corresponding parameters obtained at the last iteration of the EM algorithm for that particular model. The parameters of the remaining $K_{\max} - K$ components are all initialized by a zero value. Finally, a random permutation is drawn among the initial parameters of the $K_{\max}$ components for each of the different chains in order to encourage the presence of the label switching phenomenon in the MCMC sampler.

\section{Computational details - package in R}

\begin{figure}[ht]
\begin{tabular}{cc}
\includegraphics[scale=0.5]{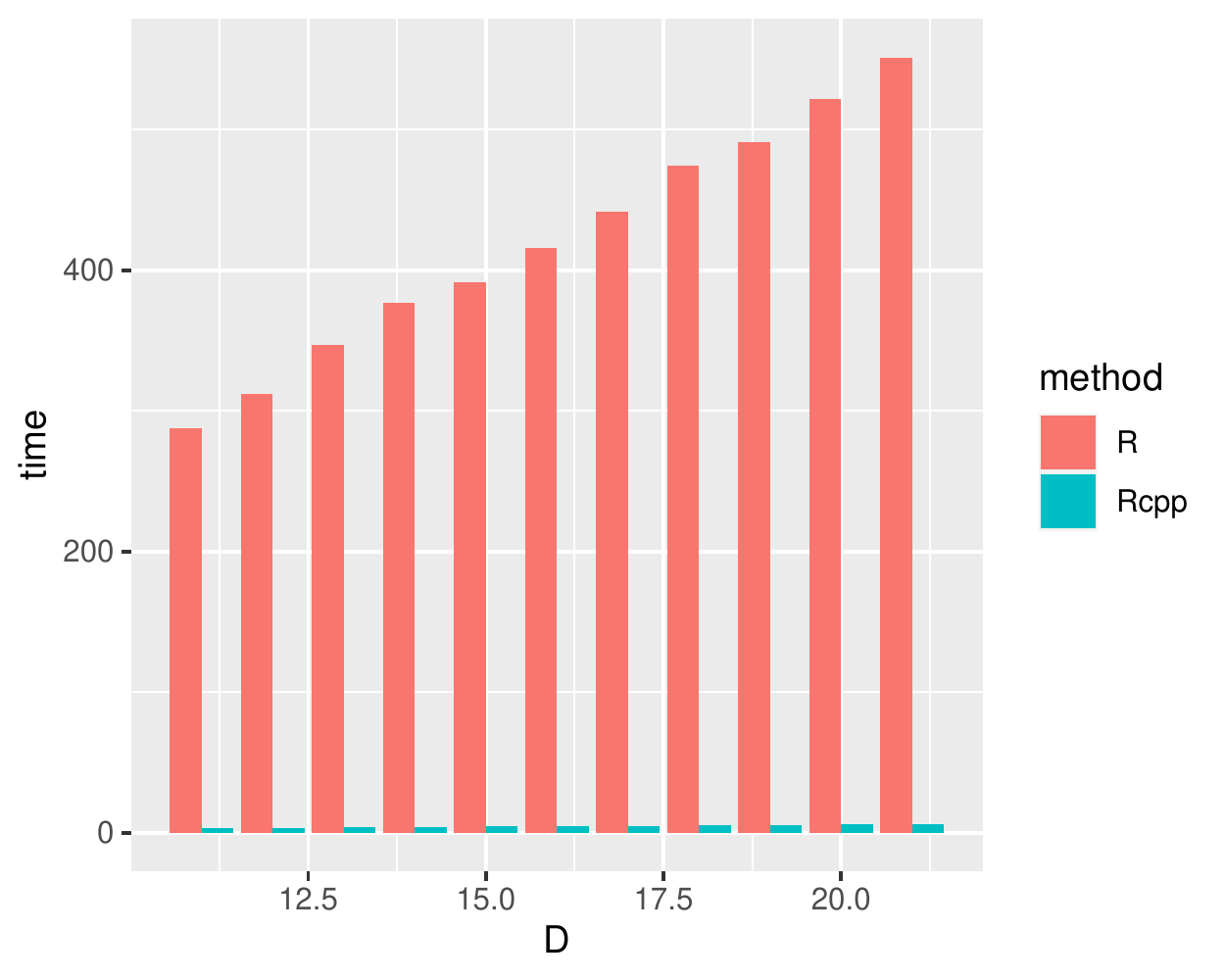}
&
\includegraphics[scale=0.5]{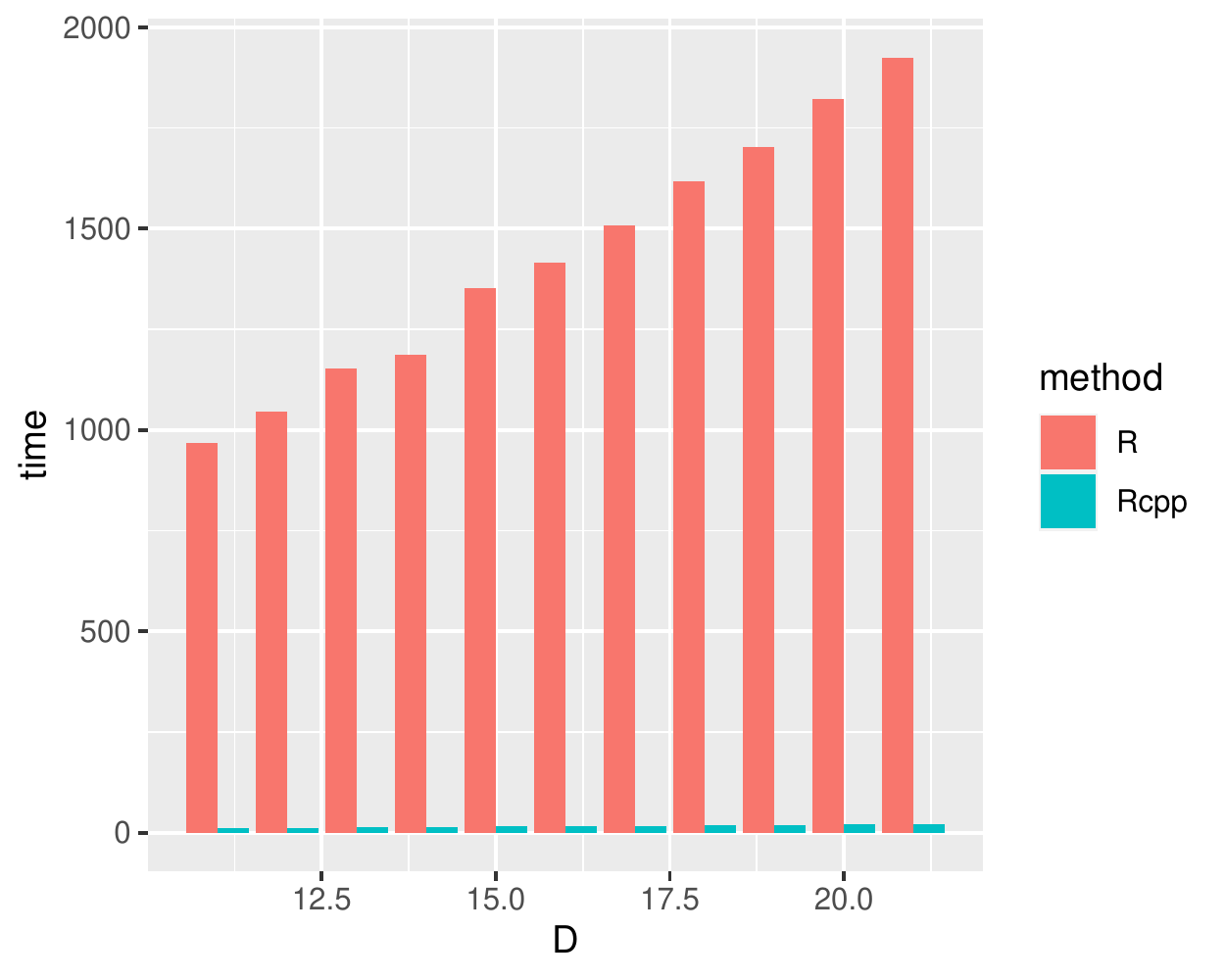}\\
$K = 3$ & $K = 10$
\end{tabular}
\caption{Benchmarking runtime for the MALA proposal (Step 3 of Algorithm \ref{alg:mala}) between {\tt R} and {\tt Rcpp} for different values of number of mixture components ($K$) and multinomial categories ($D$). }
\label{fig:timeRcpp}
\end{figure}

The computational pipeline for the proposed methodology has been implemented in {\tt R}. It is furthermore publicly available as a contributed {\tt R} package named {\tt multinomialLogitMix}, which is available at \url{https://CRAN.R-project.org/package=multinomialLogitMix}. 
The proposal scheme of the MALA sampler is implemented in the  {\tt Rcpp} \citep{rcpp1, rcpp2, rcpp3} and {\tt RcppArmadillo} \citep{RcppArmadillo} packages, which integrate R and C++. Figure \ref{fig:timeRcpp} illustrates that the gain in computing time is tremendous when replacing the R code with Rcpp. 

The basic pipeline is illustrated next. For this purpose we use a simulated dataset as the ones in Section \ref{sec:sim}.

\begin{lstlisting}
library("multinomialLogitMix")
set.seed(727)
# sample size
n <- 250
# number of covariates (constant term included)
p <- 3  
# number of multinomial categories
D <- 6  
# number of clusters
K <- 2      
# maximum number of clusters
Kmax <- 10
# generate synthetic as in Section 4.1 of the paper
simData <- simulate_multinomial_data(K = K, p= p, D = D, n = n , size = 20, prob = 0.025)   
# response (multinomial counts)
y <- simData$count_data
# design matrix
X <- simData$design_matrix
# preview multivariate response counts
> y
       [,1] [,2] [,3] [,4] [,5] [,6]
  [1,]  209  118    0    0  536  124
  [2,]  703    0    0    0    0    0
<..........(+ 246 rows of counts)..>
[249,]  819    2   47    4    3    3
[250,]  710    0    0    0    0    1
# preview design matrix
> X
       [,1]          [,2]         [,3]
  [1,]    1 -0.8469699933  0.088500232
  [2,]    1  0.5114674818  2.136154036
<..........(+ 246 rows)................>
[249,]    1  0.1812801447  0.583363993
[250,]    1  0.5725885602  1.734839847
# set the number of cores to 8
nCores <- 8
# Run both EM and MCMC with default setttings (based on the split small-EM initialization scheme)
mlm_split <- multinomialLogitMix(response = y, 
          design_matrix = X, method = "MCMC", 
          Kmax = Kmax, nCores = nCores, splitSmallEM = TRUE)

\end{lstlisting}

\vspace{2ex}

Now let's explore the output based on the EM algorithm only. At first we retrieve the selected number of clusters according to ICL. Then we display the estimated clustering conditional on the selected value. Finally, we retrieve the estimated parameters (mixing proportions and coefficients of the mixture of multinomial logits) of the model.

\begin{lstlisting}
# estimated number of clusters
> mlm_split$EM$estimated_K
[1] 2
> # estimated number of assigned observations per cluster
> table(mlm_split$EM$estimated_clustering)
  1   2 
 68 182 
> # estimated mixing proportions
> # estimated coefficients
> round(mlm_split$EM$all_runs[[2]]$beta,2)
, , 1

      [,1]  [,2]  [,3]
[1,] -0.07  0.05  9.78
[2,] -1.80  4.26  2.21
[3,]  0.01 -3.93  5.51
[4,] -0.03  3.78 -0.57
[5,] -0.04 -0.59 -0.01

, , 2

       [,1]  [,2]  [,3]
[1,]   0.16  0.01  4.71
[2,]  -0.02  0.00 -0.02
[3,] -22.73 -1.35  1.43
[4,]  -9.92  1.83 -0.13
[5,]   0.00 -1.73 -0.05
\end{lstlisting}
\vspace{2ex}

Note that in the last chunk of the output the rows correspond to multinomial categories (there are 6 categories in total so $j=1,\ldots,5$) and the columns correspond to covariates ($p=1,\ldots,3$), per cluster. For example, the estimate of the coefficient of the  second multinomial category ($j=2$) for the second covariate (so $p=3$ because the model includes a constant term) for cluster 1 ($k=1$) is equal to $\hat\beta_{kjp}=\hat\beta_{1,2,3} = 2.21$. The corresponding estimate for cluster 2 is equal to $\hat\beta_{kjp}=\hat\beta_{2,2,3} = -0.02$.

\begin{figure}[ht]
\centering
\includegraphics[scale=0.5]{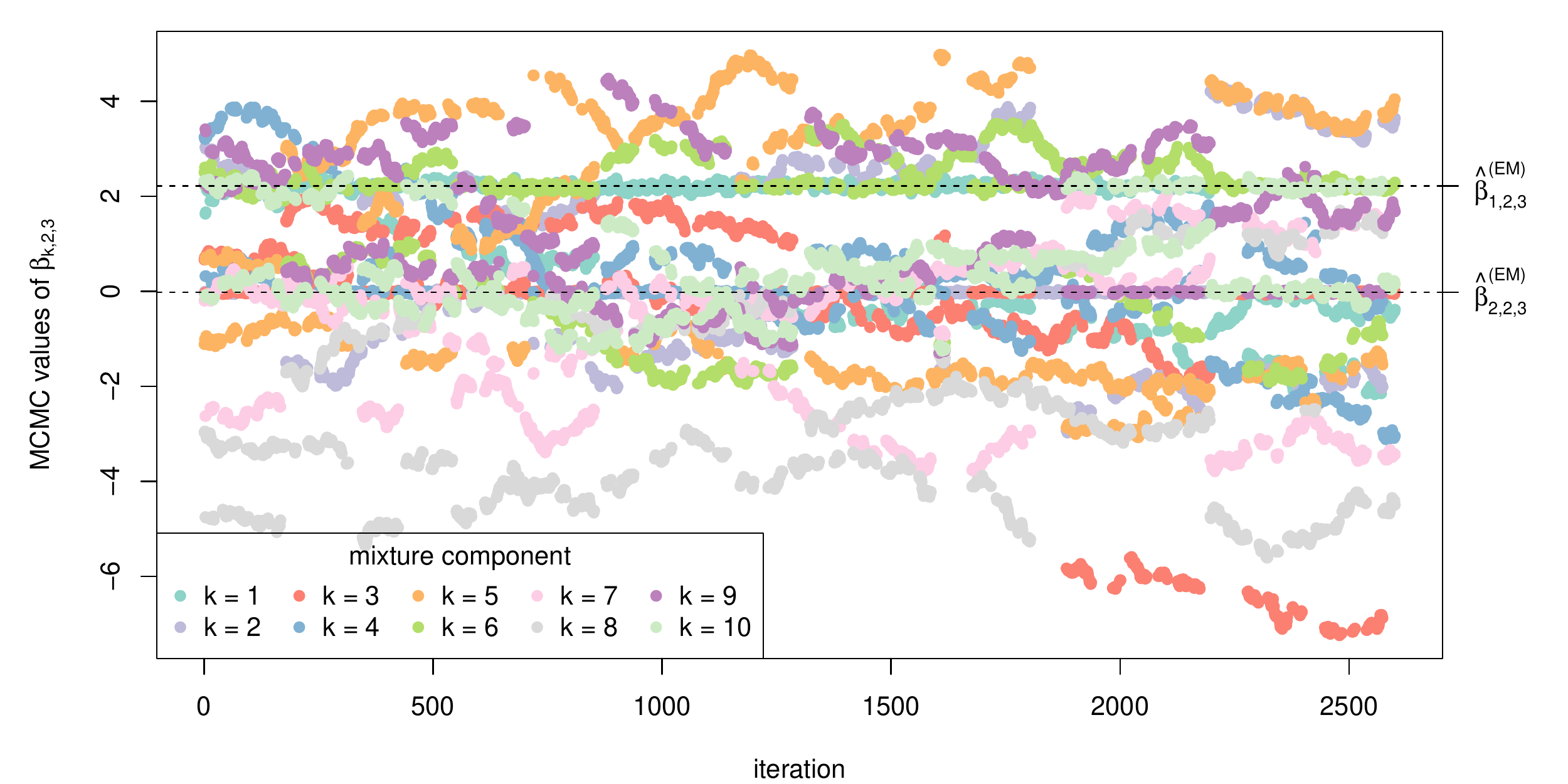}
\caption{Raw MCMC output (after an initial warm-up period) for the coefficients $\beta_{k,2,3}$ for component $k = 1,2,\ldots,10$ of the overfitting mixture model. The dotted lines correspond to the estimate of $\beta_{k,2,3}$ for $k=1,2$ when using a mixture model with $K=2$ components. }
\label{fig:mcmc_example_raw}
\end{figure}

Let's explore the MCMC output now.  We stress once again that the raw MCMC sample of the overfitting mixture is not directly interpretable due to label switching. Moreover, bear in mind that it also consists of the values of the empty mixture components which are not relevant for all practical purposes. These points are illustrated in Figure \ref{fig:mcmc_example_raw}, which displays the raw MCMC output for $\beta_{k,2,3}$ for all components $k = 1,2,\ldots,10$ of the overfitting mixture model. A careful inspection of this graph reveals that up to a switching of the labels  the sampled values of 2 (among 10 components) are concentrated around the two horizontal dotted lines which correspond the estimates of the corresponding parameters according to the EM algorithm. The remaining values which are further away from the dotted lines  correspond to the sampled values of this parameter for the remaining 8 empty mixture components (again up to switching of the labels). 

Let's proceed now by inspecting the post-processed output according to the ECR algorithm. At first we can retrieve the estimated posterior distribution of the number of clusters (which correspond to the number of non-empty mixture components across the MCMC run) as well as the number of assigned observation per cluster, conditional on the value of the most probable number of clusters.

\begin{lstlisting}
#estimated posterior distribution of the number of clusters
> mlm_split$MCMC_post_processed$nClusters_posterior
     2      3 
0.9992 0.0008 
# estimated number of assigned observations per cluster according to the MCMC sampler
> table(mlm_split$MCMC_post_processed$cluster)
  1   2 
 68 182 
\end{lstlisting}
\vspace{2ex}

\begin{figure}[ht]
\centering
\includegraphics[scale=0.4]{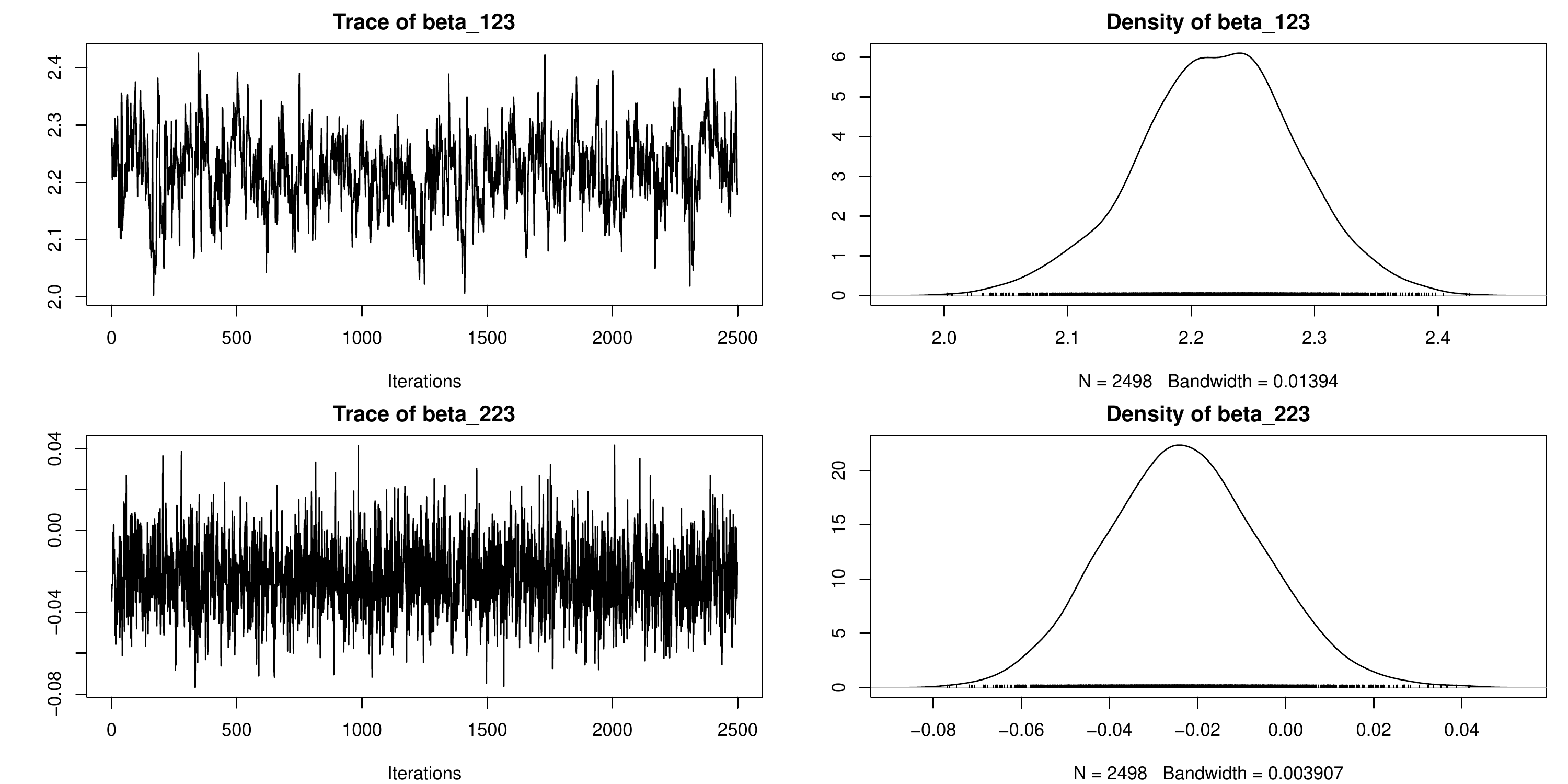}
\caption{Post-processed MCMC outputs for $\beta_{1,2,3}$ (top) and $\beta_{2,2,3}$ (bottom).}
\label{fig:mcmc_example}
\end{figure}

Now we concentrate on the post-processed output of the non-empty mixture components, so we essentially discard the sampled values of the remaining 8 empty mixture components. We can retrieve basic MCMC summaries using the {\tt coda} package. For example let us retrieve MCMC summaries for the coefficients $\beta_{1,2,3}$ and $\beta_{2,2,3}$.  
\begin{lstlisting}
> beta_123 <- mlm_split$MCMC_post_processed$mcmc[[2]][[3]][,1,2]
> beta_223 <- mlm_split$MCMC_post_processed$mcmc[[2]][[3]][,2,2]
> betas <- as.mcmc(cbind(beta_123, beta_223))
> summary(betas)

Iterations = 1:2498
Thinning interval = 1 
Number of chains = 1 
Sample size per chain = 2498 

1. Empirical mean and standard deviation for each variable,
   plus standard error of the mean:

            Mean     SD Naive SE Time-series SE
beta_123  2.2207 0.0643 0.001286       0.004478
beta_223 -0.0229 0.0178 0.000357       0.000434

2. Quantiles for each variable:

            2.5%    25%     50%     75%  97.5%
beta_123  2.0872  2.179  2.2221  2.2634 2.3436
beta_223 -0.0566 -0.035 -0.0235 -0.0114 0.0129
> plot(betas)
\end{lstlisting}
\vspace{2ex}

The last command produces Figure \ref{fig:mcmc_example}, which displays the trace of the post-processed values of $\beta_{1,2,3}$ (left) along with the corresponding estimate of the marginal posterior distribution (right).

Of particular interest is also the matrix of posterior membership probabilities for each observation. The next chunk of code shows how one can retrieve these estimates according to output from the EM and the MCMC algorithm, respectively. 
\begin{lstlisting}
# estimated posterior membership probabilities according to the EM algorithm
> round(mlm_split$EM$all_runs[[mlm_split$EM$estimated_K]]$posteriorProbabilities, 2)
       [,1] [,2]
  [1,] 0.00 1.00
  [2,] 0.38 0.62
<...(+246 rows)...>
[249,] 1.00 0.00
[250,] 0.00 1.00
# estimated posterior membership probabilities according to the MCMC algorithm
> round(mlm_split$MCMC_post_processed$posteriorProbabilities, 2)
       [,1] [,2]
  [1,] 0.00 1.00
  [2,] 0.37 0.63
<...(+246 rows)...>
[249,] 1.00 0.00
[250,] 0.00 1.00

\end{lstlisting}
\vspace{2ex}

The previous example uses the parameter setup detailed in Section \ref{sec:mcmc_details}. The user can modify these arguments by passing the desired input to the optional arguments {\tt em\_parameters} and {\tt mcmc\_parameters} of the {\tt  multinomialLogitMix()} function. Both {\tt em\_parameters} and {\tt mcmc\_parameters} should be lists, consisting of the following entries.
\paragraph{Arguments of the {\tt em\_parameters} list.\\ }
\begin{tabular}{rp{10cm}}
{\tt maxIter} & Maximum number of EM iterations. Default: 100.\\
{\tt emthreshold} & Positive real threshold for terminating the EM algorithm. The algorithm stops when the difference between two successive evaluations of the observed log-likelihood is less than this threshold. Default: $10^{-8}$. \\
{\tt maxNR} & Maximum number of Newton-Raphson iterations. Default: 10. \\
{\tt tsplit} & Number of different starts that will be used within the small-EM scheme (this quantity refers to all schemes: random, split and shake). Default: 16.\\
 {\tt msplit} & Number of iterations for each small-EM start. Default: 10.\\
 {\tt split} & Boolean denoting whether the EM algorithm will use the split-small EM scheme. Default: TRUE. In the opposite case, the small-EM scheme will use only randomly selected initial values. \\
  {\tt R0} &  The initial value for the parameter $R$ that controls the step-size of the ridge-stabilized Newton-Raphson scheme (see Equation \eqref{eq:ridge_step}). Default: 0.1. 
\end{tabular}
\paragraph{Arguments of the {\tt mcmc\_parameters} list.\\ }
\begin{tabular}{rp{10cm}}
{\tt tau} & initial value for the scale of the MALA proposal (positive, it corresponds to the parameter $\nu$ in Equation \eqref{eq:proposal}). Default: 0.00035. This parameter is adjusted in the initial (warm-up) period of the sampler in order to achieve the desirable acceptance rate of the MALA proposal. \\
{\tt nu2} &  the variance of the normal prior distribution of the logit coefficients (the parameter $\tau^2$ in Equation \eqref{eq:betaPrior}). Default: 100.\\
{\tt mcmc\_cycles} & Total number of MCMC cycles (after the initial warm-up) period of the sampler. Default: 2600. At the end of each MCMC cycle a swap between chains is attempted.\\
{\tt iter\_per\_cycle} & Number of MCMC iterations per cycle. Default: 20. \\
{\tt nChains}& Number of MCMC chains that run in parallel. Each chain uses a different prior distribution of the mixing proportions. The  inference is based on the first chain. \\
{\tt dirPriorAlphas} & The concentration parameter of the Dirichlet prior distributions per chain (see Equation \eqref{eq:dir_prior_tempering}). It should be a vector with length equal to {\tt nChains}. The default is: {\tt 5 * exp((seq(2, 14, length = nChains - 1)))/100)/(200)}, see Equations \eqref{eq:dirPriorSim1} and \eqref{eq:dirPriorSim2}.\\
{\tt warm\_up} & Initial warm-up period of the sampler, in order to adaptively tune the scale of the MALA proposal. Default: 48000.\\ 
{\tt checkAR} & Number of iterations required in order to adjust the scale of the proposal in  MALA mechanism during the initial warm-up phase of the sampler. Default: 500. \\
{\tt ar\_low} & Lowest threshold for the acceptance rate of the MALA proposal. Default: 0.15.\\
{\tt ar\_up} & Highest threshold for the acceptance rate of the MALA   proposal. Default: 0.25.\\
{\tt burn} & Number of MCMC cycles that will be discarded as burn-in. Default: 100. \\
{\tt withRandom} & Boolean value indicating whether or not to apply a random permutation in the supplied starting values for each chain. Default: true. 
\end{tabular}

\section{Further simulation results}
\label{app:more}

\subsection{Main simulation study}\label{sec:more1}

\begin{figure}[p]
\centering
\includegraphics[scale=0.45]{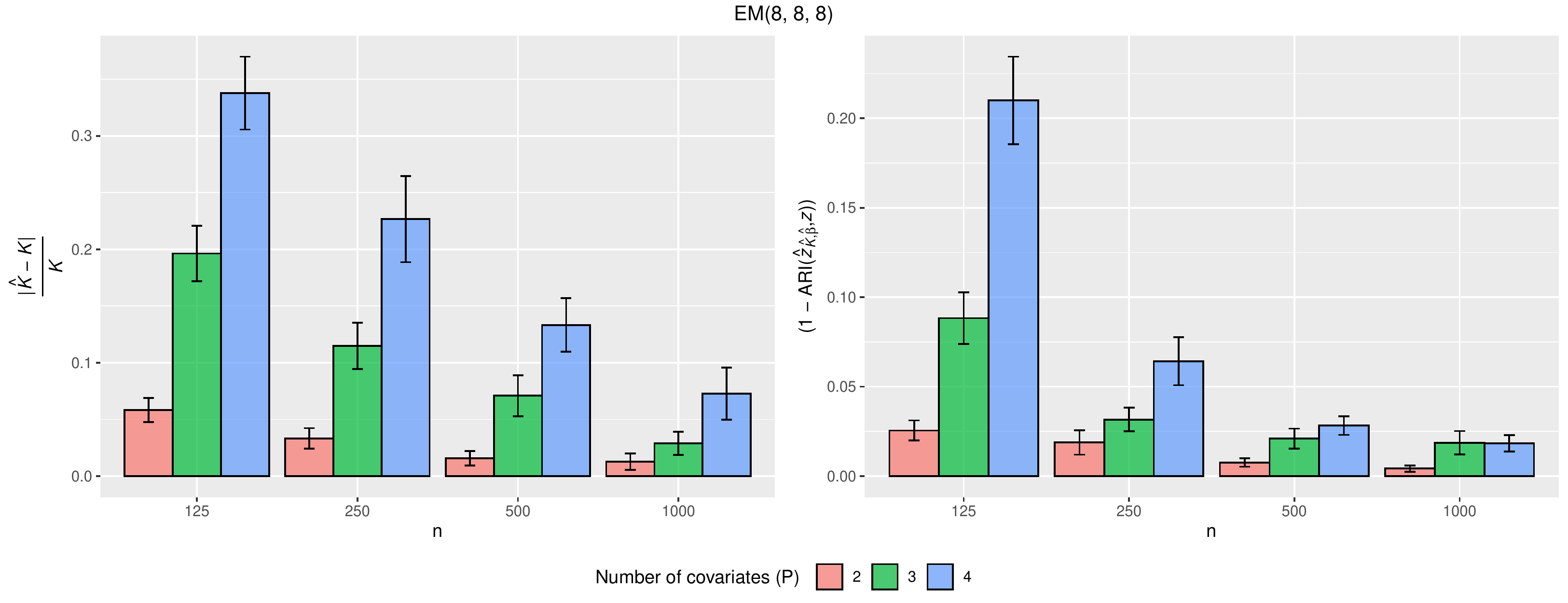}\\
\includegraphics[scale=0.45]{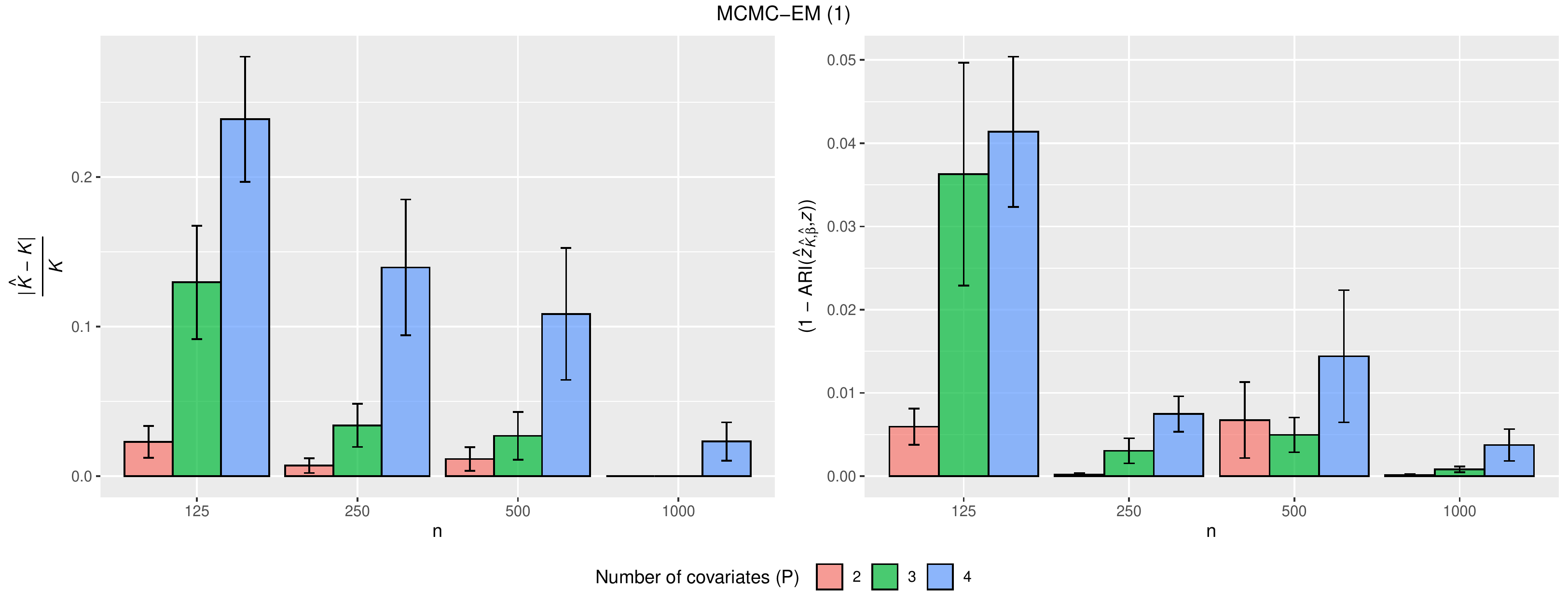}\\
\includegraphics[scale=0.45]{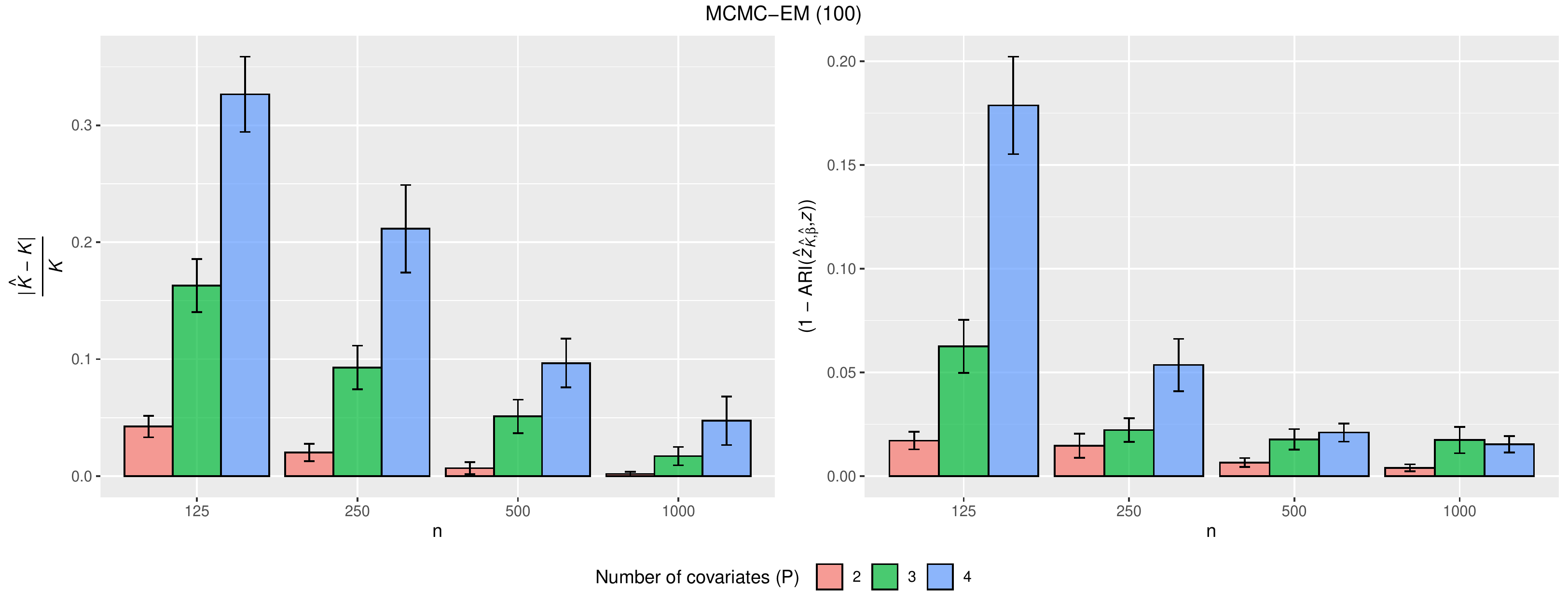}
\caption{Clustering accuracy for EM (up) and MCMC (middle and bottom) when taking into account sample size ($n$) and the number of covariates $P$ (including constant term) for the simulation study presented in Section \ref{sec:sim} of the manuscript.}
\label{fig:sim_versus_p}
\end{figure}

\begin{figure}[p]
\centering
\includegraphics[scale=0.6]{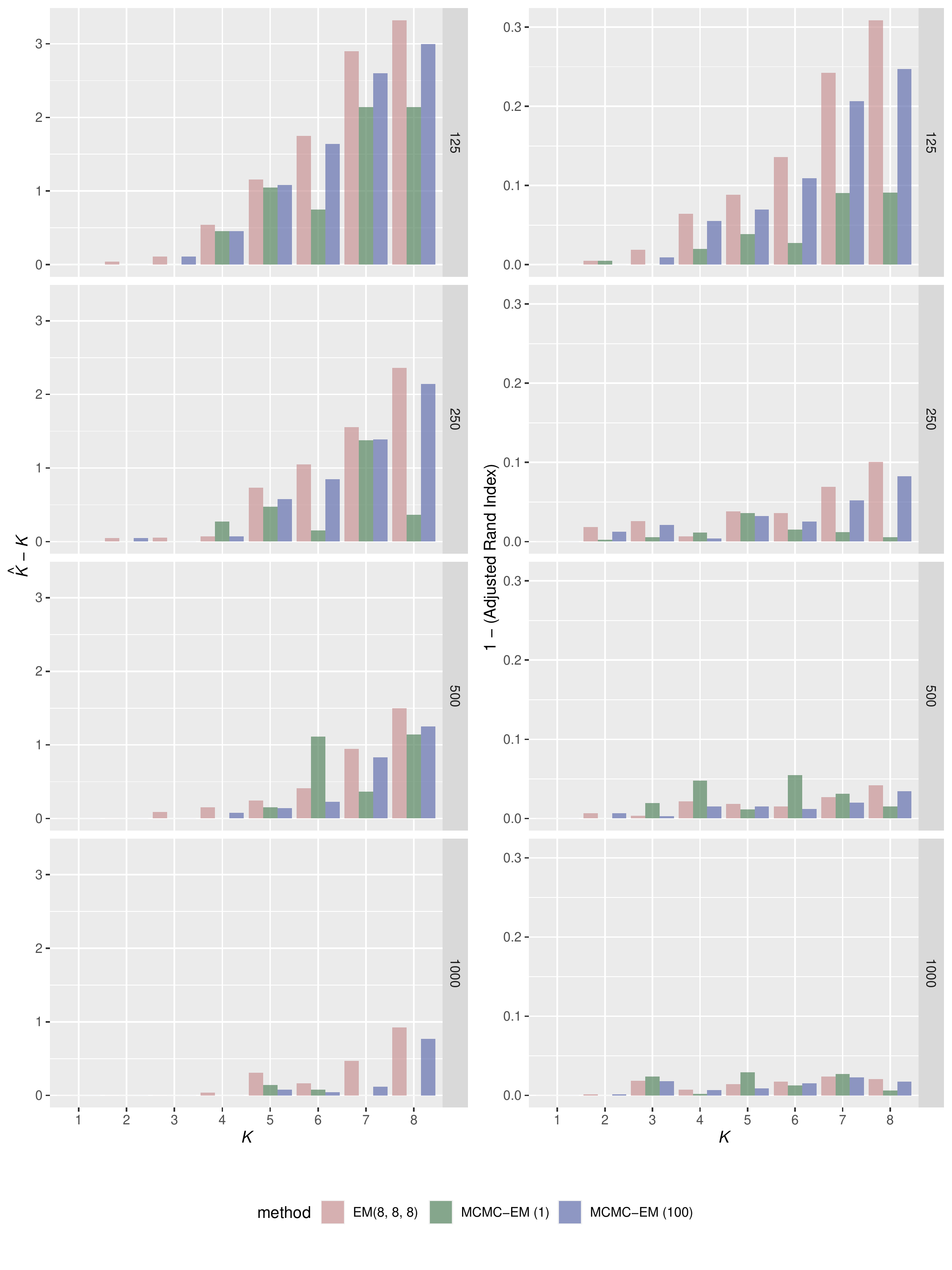}
\caption{Detailed results for our three best performing methods in the simulation study of Section \ref{sec:sim}. Left: mean difference of the estimated number of clusters ($\widehat K$) from the corresponding true value ($K$). Right: mean values of $1-\mathrm{ARI}$  between the model corresponding to the estimated number of clusters and the true partition. }
\label{fig:sim_detailed}
\end{figure}

Figure \ref{fig:sim_versus_p} displays a summary regarding the clustering accuracy versus the number of covariates (including constant) in the simulated datasets of Section \ref{sec:sim}, that is, $P\in\{2,3,4\}$ for EM(8, 8, 8), MCMC-EM (1) and MCMC-EM c(100). We observe that for both indices (estimation of the number of clusters and partition agreement as measured by the adjusted Rand index), the clustering accuracy tends to decrease as the number of covariates increase. This effect is clearly illustrated in the smaller sample  sizes $(n)$. However, as $n$ gets larger the impact of the number of covariates becomes smaller. 

Figure \ref{fig:sim_detailed} displays the difference between the estimated and true value of the numbers of clusters  $(\hat K - K)$ (left) and the adjusted Rand index (right) for each value of $K$ (horizontal axis), stratified for all different values of sample size ($n\in\{125,250,500,1000\}$).  It is evident that as the number of clusters increases, overestimates of the number of clusters occur more often and this effect is more severe for small sample sizes. Once again we note that the MCMC sampler with small prior variance (MCMC-EM (1)) outperforms the remaining implementations.

Next we are concerned with the accuracy of point estimates $\widehat\beta_{kjp}$, that is, the coefficient value at the last iteration for the EM implementation and the estimate of the posterior mean after reordering the MCMC output in order to deal with label switching for the MCMC implementation. We generated 100 synthetic datasets with $n = 250$ observations and 100 datasets with $n=500$, considering $K=4$ clusters, $J + 1 = 6$ multinomial categories and  $P=3$ covariates (including constant term). In each case, the number of components $K$ was set equal to the true number of clusters, that is, $K=4$. All other parameters were generated as described in Section \ref{sec:sim}. 
In order to meaningfully compare $\beta_{kjp}$ with the corresponding point estimates $\widehat\beta_{kjp}$ (arising either from EM(8, 8, 8) or from MCMC-EM (100)), we relabelled the resulting point estimates by applying the ECR algorithm \citep{Papastamoulis:10}, by considering that the pivot allocation vector (required in the ECR algorithm) is set to the true partition. This procedure ensures the labelling between $\beta_{kjp}$ and $\widehat\beta_{kjp}$ (for $j = 1,\ldots,J$; $p=1,\ldots,P$)  is consistent for all $k=1,\ldots,K$, and it has no impact at the quality of the estimates themselves. 

\begin{figure}[p]
\centering
\includegraphics[scale=0.55]{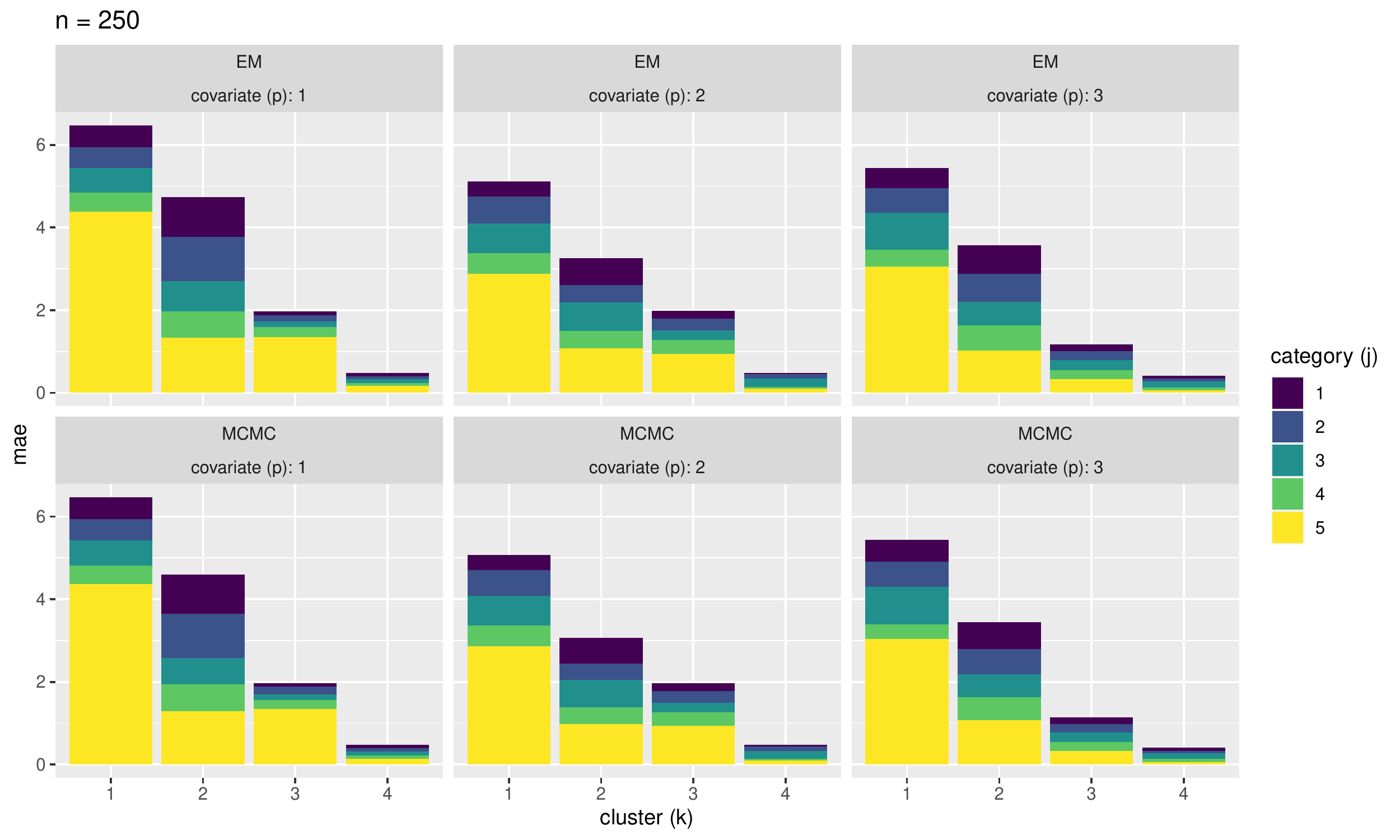}\\
\includegraphics[scale=0.55]{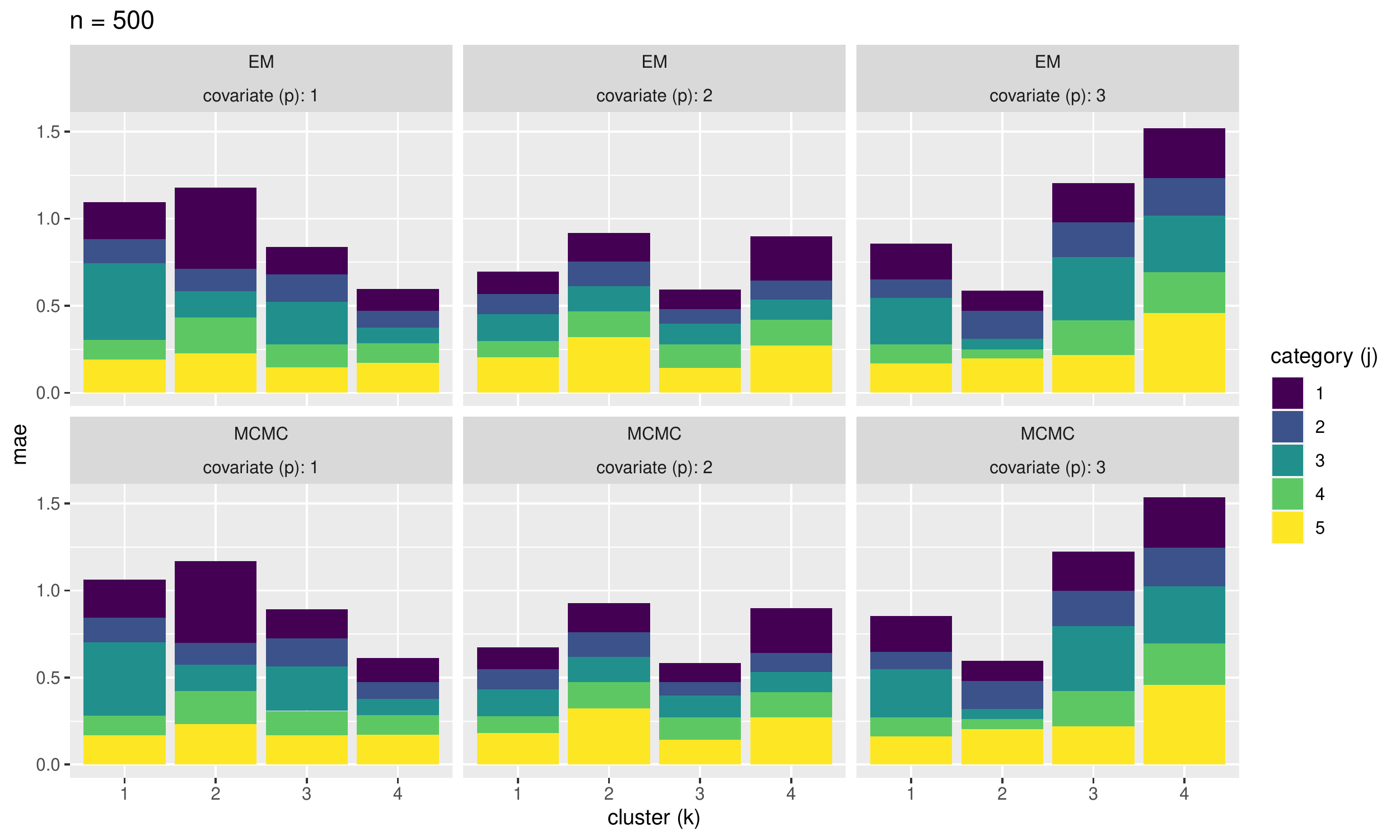}
\caption{Estimated mean absolute error (mae) of parameter estimates $\widehat{\beta}_{kjp}$ considering 100 synthetic datasets with $K=4$ clusters, $P=3$ covariates (including constant term) and $J+1 = 6$ multinomial categories, when the sample size is equal to $n = 250$ (top) and $n=500$ (bottom).}
\label{fig:sim_mae}
\end{figure}

Figure \ref{fig:sim_mae} displays the estimated Mean Absolute Error (MAE) between the point estimate $\widehat\beta_{kjp}$ and the corresponding true value $\beta_{kjp}$, for $k = 1,\ldots,K$, $j = 1,\ldots,J$ and $p=1,\ldots,P$.  As expected, the estimated MAEs are smaller on average as the sample ($n$) increases. Observe that when $n = 250$ the MAEs become smaller as $k$ increases. This is of course due to the fact that in our simulation study (see Section \ref{sec:sim}) the mixing proportions are generated according to $\pi_k \propto k$, for $k=1,\ldots,K$. Thus, in our 4-cluster scenario we have that $\pi_1 =0.1$, $\pi_2 =0.2$, $\pi_3 =0.3$ and $\pi_4 =0.4$. Naturally, the point estimates are less accurate in cases of very small clusters, a behaviour which is vividly illustrated when $n = 250$. However, when $n = 500$ we do not spot any systematic pattern in the resulting MAEs across clusters.

\subsection{Further simulations and comparison with {\tt flexmix}}\label{sec:flexmix}

In this Section we compare the proposed methods with the popular {\tt R}  package {\tt flexmix}  \citep{flexmix1, flexmix2, flexmix3} and we also explore the impact of the average number of multinomial replicates $\bar s = \sum_{i=1}^{n}S_i/n$. In the main simulation study of Section \ref{sec:sim}, the number of multinomial replicates ($S_i$) per observation is drawn from a negative binomial distribution $\mathcal{NB}(r, p)$ with number of trials equal to $r = 20$ and $p = 0.025$. This yields a potentially large of multinomial replicates: the average number is equal to 781. In order to assess the sensitivity of our results to the number of multinomial replicates we consider that $r$ varies in the set $\{2.5, 5, 10, 20\}$ and we use the $\mathcal{NB}(r, 0.025)$ distribution to simulate $S_i$ (in case that the generated number is 0 we set it to 1). It follows that the average value of $S_i$ is approximately 100, 200, 400 and 800, respectively (more precisely, the average values in our simulated datasets are 97.5, 195, 390 and 781, respectively). 

 For this task we considered the following simulation scenario.  In all cases the number of multinomial categories was set equal to $J+1 = 6$. The number of covariates (including constant term) was set equal to $P = 3$. The true values of the regression coefficients are generated as in \ref{sec:sim}. The number of clusters varied between $1\leqslant K \leqslant 5$.  We considered sample sizes of $n=250$ and $n = 500$ observations. We generated 5 synthetic datasets for each unique combination of $K$ (number of clusters), level of the average number of multinomial replicates $\bar s$ and sample size ($n$), resulting to $200$ datasets in total.

Next, we fitted mixtures of multinomial logistic regressions considering the proposed methodology, as well as the EM implementation in the {\tt R} package {\tt flexmix}. The configuration of the EM algorithm for our proposed method was EM(8, 8, 8), that is, an EM algorithm with 8 splits, 8 shakes and 8 random starts of small EM for each possible value of the number of components. Then we have used the selected model in order to initialize a run of a Bayesian overfitting mixture model. The prior variance of the regression coefficients is equal to $\nu = 100$ (vague prior distribution) using 8 chains in total. Finally, we have run {\tt flexmix} repeatedly considering 24 random starts, for each possible value of the number of components, using the {\tt stepFlexmix} function with options: {\tt k=1:Kmax} ({where \tt Kmax} denotes the maximum number of components -- see next paragraph), {\tt nrep = 24},  {\tt control = list(minprior = 0)} and {\tt model = flexmix::FLXMRmultinom()}.

\begin{figure}[ht]
\centering
\includegraphics[scale=0.6]{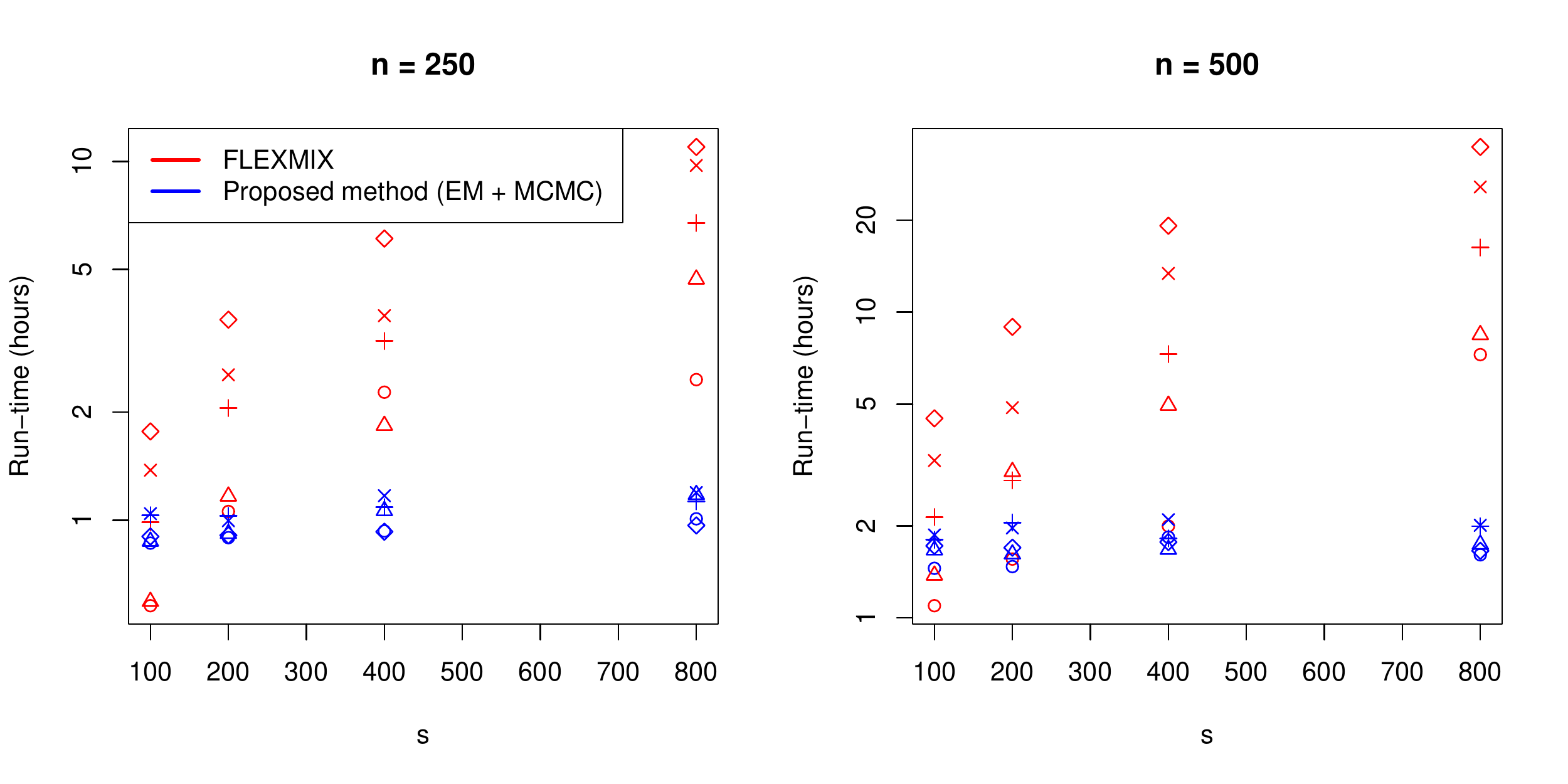}
\caption{Run-time (averaged across 5 runs) comparison of the proposed methods (EM(8, 8, 8) and MCMC)  against  the EM algorithm of {\tt flexmix} under 24 randomly selected starting values. The $x$-axis indicates the average value of multinomial replicates (rounded to the nearest multiple of 100). The $y$-axis is on log-scale. All times correspond to computations in one single CPU core. A different symbold is used to denote the maximum number of fitted mixture components: $\circ (3)$, $\triangle (4)$, $+ (5)$, $\times (6)$, $\diamond (7)$.}
\label{fig:time}
\end{figure}

In all cases, the maximum number of mixture components was set equal to $K_{\max} = K + 2$, where $K$ denotes the true value of the number of clusters for each case. This upper bound of the number of components is much more informative  regarding the number of clusters than the one in our main simulation study of Section \ref{sec:sim} (where $K_{\max} = 20$). The reason for choosing such a value was mainly to speed-up the computing time in {\tt flexmix}, where in most cases is significantly elevated (see Figure \ref{fig:time}) compared to the proposed implementation. 

\begin{figure}[ht]
\centering
\includegraphics[scale=0.6]{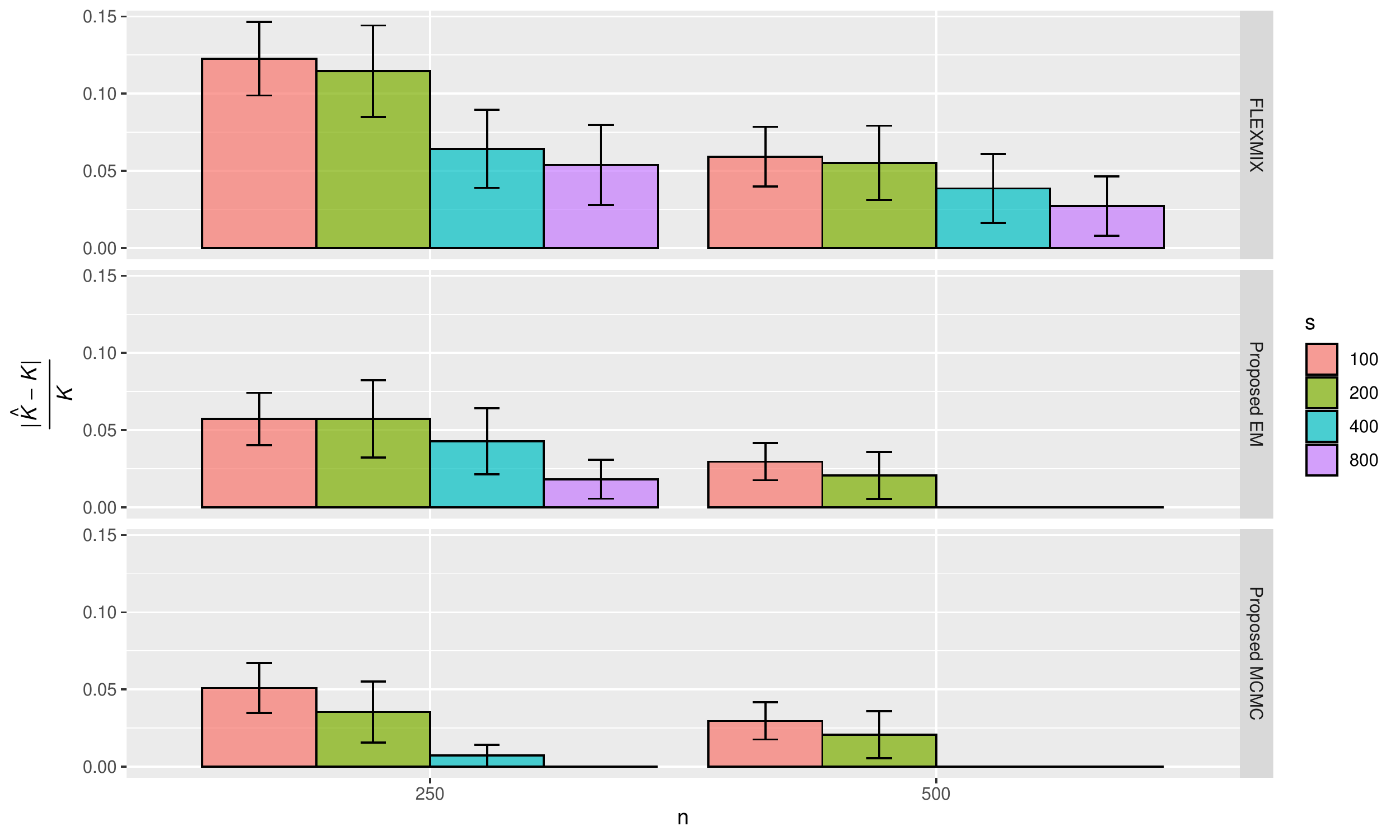}\\
\includegraphics[scale=0.6]{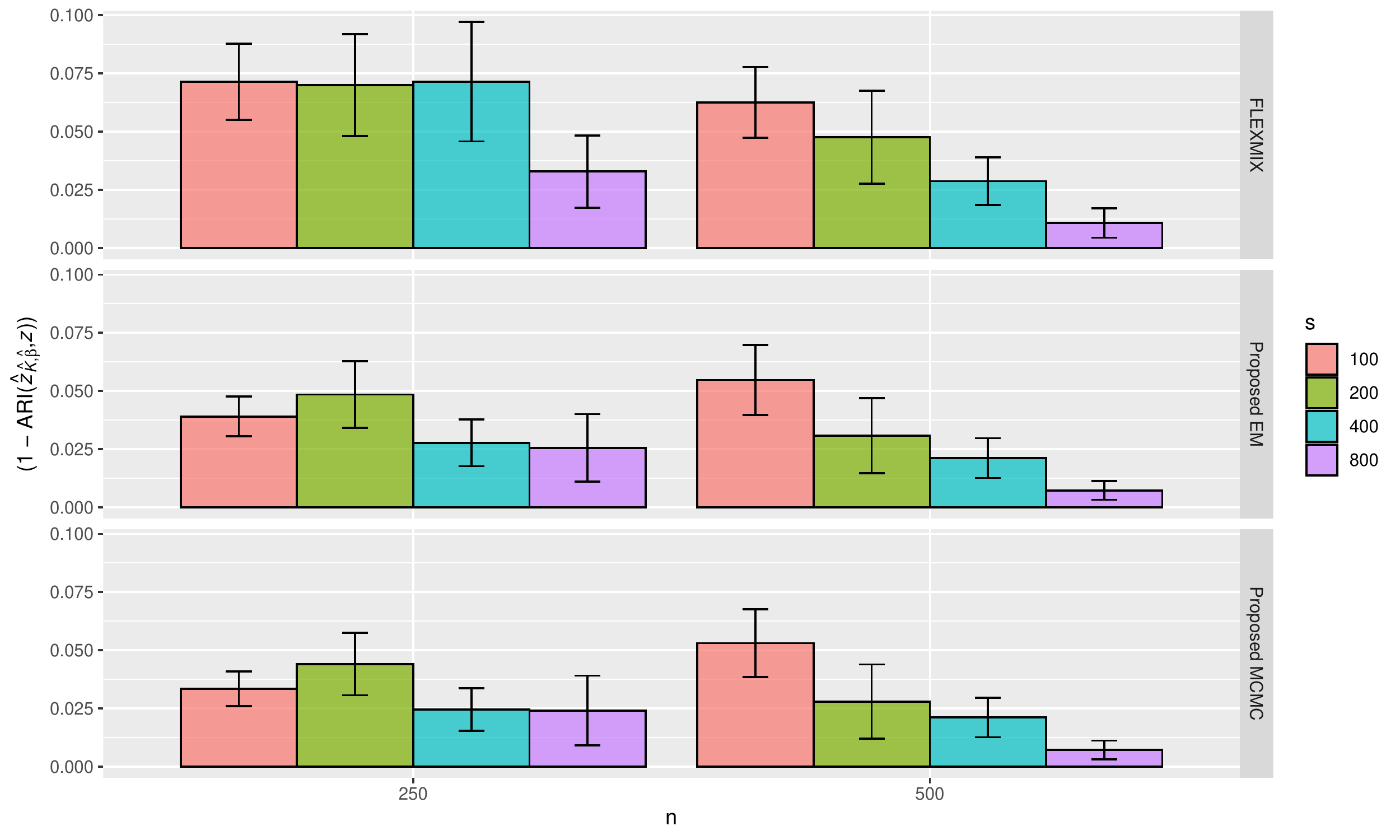}
\caption{Comparison of the proposed methods (EM and MCMC)  against  {\tt flexmix}. The average value of multinomial replicates is shown in the legend ($s$), after rounding to the nearest multiple of 100.}
\label{fig:flexmix}
\end{figure}

The resulting estimates are summarized in Figure \ref{fig:flexmix}. We conclude that in all cases the estimation becomes more challenging when the average number of multinomial replicates is smaller. Note also that our proposed methods (EM and MCMC) are better than {\tt flexmix} in terms of estimation of the number of clusters as well as classification accuracy under the adjusted Rand index.

\bibliography{sn-bibliography}

\end{document}